\documentclass[prd,aps,twocolumn,floatfix]{revtex4-2}

\usepackage{mathtools,amssymb,mathrsfs,bm}
\usepackage{booktabs,array,longtable}
\usepackage{xcolor}
\usepackage[colorlinks=true,linkcolor=blue!55!black,
  citecolor=blue!55!black,urlcolor=blue!55!black]{hyperref}
\usepackage{tikz}
\usepackage{pgfplots}
\pgfplotsset{compat=1.17}
\usetikzlibrary{arrows.meta,calc,patterns,shadings,
  decorations.pathreplacing,decorations.pathmorphing,
  decorations.markings,shapes.geometric,3d,perspective,positioning}

\newcommand{\I}{\mathrm{i}}
\newcommand{\dd}{\mathrm{d}}
\newcommand{\e}{\mathrm{e}}
\newcommand{\scri}{\mathscr{I}^{+}}
\newcommand{\hor}{\mathscr{H}^{+}}
\newcommand{\Id}{\mathbf I}
\newcommand{\Zero}{\mathbf 0}

\begin{document}
\title{Gravitational lensing of gravitational waves by a Kerr black hole}
\author{Jian-hua He}
\thanks{Corresponding author: \href{mailto:hejianhua@nju.edu.cn}{hejianhua@nju.edu.cn}}
\affiliation{School of Astronomy and Space Science, Nanjing University, Nanjing 210023, Jiangsu, P.~R.~China}
\affiliation{Key Laboratory of Modern Astronomy and Astrophysics (Nanjing University), Ministry of Education, Nanjing 210023, Jiangsu, P.~R.~China}

\begin{abstract} 
We simulate the gravitational lensing of gravitational waves (GWs) by a Kerr black hole using a $(1+1)$-dimensional time-domain mode-sum approach. To enable efficient explicit time evolution, we implement a localized hyperboloidal foliation combined with a static field redefinition, ensuring bounded characteristic speeds across an extended, uniform tortoise-coordinate grid. Incident GW initial data are constructed by projecting a plane-wave Weyl tensor onto the Kinnersley tetrad. Through local spin-dyad rotations and Wigner $D$-matrix transformations, our framework accommodates both idealized wave packets and realistic binary coalescence signals at arbitrary incidence and polarization angles. To accurately resolve the lensed waveforms, we evolve over $68,000$ coupled $(j,m)$ modes. Our results reveal that a Kerr black hole acts as a diffraction-limited thick lens, focusing GWs into a forward beam. For idealized wave packets, the primary prompt pulse exhibits a spin-independent transverse Gaussian profile, whereas the subsequent lensing coda is sensitive to the black hole's spin. For binary coalescence signals, strong-field lensing substantially modifies the post-merger ringdown. However, the lensed inspiral and merger phases in Kerr and Schwarzschild backgrounds are virtually identical, yielding waveform mismatches below $2.2\times10^{-5}$ across various inclinations and polarizations. These findings demonstrate that non-spinning Schwarzschild models provide an accurate and computationally economical surrogate to search for strong-field lensing signatures with current ground-based detectors.
\end{abstract}

\date{}
\maketitle
\section{Introduction}

The direct detection of gravitational waves (GWs) from coalescing compact binaries has inaugurated a new era of observational strong-field gravity. As the global ground-based detector network, comprising Advanced LIGO~\cite{LIGOScientific:2014pky}, Advanced Virgo~\cite{TheVirgo:2014hva}, and KAGRA~\cite{KAGRA:2020tym}, continues to accumulate transient events, and as next-generation facilities such as the Einstein Telescope~\cite{2010CQGra..27s4002P}, Cosmic Explorer~\cite{Evans:2021gyd}, and the space-borne LISA~\cite{2017arXiv170200786A} approach realization, gravitational lensing of GWs has emerged as a compelling frontier in multi-messenger astrophysics~\cite{Ezquiaga:2020gdt,Oguri:2019fix}. In contrast to electromagnetic signals, which are susceptible to absorption and scattering by interstellar dust and gas, GWs interact extremely weakly with ordinary matter and thus propagate virtually unimpeded across the cosmos. However, their propagation and waveforms can be significantly modulated by the strongly curved spacetime in the vicinity of ultra-compact objects~\cite{Yin2024PRL}. Consequently, GW lensing provides a uniquely clean probe to uncover and characterize the vast galactic population of isolated, non-accreting stellar-mass black holes (BHs)~\cite{siciliaBlackHoleMass2022}, which emit no electromagnetic radiation and thus remain invisible to traditional astronomy~\cite{Agol:2001hb,Agol:2002microlensing}.

To date, most theoretical studies of GW lensing have relied on the weak-field, thin-lens approximation~\cite{Suyama:2005mx,PhysRevLett.80.1138,PhysRevD.98.083005,Ruffa_1999,DePaolis:2002tw,Liao:2019aqq,Dai:2018enj,PhysRevD.90.062003,Christian,Meena:2019ate,2021PhRvD.103j4055W,Tambalo:2022plm,PhysRevD.111.103539}. Within this framework, the lens is idealized as a planar point-mass deflector, with wave propagation governed by the Fresnel--Kirchhoff diffraction integral. While computationally tractable, the thin-lens paradigm fundamentally omits essential strong-field gravitational phenomena, including backscattering off the photon sphere and curvature-induced diffraction tails~\cite{PhysRevD.5.2419,1975weoc.book.....F}. For stellar-mass BH lenses whose event horizon radii ($\sim 10$--$100\text{ km}$) are comparable to the wavelengths in the ground-based sensitivity band ($20$--$5000$~Hz), geometric optics breaks down and wave-optics effects dominate~\cite{Meena:2019ate,Hongsheng:2018ibg}. Observational searches across the LIGO-Virgo-KAGRA O3 and O4a observing runs based on the point-mass model have found no compelling evidence for lensing signatures predicted by weak-field templates~\cite{LIGOScientific:2021izm,LIGOScientific:2023bwz,LIGOScientific:2025cwb}, underscoring the urgent need for realistic strong-field general-relativistic calculations.

When the incident GW wavelength is comparable to the gravitational radius of the lens, an exact treatment in curved spacetime becomes necessary. In the frequency domain, the standard framework is partial-wave scattering theory~\cite{andersson_scattering_2001,Futterman:1988ni}, wherein the Teukolsky master equation~\cite{1973ApJ...185..635T}  is decomposed into radial and angular differential equations, and the lensed field is reconstructed through a sum over angular multipoles. However, this approach suffers from a long-standing divergence problem: the long-range $1/r$ Newtonian potential induces logarithmic, Coulomb-like phase shifts, $\delta_\ell\sim -2M\omega\ln\ell$. Consequently, the partial-wave coefficients of the asymptotic field grow with $\ell$ rather than decay, causing the series to diverge at future null infinity $\mathscr{I}^+$, in direct analogy with Coulomb scattering in quantum mechanics~\cite{marquez_divergence_1972}. Although summation regularization techniques, such as Ces\`{a}ro or Poisson--Abel resummation, can restore convergence away from the optical axis, they break down in the forward direction ($\theta=0$), thereby restricting existing analyses to off-axis configurations~\cite{PhysRevD.110.044054,6h6r-46cd,Dolan:2008scattering,Crispino:2009xt,9bxg-sqn5}. While semiclassical approximations in the eikonal limit~\cite{Nambu:2015aea} yield closed-form expressions, they remain valid only for high-$\ell$ modes.

An alternative approach is to place both the source and the observer at finite coordinate radii~\cite{Kanai:2013rga,PhysRevD.110.124011,Pijnenburg:2022pug,xsxj-9vdw}, where the partial-wave series converges. At any finite $r$, modes with $\ell\gtrsim\omega r$ tunnel through the centrifugal barrier and are thus exponentially suppressed, which effectively cuts off the sum over $\ell$. Consequently, \emph{any} partial-wave series converges when evaluated at finite radii. This convergence, however, does not resolve the underlying problem. In the scattering framework, the $r\to\infty$ limit does not commute with the summation over $\ell$. Asymptotic boundary conditions require that $r\to\infty$ be taken before the sum, which yields a divergent series. Moreover, even if the sum is performed first at finite $r$, the subsequent limit  $r\to\infty$ still diverges because of the long-range, Coulomb-like logarithmic phase shifts.

Direct time-domain evolution provides a natural avenue for avoiding these frequency-domain summation difficulties while accommodating finite-duration wave packets. 
This is supported by mathematical results on linear and nonlinear black hole stability. Gravitational perturbations of Schwarzschild satisfy boundedness and decay estimates modulo gauge and stationary parameter perturbations~\cite{Dafermos:2016uzj}. For subextremal Kerr ($|a|<M$), mode stability~\cite{Whiting1989ModeStability} has been complemented by boundedness and decay for the Teukolsky equation~\cite{shlapentokh-rothmanBoundednessDecayTeukolsky2020,ShlapentokhRothmanTeixeira2023Physical} and by metric-level linear stability~\cite{HaefnerHintzVasy2025FullKerr}. Nonlinear stability has been established for Schwarzschild under symmetry or suitably restricted data conditions~\cite{KlainermanSzeftel2020Schwarzschild,Dafermos:2021cbw}, for slowly rotating Kerr~\cite{KlainermanSzeftel2023Kerr,GiorgiKlainermanSzeftel2024Kerr,Shen2023GCM}, and, in a recent preprint, for every fixed subextremal Kerr background~\cite{Hintz2026NonlinearKerr}. For the sufficiently regular, small, constraint-satisfying data admitted by these theorems, nonlinear interactions remain controlled and the exterior settles to a nearby Kerr metric.

These nonlinear stability results have important practical implications. They suggest that nonlinear effects remain bounded rather than growing without bound. Hyperbolic energy estimates can constrain their deviations from linear theory. For an astrophysical observer with a finite observation window, it can be shown that nonlinear corrections scale as second order, $O(\varepsilon^2)$, in the perturbation amplitude $\varepsilon$~\cite{Ringstrom2009Cauchy,AnderssonBackdahlBlueMa2022RadiationGauge}. For an incident gravitational wave packet with a realistically small perturbation amplitude, this $O(\varepsilon^2)$ suppression ensures that the first-order linear response is the dominant physical contribution.  This behavior distinguishes GW lensing from the nonlinear regimes encountered during black hole ringdown~\cite{NonlinearRingdown1,NonlinearRingdown2}. In binary mergers, the violent coalescence generates strong dynamical nonlinearities that persist into the early ringdown, producing appreciable deviations from linear mode predictions and complicating ringdown data analysis~\cite{NonlinearRingdown1,NonlinearRingdown2}. By contrast, for gravitational waves lensed by a black hole, the small initial wave amplitude preserves a clean hierarchy between the leading-order linear waveform and subdominant nonlinear corrections throughout the transit. This inherent linearity offers a cleaner setting for testing black hole linear perturbation theory, and we therefore restrict our analysis to linear perturbations.

In the time domain, even linear waves propagating through the curved spacetime of a strong gravitational field can exhibit intricate behaviors, including trajectory deflection, gravitational focusing, the excitation of quasinormal modes (QNMs), and late-time tails associated with the breakdown of Huygens' principle~\cite{vishveshwara_scattering_1970,PhysRevD.5.2419,Andersson,PhysRevD.84.104002,PhysRevD.74.084028,He2022PRD,1975weoc.book.....F}. The first time-domain three-dimensional (3D) calculations of such phenomena were presented in our previous work~\cite{He2022PRD,Yin2024PRL}, in which linear metric perturbations were evolved directly on a Schwarzschild background. However, such direct 3D simulations are computationally expensive, requiring millions of CPU hours. To overcome this limitation, we subsequently developed a time-domain mode-sum method within a $(1+1)$-dimensional radial framework~\cite{gg33-r64q}.
This formulation retains the full 3D dynamics while improving computational efficiency by more than three orders of magnitude. In the present paper, we extend this $(1+1)$-dimensional framework from Schwarzschild to spinning Kerr black holes.

However, extending this $(1+1)$-dimensional approach to Kerr spacetime is non-trivial. Unlike the $(2+1)$ time-domain Teukolsky system~\cite{PhysRevD.56.3395}, the direct evolution of the coupled $(1+1)$ Teukolsky system under standard Boyer--Lindquist (BL) slicing suffers from numerical instabilities.  Similar spurious growth has also been reported in double-null formulations~\cite{Long:2021ufh}. This instability arises because the first-order time-derivative term in the BL-foliated Teukolsky equation develops a locally negative damping coefficient in the exterior region. This local antidamping acts as an artificial energy pump, exponentially amplifying high-frequency truncation errors and ultimately destabilizing the numerical evolution. Consequently, stable, long-term evolutions of the $(1+1)$-dimensional Teukolsky system cannot be achieved directly in BL slicing or on a uniform tortoise grid.

A numerically stable evolution can instead be obtained by transforming from BL slicing to a hyperboloidal foliation. Within this framework, the coefficient of the first-order time-derivative term remains strictly positive, introducing dissipation that guarantees numerical stability~\cite{zenginoglu_hyperboloidal_2008,zenginoglu_hyperboloidal_2011}. Typically, conventional hyperboloidal implementations rely on conformal radial compactification, which maps future null infinity $\mathscr{I}^+$ to a finite coordinate boundary~\cite{zenginogluSpacelikeMatchingNull2009,raczNumericalInvestigationLatetime2011,ansorgSpectralDecompositionBlackhole2016,csukasNumericalInvestigationDynamics2019,Markakis:2023pfh}. This compactification ensures that the characteristic speeds at $\mathscr{I}^+$ are purely outgoing and finite. This property is of paramount importance. Radiation reaches $\mathscr{I}^+$ within a finite coordinate time, and the bounded characteristic speeds prevent Courant--Friedrichs--Lewy (CFL) time-step breakdown in explicit time-stepping schemes.

Despite these advantages, a compactified radial coordinate yields an increasingly non-uniform physical resolution near $\mathscr I^+$. This degrades the radial resolution in the outer wave zone. In the lensing process, however, it is important to accurately resolve the spatial structure of the lensed wave, such as its transverse profile. A uniform tortoise coordinate is better suited to this purpose, as it maintains consistent spatial resolution for tracking an incident wave packet across an extended radial domain. 

Nevertheless, combining a hyperboloidal slicing with an extended uniform tortoise grid introduces a complication. The maximum characteristic coordinate speed grows without bound as the domain expands toward either the inner or outer boundary. As a consequence, explicit time-integration schemes such as the Runge--Kutta (RK) method face an increasingly stringent CFL stability constraint, rendering long-term evolutions computationally prohibitive. 

To address this dilemma, we develop a localized hyperboloidal foliation implemented directly on an extended, uniform tortoise grid. By smoothly capping the slope of the hyperboloidal height function in the asymptotic regions, our framework keeps the coordinate characteristic speeds bounded throughout the computational domain, thereby avoiding CFL time-step collapse while maintaining uniform spatial resolution across the wave zone. Because capping the height function leaves a small residual term that could otherwise trigger antidamping at large distances, we combine this slicing with an invertible, static field redefinition that renders the first-order time-derivative coefficients non-positive everywhere, eliminating the antidamping behavior. Together, these techniques ensure both high computational efficiency and robust numerical stability for long-term time-domain simulations of GW lensing by Kerr black holes.

This paper is organized as follows. In Sec.~\ref{sec:oneplusoneTeu}, we decompose the spin-weight $s=-2$ Teukolsky equation onto spin-weighted spherical harmonics, retaining the complete rotation-induced multipole coupling. We then identify a local antidamping mechanism in the Boyer--Lindquist representation that drives numerical error amplification. In Sec.~\ref{sec:hyperboloidal-coordinates}, we eliminate this instability by combining a localized hyperboloidal height function with an analytical field redefinition, rendering the first-order time-derivative coefficient non-positive everywhere. In Sec.~\ref{sec:initial-data}, we construct plane-wave initial data via the Weyl tensor in an intrinsic wave frame, project it onto the Kinnersley null tetrad through an intermediate ZAMO frame, and expand it in a modal basis using active Euler rotations and Wigner $D$-matrices. This formulation naturally accommodates arbitrary incidence and polarization angles $(\alpha,\iota,\gamma_{\rm E})$. In Sec.~\ref{sec:numerics}, we discretize the first-order-in-time system using fourth-order finite differences on a uniform tortoise grid, stabilized by sixth-order Kreiss--Oliger dissipation, integrated with low-storage RK4, and parallelized via MPI across decoupled azimuthal sectors. In Sec.~\ref{sec:numerical-results}, we simulate both Hanning-windowed and binary coalescence wave signals lensed by a rapidly rotating Kerr black hole. Finally, in Sec.~\ref{sec:discussion}, we discuss and conclude this work.

\section{\texorpdfstring{$(1+1)$}{(1+1)} Formalism of the Teukolsky Master Equation}
\label{sec:oneplusoneTeu}
\subsection{Teukolsky Master Equation and Notation}

In Boyer--Lindquist (BL) coordinates $(t,r,\theta,\phi)$, the vacuum Teukolsky master equation
in Kerr spacetime reads~\cite{1973ApJ...185..635T}
\begin{align}
{}&\Bigl[\frac{(r^2+a^2)^2}{\Delta_r}-a^2\sin^2\theta\Bigr]
     \partial_t^2\psi^{(s)}                                         \notag\\
   &+\frac{4Mar}{\Delta_r}\partial_t\partial_\phi\psi^{(s)}
     +\Bigl(\frac{a^2}{\Delta_r}-\frac{1}{\sin^2\theta}\Bigr)
     \partial_\phi^2\psi^{(s)}                                      \notag\\
   &-\Delta_r^{-s}\partial_r\!\bigl(\Delta_r^{s+1}
     \partial_r\psi^{(s)}\bigr)                                     \notag\\
   &-\frac{1}{\sin\theta}\partial_\theta\!\bigl(
     \sin\theta\,\partial_\theta\psi^{(s)}\bigr)
     +(s^2\cot^2\theta-s)\psi^{(s)}                                 \notag\\
   &-2s\Bigl[\frac{M(r^2-a^2)}{\Delta_r}-r-\I a\cos\theta\Bigr]
     \partial_t\psi^{(s)}                                           \notag\\
   &-2s\Bigl[\frac{a(r-M)}{\Delta_r}
     +\frac{\I\cos\theta}{\sin^2\theta}\Bigr]
     \partial_\phi\psi^{(s)}=0,
\label{eq:teukolsky-master}
\end{align}
where
\begin{equation}
 \Delta_r=r^2-2Mr+a^2=(r-r_+)(r-r_-),
\end{equation}
with the outer and inner horizon radii
\begin{equation}
 r_\pm=M\pm\sqrt{M^2-a^2}.
\end{equation}
We adopt geometric units with $G=c=1$, where $M$ is the black hole mass, $a=J_{\rm BH}/M$
is the angular momentum per unit mass, with $|a|<M$. The spin weight $s=0,\pm1,\pm2$
covers scalar, electromagnetic, and gravitational perturbations, respectively.

\subsection{Expansions in spin-weighted spherical harmonics}
For each fixed spin weight $s$, the spin-weighted spherical harmonics ${}_sY_{jm}(\theta,\phi)$ form a complete orthonormal basis on the two-sphere, so that any spin-$s$ field can be expanded as~\cite{Markakis:2023pfh,PhysRevD.61.024026,PhysRevD.95.104033,zenginogluSpacelikeMatchingNull2009,raczNumericalInvestigationLatetime2011,harmsNumericalSolutionTeukolsky2013,csukasNumericalInvestigationDynamics2019,fieldGPUAcceleratedMixedPrecisionWENO2021}
\begin{equation} 
\psi^{(s)}(t,r,\theta,\phi) =\sum_{j=|s|}^{\infty}\sum_{m=-j}^{j} R_{jm}^{(s)}(t,r)\,e^{\I m\phi}\,{}_sy_{jm}(\theta)\,, \label{eq:bl-spherical-expansion-a} \end{equation}
where ${}_sY_{jm}(\theta,\phi)=e^{\I m\phi}\,{}_sy_{jm}(\theta)$ separates the polar and azimuthal dependencies.
Unlike in the Schwarzschild background, the explicit $\cos\theta$ and $\sin^{2}\theta$ factors in the Kerr Teukolsky master equation couple different multipole modes $j$ within a given azimuthal sector $m$. However, because the Kerr background is axisymmetric, distinct $m$ sectors evolve independently, restricting the mode coupling entirely to the $j$ index. For a given $m$, the minimum allowable angular index is
\begin{equation} 
j_{\min}=\max(|s|,|m|),\qquad j=j_{\min},\,j_{\min}+1,\,\ldots. \label{eq:jmin-definition} 
\end{equation}
Since the constraints $|m|\le j$ and $|s|\le j$ must hold simultaneously, the summation order over the lattice of $(j,m)$ pairs can be interchanged as
\begin{equation}
 \sum_{j=|s|}^{\infty}\sum_{m=-j}^{j} =\sum_{m=-\infty}^{\infty}\ \sum_{j=\max(|s|,|m|)}^{\infty}. \label{eq:swsh-sum-order} 
 \end{equation} 
This reordering isolates each decoupled $m$ sector while allowing $j$ to index the coupled modes within it, giving
 \begin{equation} 
 \psi^{(s)}(t,r,\theta,\phi) =\sum_{m=-\infty}^{\infty}e^{\I m\phi} \sum_{j=j_{\min}}^{\infty} R_{jm}^{(s)}(t,r)\,{}_sy_{jm}(\theta). \label{eq:bl-spherical-expansion} 
 \end{equation}
Each $m$ sector can then be evolved independently as a $(1+1)$-dimensional system in $(t,r)$ for the coupled tower of radial functions $\{R_{jm}^{(s)}\}_{j\ge j_{\min}}$. In numerical simulations, the infinite $j$ tower is truncated at a finite cutoff $J$, with $m$ restricted to $-J\le m \le J$.

\subsection{The coupled \texorpdfstring{$(1+1)$}{(1+1)}-dimensional PDE system}
We begin by formulating the system in BL coordinates. For a fixed azimuthal mode number $m$, the decomposed field is denoted by
\begin{equation}
 \psi_m^{(s)}(t,r,\theta,\phi)
 =e^{\I m\phi}\sum_{j=j_{\min}}^\infty
 R_j(t,r){}_sy_{jm}(\theta)\,.
\label{eq:fixed-m-bl-expansion}
\end{equation}
Substituting Eq.~\eqref{eq:fixed-m-bl-expansion} into Eq.~\eqref{eq:teukolsky-master} and factoring out $e^{\I m\phi}$, we obtain
\begin{align}
\sum_{j=j_{\min}}^\infty\Bigl\{
 &\left[\mathcal A_0-a^2\sin^2\theta\right]\partial_t^2R_j
 +\left[\mathcal B_0+2\I sa\cos\theta\right]\partial_tR_j             \notag\\
 &+\left[\mathcal D_r+\mathcal C_0+\lambda_j\right]R_j
 \Bigr\}{}_sy_{jm}(\theta)=0,
\label{eq:bl-before-projection}
\end{align}
where the radial coefficients $\mathcal A_0$, $\mathcal B_0$, $\mathcal C_0$ and the radial differential operator $\mathcal D_r$ are defined by
\begin{align}
\begin{aligned}
 \mathcal A_0(r)&\equiv\frac{(r^2+a^2)^2}{\Delta_r},                  \\
 \mathcal B_0(r)&\equiv
 \frac{4\I mMar-2sM(r^2-a^2)}{\Delta_r}+2sr,                         \\
 \mathcal C_0(r)&\equiv
 \frac{-a^2m^2+2\I Mams-2\I amrs}{\Delta_r},                         \\
 \mathcal D_r&\equiv-\Delta_r^{-s}\partial_r
 \left(\Delta_r^{s+1}\partial_r\,\cdot\,\right)\,.
\end{aligned}
\label{eq:bl-radial-operators}
\end{align}
Here we have used the eigenvalue equation for ${}_sy_{jm}$,
\begin{equation}
\mathcal L_{sm} \,{}_sy_{jm}=\lambda_j\,{}_sy_{jm}\,,
\end{equation}
with the angular operator
\begin{equation}
 \mathcal L_{sm}\equiv
 -\frac{1}{\sin\theta}\partial_\theta
  (\sin\theta\,\partial_\theta)
 +\frac{(m+s\cos\theta)^2}{\sin^2\theta}-s\,,
\label{eq:pure-angular-operator}
\end{equation}
and eigenvalues
\begin{equation}
 \lambda_j=(j-s)(j+s+1).
\label{eq:lambda-def}
\end{equation}
To eliminate the explicit $\theta$ dependence, we project Eq.~\eqref{eq:bl-before-projection} onto ${}_sy_{\ell m}(\theta)$ by multiplying by $2\pi\,{}_sy_{\ell m}^{*}(\theta)\sin\theta$ and integrating over $\theta\in[0,\pi]$. This yields the coupled radial system
\begin{equation}
\begin{aligned}
\sum_{j=j_{\min}}^\infty\Bigl\{
 &\left[(\mathcal A_0-a^2)\delta_{\ell j}
       +a^2M^{(2)}_{\ell j}\right]\partial_t^2R_j                    \\
 &+\left[\mathcal B_0\delta_{\ell j}
       +2\I saM^{(1)}_{\ell j}\right]\partial_tR_j                   \\
 &+\delta_{\ell j}
  \left[\mathcal D_r+\mathcal C_0+\lambda_j\right]R_j
\Bigr\}=0,
\end{aligned}
\label{eq:bl-projected-system}
\end{equation}
where we have used the orthogonality and projection identities
\begin{align}
\begin{aligned}
 2\pi\int_0^\pi{}_sy_{\ell m}^{*}{}_sy_{jm}\sin\theta\,\dd\theta
 &=\delta_{\ell j},                                                    \\
 2\pi\int_0^\pi{}_sy_{\ell m}^{*}\cos\theta\,{}_sy_{jm}
       \sin\theta\,\dd\theta
 &=M^{(1)}_{\ell j},                                                    \\
 2\pi\int_0^\pi{}_sy_{\ell m}^{*}\sin^2\theta\,{}_sy_{jm}
       \sin\theta\,\dd\theta
 &=\delta_{\ell j}-M^{(2)}_{\ell j}\,.
\end{aligned}
\label{eq:projection-identities}
\end{align}
The matrices $M^{(1)}_{\ell j}$ and $M^{(2)}_{\ell j}$ parameterize the angular coupling for a fixed $m$, defined by
\begin{align}
\begin{aligned}
 M^{(1)}_{\ell j}
 &\equiv2\pi\int_0^\pi{}_sy_{\ell m}^{*}\cos\theta\,{}_sy_{jm}
                  \sin\theta\,\dd\theta,                              \\
 M^{(2)}_{\ell j}
 &\equiv2\pi\int_0^\pi{}_sy_{\ell m}^{*}\cos^2\theta\,{}_sy_{jm}
                  \sin\theta\,\dd\theta\,.
\end{aligned}
\label{eq:angular-coupling-matrices}
\end{align}

These coupling matrices can be expressed analytically using the recurrence relation for spin-weighted spherical harmonics,
\begin{equation}
 \cos\theta\,{}_sy_{jm}
 ={}_sC_{j+1}^m{}_sy_{j+1,m}
 +\beta_j{}_sy_{jm}
 +{}_sC_j^m{}_sy_{j-1,m}\,,
\label{eq:cos-recurrence}
\end{equation}
with coefficients given by
\begin{equation}
 {}_sC_j^m=
 \sqrt{\frac{(j^2-m^2)(j^2-s^2)}{j^2(4j^2-1)}},
 \quad
 \beta_j=-\frac{sm}{j(j+1)}\,.
\label{eq:C-beta}
\end{equation}
Here we adopt the convention that ${}_sC_0^m=\beta_0=0$, and all coefficients below $j_{\min}$ vanish.

Using Eq.~\eqref{eq:cos-recurrence}, the explicit elements of $M^{(1)}_{\ell j}$ read
\begin{equation}
 M^{(1)}_{\ell j}
 ={}_sC_{j+1}^m\delta_{\ell,j+1}
 +\beta_j\delta_{\ell j}
 +{}_sC_j^m\delta_{\ell,j-1}.
\label{eq:M1-elements}
\end{equation}
Applying the recurrence relation a second time yields the elements of $M^{(2)}_{\ell j}$,
\begin{equation}
\begin{aligned}
M^{(2)}_{\ell j}={}&
 {}_sC_{j+1}^m{}_sC_{j+2}^m\delta_{\ell,j+2}                           \\
&+{}_sC_{j+1}^m(\beta_j+\beta_{j+1})\delta_{\ell,j+1}                  \\
&+\bigl[({}_sC_j^m)^2+\beta_j^2+({}_sC_{j+1}^m)^2\bigr]\delta_{\ell j}\\
&+{}_sC_j^m(\beta_j+\beta_{j-1})\delta_{\ell,j-1}
 +{}_sC_j^m{}_sC_{j-1}^m\delta_{\ell,j-2}.
\end{aligned}
\label{eq:M2-elements}
\end{equation}
Setting $L=j_{\min}$, the matrices exhibit explicit banded structures
\begin{align}
\mathbf M^{(1)}&=
\begin{pmatrix}
\beta_L&{}_sC_{L+1}^m&0&0&\cdots\\
{}_sC_{L+1}^m&\beta_{L+1}&{}_sC_{L+2}^m&0&\cdots\\
0&{}_sC_{L+2}^m&\beta_{L+2}&{}_sC_{L+3}^m&\ddots\\
0&0&{}_sC_{L+3}^m&\beta_{L+3}&\ddots\\
\vdots&\vdots&\ddots&\ddots&\ddots
\end{pmatrix},
\label{eq:M1-matrix}\\[4pt]
\mathbf M^{(2)}&=
\begin{pmatrix}
d_L&e_L&f_L&0&\cdots\\
e_L&d_{L+1}&e_{L+1}&f_{L+1}&\cdots\\
f_L&e_{L+1}&d_{L+2}&e_{L+2}&\ddots\\
0&f_{L+1}&e_{L+2}&d_{L+3}&\ddots\\
\vdots&\vdots&\ddots&\ddots&\ddots
\end{pmatrix},
\label{eq:M2-matrix}
\end{align}
where
\begin{align}
\begin{aligned}
d_j&=({}_sC_j^m)^2+\beta_j^2+({}_sC_{j+1}^m)^2,\\
e_j&={}_sC_{j+1}^m(\beta_j+\beta_{j+1}),\\
f_j&={}_sC_{j+1}^m{}_sC_{j+2}^m.
\end{aligned}
\end{align}
In the infinite-dimensional basis, $\mathbf M^{(2)}=(\mathbf M^{(1)})^2$.
However, for a finite numerical truncation, this identity breaks down near the truncation boundary. Hence, $\mathbf M^{(2)}$ should be constructed directly from Eq.~\eqref{eq:M2-elements}.

Note that $\mathbf M^{(1)}$ is tridiagonal and $\mathbf M^{(2)}$ is pentadiagonal.
This implies that $\partial_tR_\ell$ couples up to $\ell\pm1$, and
$\partial_t^2R_\ell$ up to $\ell\pm2$. The equations for different multipole modes $j$ decouple entirely only in the non-rotating limit $a=0$.

\subsection{Local antidamping and numerical instabilities}
For a fixed value of $m$, we truncate the $j$-tower at $j=J$. Equation~\eqref{eq:bl-projected-system} can then be written in matrix form as
\begin{equation}
 \mathbf A_{\rm BL}\,\partial_t^2\mathbf R
 +\mathbf B_{\rm BL}\,\partial_t\mathbf R
 +\mathcal D_r\mathbf R
 +\mathbf V_{\rm BL}\mathbf R=0,
\label{eq:bl-energy-matrix-system}
\end{equation}
where
\begin{equation}
 \mathbf R=(R_{j_{\min}},\ldots,R_J)^{\mathsf T}.
\end{equation}
The coefficient matrices take the form
\begin{align}
 \mathbf A_{\rm BL}&=(\mathcal A_0-a^2)\mathbf I+a^2\mathbf M^{(2)},\\
 \mathbf B_{\rm BL}&=\mathcal B_0\mathbf I+2\I sa\,\mathbf M^{(1)},\\
 \mathbf V_{\rm BL}&=\operatorname{diag}(\lambda_j)+\mathcal C_0\mathbf I\,.
\label{eq:bl-energy-blocks}
\end{align}
Equation~\eqref{eq:bl-energy-matrix-system} contains a first-order time-derivative term $\partial_t\mathbf R$ whose Hermitian coefficient gives a local damping or antidamping contribution in the chosen variables. We use this contribution to diagnose numerical error amplification, rather than as a criterion for physical or continuum stability. Boundedness and decay of subextremal Kerr perturbations are established by global estimates for the full equation~\cite{shlapentokh-rothmanBoundednessDecayTeukolsky2020,ShlapentokhRothmanTeixeira2023Physical,HaefnerHintzVasy2025FullKerr}.

Outside the event horizon ($r>r_+$), $\mathbf A_{\rm BL}$ is Hermitian and positive definite. Since $\mathbf M^{(1)}$ is real and symmetric, the Hermitian part of $\mathbf B_{\rm BL}$ reduces to
\begin{equation}
 \mathbf B_{{\rm BL},{\rm H}}
 :=\frac{\mathbf B_{\rm BL}+\mathbf B_{\rm BL}^\dagger}{2}
 =\operatorname{Re}(\mathcal B_0)\mathbf I\,.
\label{eq:bl-damping-hermitian-part}
\end{equation}
The matrix $\mathbf B_{{\rm BL},{\rm H}}$ governs the quadratic energy production rate associated with the $\partial_t\mathbf R$ term.

To illustrate this, we decompose $\mathbf V_{\rm BL}=\mathbf V_{{\rm BL},{\rm H}}+\mathbf V_{{\rm BL},{\rm A}}$ into its Hermitian and anti-Hermitian parts and introduce the formal quadratic energy functional
\begin{equation}
\begin{aligned}
 \mathscr E_I(t)=\frac12\int_I\Bigl[
 &w\,(\partial_t\mathbf R)^\dagger\mathbf A_{\rm BL}\,\partial_t\mathbf R
 +\Delta_r^{s+1}
   (\partial_r\mathbf R)^\dagger\partial_r\mathbf R\\
 &+w\,\mathbf R^\dagger\mathbf V_{{\rm BL},{\rm H}}\mathbf R
 \Bigr]\,\dd r \,,
\end{aligned}
\label{eq:bl-formal-energy}
\end{equation}
where $w(r)=\Delta_r^s$. This representation-dependent functional is a formal diagnostic; the potential term is not assumed to be positive, so $\mathscr E_I$ is not asserted to be a coercive physical energy. 

Multiplying Eq.~\eqref{eq:bl-energy-matrix-system} by $w(\partial_t\mathbf R)^\dagger$, integrating over a radial interval $I=[r_a,r_b]\subset(r_+,\infty)$, and taking the real part yields the energy balance equation
\begin{equation}
\begin{aligned}
 \frac{\dd\mathscr E_I}{\dd t}
 ={}&
 \operatorname{Re}\!\left[
  (\partial_t\mathbf R)^\dagger
\Delta_r^{s+1}\partial_r\mathbf R
 \right]_{r_a}^{r_b}\\
 &-\int_I w\,(\partial_t\mathbf R)^\dagger
       \mathbf B_{{\rm BL},{\rm H}}\partial_t\mathbf R\,\dd r\\
 &+\mathscr S_{\rm A},
\end{aligned}
\label{eq:bl-energy-balance}
\end{equation}
where
\begin{equation}
 \mathscr S_{\rm A}
 =-\operatorname{Re}\int_I
 w\,(\partial_t\mathbf R)^\dagger\mathbf V_{{\rm BL},{\rm A}}\mathbf R\,\dd r
\end{equation}
represents the energy exchange with the anti-Hermitian potential $\mathbf V_{{\rm BL},{\rm A}}$. 

The specific contribution to the energy rate from the first-order time-derivative term $\partial_t\mathbf R$ is
\begin{equation}
 \left.\frac{\dd\mathscr E_I}{\dd t}\right|_{\partial_t\mathbf R}
 =-\int_I w\,b_s(r)\,
   \|\partial_t\mathbf R\|_2^2\,\dd r,
 \quad
 b_s(r):=\operatorname{Re}(\mathcal B_0).
\label{eq:bl-damping-energy-rate}
\end{equation}
Using Eq.~\eqref{eq:bl-radial-operators}, $b_s(r)$ can be written explicitly as 
\begin{equation}
\begin{aligned}
 b_s(r)
 &=2s\left[
 r-\frac{M(r^2-a^2)}{\Delta_r}
 \right]\\
 &=\frac{2s}{\Delta_r}
 \left(r^3-3Mr^2+a^2r+Ma^2\right).
\end{aligned}
\label{eq:bl-local-damping-coefficient}
\end{equation}
For gravitational perturbations ($s=-2$), this yields
\begin{equation}
 b_{-2}(r)
 =-\frac{4}{\Delta_r}
 \left(r^3-3Mr^2+a^2r+Ma^2\right).
\label{eq:bl-gravitational-antidamping}
\end{equation}
Because $\Delta_r>0$ outside the event horizon, the $\partial_t\mathbf R$ term becomes locally antidamping ($b_{-2}(r)<0$) wherever
\begin{equation}
 r^3-3Mr^2+a^2r+Ma^2>0.
\label{eq:bl-antidamping-region}
\end{equation}
The polynomial in Eq.~\eqref{eq:bl-antidamping-region} is negative at $r=r_+$ and positive for sufficiently large $r$. Hence, its largest real root marks a spatial transition from local damping to local antidamping. In the Schwarzschild limit ($a=0$), this transition occurs at $r=3M$, with $b_{-2}>0$ for $2M<r<3M$ (damping) and $b_{-2}<0$ for $r>3M$ (antidamping). Perturbations localized in the region $r>3M$ therefore give a positive contribution to Eq.~\eqref{eq:bl-damping-energy-rate}. 

The sign of this local contribution does not imply an instability of the continuum Teukolsky equation. The full balance includes boundary fluxes and potential exchange, and the mathematical stability proofs use global estimates that account for the complete operator~\cite{ShlapentokhRothmanTeixeira2023Physical,HaefnerHintzVasy2025FullKerr}. At the discrete level, however, numerical schemes need not preserve these continuum cancellations and bounds. Consequently, the amplification of truncation and round-off errors can trigger spurious exponential growth, a phenomenon previously observed in double-null Teukolsky evolutions~\cite{Long:2021ufh}. Throughout this work, \emph{antidamping instability} refers to this numerical behavior, not to a growing physical mode.

\section{Hyperboloidal Coordinates}
\label{sec:hyperboloidal-coordinates}
To suppress this local antidamping, we replace the Boyer--Lindquist time foliation with a hyperboloidal slicing. However, a hyperboloidal transformation does not automatically ensure numerical stability. The primary diagnostic remains the sign of the Hermitian coefficient associated with the first-order time derivative (analogous to $\mathbf B_{{\rm BL},{\rm H}}$ in BL coordinates). In conformally compactified formulations~\cite{csukasNumericalInvestigationDynamics2019,2019arXiv191013452P,2014JCoPh.276..357P,Markakis:2023pfh}, the asymptotic behavior of this coefficient provides a strictly dissipative (damping) contribution for spin weight $s=-2$, thereby stabilizing the discrete evolution.

Unlike compactified schemes that map the future event horizon $\mathscr H^+$ and future null infinity $\mathscr I^+$ to finite coordinate locations, we implement a uniform tortoise grid. Although conformal compactification provides direct computational access to asymptotic boundaries, it produces an increasingly nonuniform physical radial resolution near $\mathscr I^+$. A uniform tortoise coordinate is better suited to our lensing setup, as it maintains consistent spatial resolution for tracking an incident wave packet across an extended radial domain.

To this end, we first define the physical Kerr tortoise coordinate $r_*^{\rm phys}$ by
\begin{equation}
 \frac{\dd r_*^{\rm phys}}{\dd r}=\frac{r^2+a^2}{\Delta_r}.
\end{equation}
Integrating yields
\begin{equation}
 r_*^{\rm phys}=r+\frac{2Mr_+}{r_+-r_-}
       \log\left|\frac{r-r_+}{2M}\right|
      -\frac{2Mr_-}{r_+-r_-}
       \log\left|\frac{r-r_-}{2M}\right|\,,
\label{eq:physical-kerr-tortoise-explicit}
\end{equation}
where the integration constant is chosen to recover the standard Schwarzschild tortoise coordinate when $a\to0$. We also introduce the horizon-regular azimuthal angle
\begin{align}
 \phi_\star&=\phi+F(r),&
 F(r)&=\frac{a}{r_+-r_-}
 \log\!\left(\frac{r-r_+}{r-r_-}\right).
\label{eq:horizon-regular-azimuth}
\end{align}
To ensure that the evolution variables remain regular at both future null infinity $\scri$ and the future event horizon $\hor$, we expand the Teukolsky field in terms of rescaled modal functions $\varphi_{jm}^{(s)}$ as
\begin{equation}
 \psi^{(s)}
 =\frac{\Delta_r^{-s}}{r}
 \sum_{m=-\infty}^{\infty}\sum_{j=j_{\min}}^\infty
 \varphi_{jm}^{(s)}(t,r)\,
 \e^{\I m\phi_\star}{}_sy_{jm}(\theta),
\label{eq:multipole-expansion}
\end{equation}
so that the relation to the BL radial functions $R_j(t,r)$ is
\begin{equation}
 R_j(t,r)=\frac{\Delta_r^{-s}}{r}\,
          \e^{\I mF(r)}\varphi_j(t,r).
\label{eq:R-to-varphi}
\end{equation}
For gravitational perturbations ($s=-2$), using the Kinnersley tetrad spin coefficient $\rho_{\rm NP}=-(r-\I a\cos\theta)^{-1}$, the Teukolsky master variable $\psi^{(-2)}=\rho_{\rm NP}^{-4}\Psi_4$ takes the form
\begin{equation}
 r\Psi_4=\rho_{\rm NP}^{4}\Delta_r^2
 \sum_{m=-\infty}^{\infty}\sum_{j=j_{\min}}^\infty
 \varphi_{jm}^{(-2)}(t,r)
 e^{\I m\phi_\star}{}_{-2}y_{jm}(\theta).
\label{eq:rpsi4-varphi}
\end{equation}

For numerical evolution, we scale variables by the black hole mass $M$ to define the dimensionless quantities
\begin{align}
 T&=\frac{t}{M},&
 x&=\frac{r_*^{\rm phys}}{M},&
 R&=\frac{r}{M},&
 \widehat a&=\frac{a}{M},
\label{eq:rstar-dimensionless-variables}
\end{align}
along with the dimensionless metric functions
\begin{align}
 \widehat\Delta&=R^2-2R+\widehat a^2,&
 \mathcal R&=R^2+\widehat a^2,\notag\\
 R_\pm&=1\pm\sqrt{1-\widehat a^2},&
 \frac{\dd x}{\dd R}&=\frac{\mathcal R}{\widehat\Delta}.
\label{eq:dimensionless-kerr-functions}
\end{align}
Here $R(x)$ denotes the inverse map from the tortoise coordinate $x$ to the BL radial coordinate $R$. Near the event horizon, it is numerically advantageous to treat $\delta R:=R-R_+$ as an independent variable and evaluate $\widehat\Delta=\delta R(R_+-R_-+\delta R)$, avoiding catastrophic loss of significance near the horizon boundary.

Truncating the multipole expansion at $j=J$, we collect the evolved modes into the vector $\bm\varphi=(\varphi_{j_{\min}},\ldots,\varphi_J)^T$. Dividing by $\mathcal R^2/\widehat\Delta$, the evolution system in $(T,x)$ coordinates becomes
\begin{equation}
 \mathbf M\,\partial_T^2\bm\varphi
 -\partial_x^2\bm\varphi-p\,\partial_x\bm\varphi
 +\mathbf B\,\partial_T\bm\varphi
 +\mathbf W\bm\varphi=0.
\label{eq:regular-rstar-system}
\end{equation}
The coefficient blocks are given by
\begin{align}
 \mu&=\frac{\widehat a^2\widehat\Delta}{\mathcal R^2},\qquad
 \mathbf M=(1-\mu)\Id+\mu\mathbf M^{(2)},\notag\\
 p_0&=\frac{2s(R-1)}{\mathcal R}
       +\frac{2R\widehat\Delta}{\mathcal R^2},\notag\\
 \mathbf B&=b_0\Id+\I b_c\mathbf M^{(1)},\qquad
 b_c=\frac{2s\widehat a\widehat\Delta}{\mathcal R^2},\notag\\
 b_0&=\frac{-2s(R^2-\widehat a^2)+2sR\widehat\Delta
                 +4\I m\widehat a R}{\mathcal R^2},\notag\\
 v_j&=\frac{\lambda_j\widehat\Delta-\widehat a^2m^2
             +2\I\widehat a ms(1-R)}{\mathcal R^2},\notag\\
 \mathbf V&=\operatorname{diag}_{j=j_{\min}}^J(v_j).
\label{eq:rstar-coefficient-blocks}
\end{align}
Defining the tortoise logarithmic derivative of the field factor as
\begin{equation}
 \ell:=\partial_x\log\!\left(
 \frac{\widehat\Delta^{-s}}{R}\e^{\I mF}\right)
 =-\frac{2s(R-1)}{\mathcal R}
  -\frac{\widehat\Delta}{R\mathcal R}
  +\frac{\I m\widehat a}{\mathcal R},
\label{eq:rstar-log-derivative}
\end{equation}
the remaining radial coefficients are
\begin{equation}
 p=p_0+2\ell,\qquad
 \mathbf W=\mathbf V-(\partial_x\ell+\ell^2+p_0\ell)\Id.
\label{eq:rstar-regular-coefficients}
\end{equation}

For the hyperboloidal time foliation, we adopt the minimal gauge transformation~\cite{Zenginoglu:2008wc,zenginogluIntermediateBehaviorKerr2014,Markakis:2023pfh}
\begin{equation}
 \tau=T+H(x),\qquad u(x):=H'(x),\qquad H(0)=0\,,
\label{eq:localized-hyperboloidal-time}
\end{equation}
where $H(x)$ is the height function. The standard minimal gauge is formulated using the inverted compactifying radial coordinate $\sigma=2/R$. In terms of $\sigma$, the dimensionless Kerr tortoise coordinate in Eq.~\eqref{eq:physical-kerr-tortoise-explicit} is represented as $x=4g(\sigma)-2\log(1+\kappa^2)$, with
\begin{align}
g(\sigma)&=\frac12\biggl[
  \frac1\sigma
  +\frac{1}{1-\kappa^2}
   \log\!\Bigl(\frac{1+\kappa^2}{\sigma}-1\Bigr)\notag\\
 &\qquad-\frac{\kappa^2}{1-\kappa^2}
   \log\!\Bigl(\frac{1+\kappa^2}{\sigma}-\kappa^2\Bigr)
 \biggr]\,,\\
 \kappa&=\frac{\widehat a}{1+\sqrt{1-\widehat a^2}}\,.
\label{eq:minimal-height}
\end{align}
With $\sigma_0=\sigma(x=0)$, the height function in this gauge is normalized by $H_{\min}(0)=0$ and satisfies
\begin{align}
\begin{aligned}
H_{\min}(x)&=4[h(\sigma(x))-h(\sigma_0)],\\
h(\sigma)&=g(\sigma)-\frac1\sigma+\log\sigma\,.
\end{aligned}
\end{align}
The minimal gauge is minimal in the sense that the singular terms required for constant-$\tau$ hypersurfaces to penetrate the future horizon and extend smoothly to $\scri$ are subtracted directly from the tortoise function, without adding an arbitrary regular function to $h$. Consequently, its constant-time hypersurfaces become ingoing null at the future horizon and outgoing null at $\scri$. Under this choice, the transformed Teukolsky equation takes an exceptionally concise form with simple coefficients, and the asymptotic boundaries correspond to pure outflow in globally compactified formulations. The slope of the minimal height function is
\begin{equation}
 u_{\min}(x)=\frac{\dd h}{\dd g}
 =1+\frac{\sigma^{-2}+\sigma^{-1}}{g'(\sigma)},
\label{eq:minimal-slope}
\end{equation}
which satisfies $u_{\min}\to+1$ at the horizon and $u_{\min}\to-1$ at future null infinity.

On an extended uniform tortoise grid, however, this limiting behavior introduces severe numerical challenges. For the asymptotically flat wave operator, the outgoing and ingoing radial characteristic speeds with respect to the foliation time $\tau$ are
\begin{equation}
 c_+=\frac{1}{1+u},\qquad
 c_-=-\frac{1}{1-u}.
\label{eq:hyperboloidal-characteristic-speeds}
\end{equation}
Thus $|c_-|$ diverges as $u_{\min}\to+1$ near the horizon, while $c_+$ diverges as $u_{\min}\to-1$ near $\scri$. Extending the computational domain toward either boundary increases the maximum coordinate speed without bound. Any explicit time-integration scheme (such as RK4) must then satisfy the increasingly restrictive Courant–Friedrichs–Lewy (CFL) stability condition $\Delta\tau\lesssim\Delta x/\max|c_\pm|$, rendering long-term evolutions computationally prohibitive. Although implicit integrators bypass this time-step restriction, the underlying coordinate degeneracy remains, as the coefficient of the second-order time derivative approaches singularity and produces severely ill-conditioned linear systems that substantially degrade solver convergence.

To resolve this issue, we introduce a hybrid framework combining a \emph{localized minimal gauge} with a time-independent radial field redefinition. The localized gauge coincides identically with the standard minimal gauge in the strong-field interior region ($|u_{\min}|\le0.7$), but smoothly caps the asymptotic approach to $\pm1$ through a $C^2$ transition to bounded plateaus at $\pm0.8$. Specifically, we construct the boost profile as
\begin{equation}
u(x)=\operatorname{sgn}(u_{\min})\,S(|u_{\min}|)\,,
\label{eq:localized-minimal-slope}
\end{equation}
where the smoothing function $S(y)$ is defined by
\begin{equation}
 S(y)=
 \begin{cases}
  y, & y\le0.7,\\
  0.7+0.2z-0.2z^3+0.1z^4, &0.7<y<0.9,\\
  0.8, & y\ge0.9,
 \end{cases}
\label{eq:slope-cap-polynomial}
\end{equation}
with $z=5(y-0.7)$. Once the slope $u(x)$ in Eq.~\eqref{eq:localized-minimal-slope} is prescribed, the localized height function is defined uniquely (up to the normalization $H_{\rm loc}(0)=0$) by
\begin{equation}
 H_{\rm loc}(x)=\int_0^x u(\xi)\,\dd\xi.
\label{eq:localized-height-function}
\end{equation}
Explicitly, let the four ordered transition points $x_1<x_2<x_3<x_4$ be determined by
\begin{align}
 u_{\min}(x_1)&=0.9, & u_{\min}(x_2)&=0.7,\notag\\
 u_{\min}(x_3)&=-0.7, & u_{\min}(x_4)&=-0.9,
\label{eq:localized-transition-points}
\end{align}
and denote $\sigma_0=\sigma(0)$ and $H_k=H_{\rm loc}(x_k)$. Continuity and the normalization $H_{\rm loc}(0)=0$ give
\begin{align}
 H_2&=4[h(\sigma(x_2))-h(\sigma_0)],\notag\\
 H_3&=4[h(\sigma(x_3))-h(\sigma_0)],\notag\\
 H_1&=H_2-\int_{x_1}^{x_2}S(u_{\min}(\xi))\,\dd\xi,\notag\\
 H_4&=H_3-\int_{x_3}^{x_4}S(|u_{\min}(\xi)|)\,\dd\xi.
\label{eq:localized-height-matching-values}
\end{align}
The height function is then expressed piecewise across five radial zones as
\begin{equation}
 H_{\rm loc}(x)=
 \begin{cases}
  H_1+0.8(x-x_1), & x\le x_1,\\
  H_2-\displaystyle\int_x^{x_2}S(u_{\min}(\xi))\,\dd\xi,
      &x_1<x<x_2,\\
  4[h(\sigma(x))-h(\sigma_0)], &x_2\le x\le x_3,\\
  H_3-\displaystyle\int_{x_3}^{x}S(|u_{\min}(\xi)|)\,\dd\xi,
      &x_3<x<x_4,\\
  H_4-0.8(x-x_4), &x\ge x_4.
 \end{cases}
\label{eq:localized-height-piecewise}
\end{equation}
The inner ($x\le x_1$) and outer ($x\ge x_4$) regions have constant slopes $u=0.8$ and $u=-0.8$, respectively. The intermediate zones provide $C^2$ transitions, while the central zone ($x_2\le x\le x_3$) reproduces the exact minimal gauge.

Because $|u|\le0.8$ throughout the domain, the characteristic speeds remain bounded, satisfying
\begin{equation}
 \max\{|c_+|,|c_-|\}\le 5\,.
\label{eq:localized-speed-bound}
\end{equation}
In the inner constant-slope region ($u=0.8$), the speeds are $c_+=5/9$ and $c_-=-5$, whereas in the outer constant-slope region ($u=-0.8$), they are $c_+=5$ and $c_-=-5/9$. The localized gauge therefore preserves the geometric benefits of the minimal gauge in the central lensing region while imposing a uniform upper bound on all characteristic speeds.

Transforming coordinates from $(T,x)$ to $(\tau,x)$ via $\partial_x|_T=\partial_x|_\tau+u\partial_\tau$, the wave equation~\eqref{eq:regular-rstar-system} becomes
\begin{align}
 (\mathbf M-u^2\Id)\partial_\tau^2\bm\varphi
 ={}&2u\,\partial_x\partial_\tau\bm\varphi
+\partial_x^2\bm\varphi+p\,\partial_x\bm\varphi\notag\\
 &-[\mathbf B-(pu+u')\Id]\partial_\tau\bm\varphi
  -\mathbf W\bm\varphi.
\label{eq:hyper-teuk-matrix}
\end{align}
The evolution equation depends only on $u=H_{\rm loc}'$ and $u'=H_{\rm loc}''$. The height function $H_{\rm loc}(x)$ itself is needed only to map initial data and reconstructed observables between constant-$\tau$ and constant-$T$ hypersurfaces.

The effect of the hyperboloidal slicing on antidamping can be diagnosed directly from the coefficient multiplying $\partial_\tau\bm\varphi$, namely $[\mathbf B-(pu+u')\Id]$. Because $\I b_c\mathbf M^{(1)}$ is anti-Hermitian, it does not contribute to the quadratic energy production rate. The relevant Hermitian coefficient on the right-hand side of Eq.~\eqref{eq:hyper-teuk-matrix} is the scalar
\begin{equation}
 D(x):=u'+u\operatorname{Re}(p)-\operatorname{Re}(b_0)\,,
\label{eq:rstar-antidamping-coefficient}
\end{equation}
which contributes to the kinetic energy growth rate as
\begin{equation}
 \mathcal P_D
 =2D\,\|\partial_\tau\bm\varphi\|_2^2\,.
\label{eq:hyperboloidal-antidamping-power}
\end{equation}
Thus $D\le0$ gives a non-positive contribution from this term in the chosen quadratic balance, whereas $D>0$ gives a positive local contribution. These signs describe the evolution variables and provide a numerical diagnostic; they are not a statement about net physical energy dissipation or continuum instability.

For gravitational perturbations ($s=-2$), the coefficients in the far field behave as
\begin{equation}
 \operatorname{Re}(p)=\frac{4}{R}+O(R^{-2}),\qquad
 \operatorname{Re}(b_0)=-\frac{4}{R}+O(R^{-2}),
\end{equation}
which yields
\begin{equation}
 D=u'+\frac{4(1+u)}{R}+O(R^{-2}).
\label{eq:far-field-antidamping-coefficient}
\end{equation}
In the global minimal gauge, $u_{\min}\to-1$ asymptotically, so the leading BL antidamping $4/R$ is exactly cancelled. In our localized gauge, however, $u$ saturates at $-\beta_{\rm cap}=-0.8$ (with $u'=0$) in the outer capped region, leaving a small positive residual,
\begin{equation}
 D=\frac{4(1-\beta_{\rm cap})}{R}+O(R^{-2})>0,
 \qquad \beta_{\rm cap}=0.8.
\label{eq:localized-residual-antidamping}
\end{equation}
This residual antidamping can induce long-term numerical instability if left uncorrected.

To remove this residual while preserving the bounded characteristic speeds, we introduce the time-independent field redefinition
\begin{equation}
 \bm\varphi=\e^{q(x)}\bm\chi,\qquad
\partial_\tau\bm\varphi=\e^{q(x)}\bm\Pi,\qquad q(x_{\min})=0.
\label{eq:static-antidamping-rescaling}
\end{equation}
We define the regularized positive part of $D$ as
\begin{align}
D_+&=\frac12\left(D+\sqrt{D^2+\epsilon(x)^2}\right),
\label{eq:antidamping-qprime}
\end{align}
where the smoothing profile is
\begin{equation}
 \epsilon(x)=10^{-8}\mathcal S_5\!\left(\frac{-u-0.75}{0.04}\right),
\end{equation}
and $\mathcal S_5(z)$ is the standard $C^2$-smooth quintic transition function
\begin{equation}
 \mathcal S_5(z)=
 \begin{cases}
  0,&z\le0\,,\\
  z^3(10-15z+6z^2),&0<z<1\,,\\
  1,&z\ge 1 \,.
 \end{cases}
\label{eq:antidamping-activation}
\end{equation}
The rescaling exponent $q(x)$ is then determined by its spatial derivative
\begin{align}
 q'&=\begin{cases}
 0,&D_+=0\,,\\
 -D_+/(2u),&D_+>0\,,
 \end{cases}
\end{align}
which removes the positive part of the local antidamping coefficient. Where $\epsilon=0$, $D_+=\max(D,0)$, so the modification vanishes for $D\le0$. In the outer transition and capped region, the smoothing profile can be nonzero, and then $D_+>0$ even when $D<0$. The rescaling therefore has a small smooth extension into that region. Under this transformation, the effective damping coefficient is $\widetilde D=D-D_+$ and satisfies
\begin{equation}
\widetilde D=D+2uq'\le0\,,
\label{eq:corrected-antidamping}
\end{equation}
and the corresponding power density obeys
\begin{equation}
 \widetilde{\mathcal P}_D
 =2\widetilde D\,\|\bm\Pi\|_2^2\le0\,.
\label{eq:corrected-antidamping-power}
\end{equation}
Consequently, the contribution of the first-order time-derivative term is non-positive throughout the computational domain. This removes the local source of error amplification diagnosed above; it is not, by itself, a proof that the total energy decreases or that the complete discrete scheme is stable.

Because $q(x)$ is time-independent, this transformation is analytically invertible on the finite computational domain and preserves the continuum solutions after reconstruction. In addition to the change in the first-order time-derivative coefficient, it modifies the lower-order spatial coefficients,
\begin{equation}
 \widetilde p=p+2q',\qquad
 \widetilde{\mathbf W}
 =\mathbf W-[q''+(q')^2+pq']\Id.
\label{eq:rescaled-rstar-coefficients}
\end{equation}
The principal part of Eq.~\eqref{eq:hyper-teuk-matrix} and the characteristic speed bound in Eq.~\eqref{eq:localized-speed-bound} remain unchanged. The localized gauge ensures bounded coordinate speeds and a uniform CFL condition, while the static field redefinition removes the residual antidamping. Operationally, $\bm\chi$ and $\bm\Pi$ serve as internal variables during time integration, with all physical initial data and reconstructed observables expressed in terms of the regular variables $\bm\varphi$ and $\partial_\tau\bm\varphi$.

\section{Initial Conditions}
\label{sec:initial-data}
To model the incident gravitational radiation, we consider a plane gravitational wave propagating in flat spacetime, characterized by its plane-wave Weyl tensor. We first project this tensor onto the local Zero Angular Momentum Observer (ZAMO) orthonormal frame of the Kerr background, and then contract it with the Kinnersley tetrad to extract the incoming Weyl scalar $\Psi_4$. The resulting expression is subsequently mapped onto the initial hyperboloidal slice $\tau=\tau_0$ and converted into the regularized evolution variables.

In addition to the Boyer--Lindquist (BL) coordinates $(t,r,\theta,\phi)$ and the hyperboloidal coordinates $(\tau,x,\theta,\phi_\star)$ introduced in Sec.~\ref{sec:hyperboloidal-coordinates}, we introduce a dedicated wave frame. Quantities defined in this wave frame are denoted with a superscript `$\mathrm{w}$'. In particular, $(x^{\rm w},y^{\rm w},z^{\rm w})$ and $(r,\theta^{\rm w},\phi^{\rm w})$ represent the Cartesian and spherical coordinates of the wave frame, respectively. The radial coordinate $r$ is identical in both the Kerr BL frame and the wave frame ($r = \sqrt{(x^{\rm w})^2+(y^{\rm w})^2+(z^{\rm w})^2}$). Regarding tensor components, Greek indices denote BL coordinates, unhatted Latin indices denote wave Cartesian axes, and hatted indices denote ZAMO orthonormal axes.

\subsection{Plane-Wave Data in the Wave Frame}
Let $(\mathbf e_1^{\rm w},\mathbf e_2^{\rm w},\widehat{\mathbf k}^{\rm w})$ denote a right-handed orthonormal triad associated with the wave frame, oriented such that the propagation unit vector $\widehat{\mathbf k}^{\rm w}$ points along the $+z^{\rm w}$ direction,
\begin{equation}
 \mathbf e_1^{\rm w}\cdot\widehat{\mathbf k}^{\rm w}
 =\mathbf e_2^{\rm w}\cdot\widehat{\mathbf k}^{\rm w}=0\,,\qquad
 \mathbf e_1^{\rm w}\times\mathbf e_2^{\rm w}=\widehat{\mathbf k}^{\rm w}\,.
 \label{gw:eq:wave-frame}
\end{equation}
The transverse unit vectors $(\mathbf e_1^{\rm w},\mathbf e_2^{\rm w})$ define the standard plus ($+$) and cross ($\times$) polarization basis tensors,
\begin{equation}
\begin{aligned}
 e_{ij}^{+,{\rm w}}&=e_{1i}^{\rm w}e_{1j}^{\rm w}
                    -e_{2i}^{\rm w}e_{2j}^{\rm w}\,,\\
 e_{ij}^{\times,{\rm w}}&=e_{1i}^{\rm w}e_{2j}^{\rm w}+e_{2i}^{\rm w}e_{1j}^{\rm w}\,.
\end{aligned}
 \label{gw:eq:polarization-tensors}
\end{equation}
Let $\mathbf X^{\rm w}=(x^{\rm w},y^{\rm w},z^{\rm w})$ denote the Cartesian position vector and $t^{\rm w}$ the coordinate time in the wave frame. We introduce the retarded phase coordinate
\begin{equation}
 \zeta^{\rm w}=\widehat{\mathbf k}^{\rm w}\cdot \mathbf X^{\rm w}-c^{\rm w}t^{\rm w}=z^{\rm w}-c^{\rm w}t^{\rm w},
 \qquad c^{\rm w}=1\,,
 \label{gw:eq:q-and-strain}
\end{equation}
where the propagation speed $c^{\rm w}=1$ is retained symbolically to trace dimensions and explicit time derivatives.

In transverse-traceless (TT) gauge, the metric perturbation reads
\begin{equation}
 h_{ij}^{{\rm TT},{\rm w}}
 =h_+^{\rm w}(\zeta^{\rm w})e_{ij}^{+,{\rm w}}
  +h_\times^{\rm w}(\zeta^{\rm w})e_{ij}^{\times,{\rm w}}\,,
 \label{gw:eq:plane-TT-metric}
\end{equation}
satisfying the gauge conditions $h_{0\mu}^{{\rm TT},{\rm w}}=0$, $\partial^i h_{ij}^{{\rm TT},{\rm w}}=0$, and $\delta^{ij}h_{ij}^{{\rm TT},{\rm w}}=0$. Here $h_+^{\rm w}$ and $h_\times^{\rm w}$ denote the dimensionless strain waveforms of the two polarization states.

In linearized vacuum gravity, the perturbed Weyl tensor of the wave is given by
\begin{equation}
 \delta C_{abcd}^{\rm w}
 =\frac12\bigl(h_{ad,bc}^{\rm w}+h_{bc,ad}^{\rm w}
 -h_{ac,bd}^{\rm w}-h_{bd,ac}^{\rm w}\bigr)\,,
\label{gw:eq:linearized-Riemann}
\end{equation}
where all derivatives are taken with respect to $(t^{\rm w},\mathbf X^{\rm w})$.
We adopt standard definitions for the electric and magnetic parts of the Weyl tensor,
\begin{equation}
 E_{ij}^{\rm w}:=\delta C_{0i0j}^{\rm w}\,,\qquad
 B_{ij}^{\rm w}:=\frac12\varepsilon_i{}^{kl}\delta C_{klj0}^{\rm w}\,,
 \label{gw:eq:electric-magnetic-def}
\end{equation}
with the spatial orientation convention $\varepsilon_{123}=+1$. Both $E_{ij}^{\rm w}$ and $B_{ij}^{\rm w}$ are symmetric and trace-free, and they completely determine the remaining components of the Weyl tensor via
\begin{equation}
\begin{aligned}
 \delta C_{0i0j}^{\rm w}&=E_{ij}^{\rm w}\,,\\
 \delta C_{klj0}^{\rm w}&=\varepsilon_{kl}{}^p B_{pj}^{\rm w}\,,\\
 \delta C_{ijkl}^{\rm w}
 &=-\varepsilon_{ij}{}^p\varepsilon_{kl}{}^q E_{pq}^{\rm w}\,.
\end{aligned}
 \label{kp:eq:weyl-from-EB}
\end{equation}
For a plane wave propagating along $+z^{\rm w}$, these reduce to
\begin{equation}
 E_{ij}^{\rm w}=-\frac12\partial_{t^{\rm w}}^2 h_{ij}^{{\rm TT},{\rm w}}\,,
 \qquad
 B_{ij}^{\rm w}=\varepsilon_i{}^{kl}\widehat k_k^{\rm w}E_{lj}^{\rm w}\,.
 \label{gw:eq:electric-from-strain}
\end{equation}
For idealized packets with a fixed polarization ratio, we define a constant complex curvature amplitude ${\cal A}^{\rm w}$ and a normalized real profile function $W^{\rm w}(\zeta^{\rm w})$ through
\begin{equation}
 {\cal A}^{\rm w}=A_+^{\rm w}-\I A_\times^{\rm w}\,,
 \qquad A_+^{\rm w},A_\times^{\rm w}\in\mathbb R\,,
 \label{gw:eq:complex-amplitude}
\end{equation}
\begin{equation}
 {\cal A}^{\rm w}W^{\rm w}(\zeta^{\rm w})
 =\partial_{t^{\rm w}}^2
   \bigl[h_+^{\rm w}(\zeta^{\rm w})-\I h_\times^{\rm w}(\zeta^{\rm w})\bigr]\,.
 \label{gw:eq:amplitude-from-strain}
\end{equation}
The electric part of the Weyl tensor then becomes
\begin{equation}
 E_{ij}^{\rm w}
 =-\frac{W^{\rm w}(\zeta^{\rm w})}{2}
  \bigl(A_+^{\rm w}e_{ij}^{+,{\rm w}}
       +A_\times^{\rm w}e_{ij}^{\times,{\rm w}}\bigr)\,,
 \label{gw:eq:plane-weyl-EM}
\end{equation}
with $B_{ij}^{\rm w}$ following directly from Eq.~\eqref{gw:eq:electric-from-strain}. A prime on $W^{\rm w}$ denotes differentiation with respect to $\zeta^{\rm w}$, so that $\partial_{t^{\rm w}}W^{\rm w}=-c^{\rm w}(W^{\rm w})'$.

For a general incident signal, the two polarizations have independent time dependence. We use the complex profile
\begin{equation}
\mathcal F^{\rm w}(\zeta^{\rm w})
 :=\partial_{t^{\rm w}}^2
   [h_+^{\rm w}(\zeta^{\rm w})-\I h_\times^{\rm w}(\zeta^{\rm w})]\,,
 \label{gw:eq:general-complex-curvature}
\end{equation}
with $\partial_{t^{\rm w}}\mathcal F^{\rm w}=-c^{\rm w}(\mathcal F^{\rm w})'$. The electric part of the Weyl tensor $E_{ij}^{\rm w}$ becomes
\begin{equation}
E_{ij}^{\rm w}=-\frac12\left[
  \operatorname{Re}(\mathcal F^{\rm w})e_{ij}^{+,{\rm w}}
  -\operatorname{Im}(\mathcal F^{\rm w})e_{ij}^{\times,{\rm w}}\right]\,.
 \label{gw:eq:general-plane-electric}
\end{equation}
All subsequent projection formulas extend to this case through the replacements
\begin{equation}
\begin{aligned}
 \mathcal A^{\rm w}W^{\rm w}&\longrightarrow\mathcal F^{\rm w}\,,\\
 \mathcal A^{\rm w}(W^{\rm w})'&\longrightarrow(\mathcal F^{\rm w})'\,.
\end{aligned}
 \label{gw:eq:general-profile-replacement}
\end{equation}
\subsection{ZAMO Representation and Kinnersley Projection}
\subsubsection{ZAMO orthonormal frame}
To connect the incident plane wave in the asymptotically flat region to the Teukolsky variables in the Kerr BL frame, we introduce the local ZAMO orthonormal frame $(\mathbf e_{\hat0},\mathbf e_{\hat r},\mathbf e_{\hat\theta},\mathbf e_{\hat\phi})$ and its dual coframe $(\vartheta^{\hat0},\vartheta^{\hat r},\vartheta^{\hat\theta},\vartheta^{\hat\phi})$, with
\begin{equation}
\begin{aligned}
 \vartheta^{\hat0}&=\alpha_{\rm Z}\,\dd t,\\
 \vartheta^{\hat r}&=\sqrt{\frac{\Sigma}{\Delta_r}}\,\dd r,\\
 \vartheta^{\hat\theta}&=\sqrt{\Sigma}\,\dd\theta,\\
 \vartheta^{\hat\phi}
 &=\sqrt{\frac{{\cal A}_{\rm K}}{\Sigma}}\sin\theta
   (\dd\phi-\omega\dd t)\,,
\end{aligned}
\label{kp:eq:zamo-coframe}
\end{equation}
where
\begin{align}
\Sigma&=r^2+a^2\cos^2\theta,\\
 {\cal A}_{\rm K}
 &=(r^2+a^2)^2-a^2\Delta_r\sin^2\theta .
\label{kp:eq:kerr-functions}
\end{align}
The lapse function $\alpha_{\rm Z}$ and the frame-dragging angular velocity $\omega$ are given by
\begin{equation}
 \alpha_{\rm Z}=\sqrt{\frac{\Sigma\Delta_r}{{\cal A}_{\rm K}}},
 \qquad
 \omega=\frac{2Mar}{{\cal A}_{\rm K}} .
\label{kp:eq:lapse-drag}
\end{equation}
The reference $\hat z$-axis of the ZAMO frame is aligned with the black-hole spin axis. 

The coframe satisfies the orthonormality condition
\begin{equation}
 g_{\mu\nu}
 =\eta_{\hat a\hat b}
 \vartheta^{\hat a}{}_\mu\vartheta^{\hat b}{}_\nu\,,
 \qquad
 \eta_{\hat a\hat b}=\operatorname{diag}(-1,1,1,1)\,.
\label{kp:eq:zamo-orthonormality}
\end{equation}
The dual orthonormal frame vectors are
\begin{equation}
\begin{aligned}
 e_{\hat0}
 &=\alpha_{\rm Z}^{-1}(\partial_t+\omega\partial_\phi)\,,\\
 e_{\hat r}
 &=\sqrt{\frac{\Delta_r}{\Sigma}}\,\partial_r\,,\\
 e_{\hat\theta}
 &=\Sigma^{-1/2}\partial_\theta\,,\\
 e_{\hat\phi}
 &=\sqrt{\frac{\Sigma}{{\cal A}_{\rm K}}}
   \frac1{\sin\theta}\partial_\phi\,.
\end{aligned}
\label{kp:eq:zamo-frame}
\end{equation}
As $r\to\infty$, the ZAMO frame smoothly approaches the standard spherical orthonormal frame of Minkowski spacetime.

\subsubsection{Kinnersley tetrad and Weyl contraction}
In BL coordinates, the Kinnersley null tetrad is given by
\begin{align}
 \ell_{\rm K}^{\mu}
 &=\left(\frac{r^2+a^2}{\Delta_r},
          1,0,\frac{a}{\Delta_r}\right),
\label{kp:eq:kinnersley-l}\\
 n_{\rm K}^{\mu}
 &=\frac1{2\Sigma}
   \left(r^2+a^2,-\Delta_r,0,a\right),
\label{kp:eq:kinnersley-n}\\
 m_{\rm K}^{\mu}
 &=\frac1{\sqrt2(r+\I a\cos\theta)}
   \left(\I a\sin\theta,0,1,\frac{\I}{\sin\theta}\right)\,,
\label{kp:eq:kinnersley-m}
\end{align}
satisfying $\ell_{\rm K}\cdot n_{\rm K}=-1$, $m_{\rm K}\cdot\overline m_{\rm K}=1$, with all other contractions vanishing.

Because the background Kerr metric is Petrov type~D, its background Weyl scalar vanishes identically, $\Psi_4^{\rm K}=0$. We use $\Psi_4^{\rm K}$ below to denote the linear perturbation
\begin{equation}
\begin{aligned}
 \Psi_4^{\rm K}
 &:=-\delta C_{\mu\nu\rho\sigma}
    n_{\rm K}^{\mu}\overline m_{\rm K}^{\,\nu}
    n_{\rm K}^{\rho}\overline m_{\rm K}^{\,\sigma}\\
 &=-\delta C_{\hat a\hat b\hat c\hat d}^{\rm Z}
    N^{\hat a}M^{\hat b}N^{\hat c}M^{\hat d},
\end{aligned}
 \label{kp:eq:kerr-direct-psi4}
\end{equation}
where $N^{\hat a}$ and $M^{\hat a}$ denote the projections of the Kinnersley null vectors onto the ZAMO coframe,
\begin{equation}
 N^{\hat a}:=\vartheta^{\hat a}{}_\mu n_{\rm K}^{\mu}\,,\qquad
 M^{\hat a}:=\vartheta^{\hat a}{}_\mu\overline m_{\rm K}^{\,\mu}\,.
\label{kp:eq:projected-kinnersley}
\end{equation}
The superscript `$\mathrm{Z}$' denotes tensor components evaluated in the orthonormal ZAMO frame. The corresponding electric and magnetic parts of the perturbed Weyl tensor, $E_{\hat i\hat j}^{\rm Z}$ and $B_{\hat i\hat j}^{\rm Z}$, are defined via standard spatial decomposition (analogous to Eq.~\eqref{gw:eq:electric-magnetic-def}) with respect to the ZAMO timelike normal $e_{\hat0}$, with orientation $\varepsilon_{\hat r\hat\theta\hat\phi}=+1$. Consequently, the algebraic reconstruction formula~\eqref{kp:eq:weyl-from-EB} holds directly for ZAMO indices.

To simplify the contractions, we define the spatial vectors
\begin{equation}
\begin{aligned}
p^{\hat i}&:=N^{\hat0}M^{\hat i}-N^{\hat i}M^{\hat0},\\
q^{\hat i}&:=\varepsilon^{\hat i}{}_{\hat j\hat k}
N^{\hat j}M^{\hat k}\,.
\end{aligned}
 \label{kp:eq:pq-definition}
\end{equation}
In terms of these vectors, Eq.~\eqref{kp:eq:kerr-direct-psi4} decomposes as
\begin{equation}
 \Psi_4^{\rm K}
 =-E_{\hat i\hat j}^{\rm Z}p^{\hat i}p^{\hat j}
     +E_{\hat i\hat j}^{\rm Z}q^{\hat i}q^{\hat j}
    +2B_{\hat i\hat j}^{\rm Z}p^{\hat i}q^{\hat j}.
 \label{kp:eq:psi4-EB-pq}
\end{equation}
For the Kinnersley tetrad, these spatial vectors satisfy the exact identity $q^{\hat i}=-\I p^{\hat i}$, where the components of $p^{\hat i}$ read explicitly
\begin{equation}
 \begin{pmatrix}
  p^{\hat r}\\[1mm]
  p^{\hat\theta}\\[1mm]
  p^{\hat\phi}
 \end{pmatrix}
 =
 \frac{1}
 {2\sqrt2\,(r-\I a\cos\theta)\sqrt{{\cal A}_{\rm K}}}
 \begin{pmatrix}
  -\I a\Delta_r\sin\theta\\[1mm]
  (r^2+a^2)\sqrt{\Delta_r}\\[1mm]
  -\I\sqrt{{\cal A}_{\rm K}\Delta_r}
 \end{pmatrix}.
\label{kp:eq:explicit-p-kerr}
\end{equation}
Substituting $q^{\hat i}=-\I p^{\hat i}$ into Eq.~\eqref{kp:eq:psi4-EB-pq} yields the compact formula
\begin{equation}
 \Psi_4^{\rm K}
 =-2p^{\hat i}\bigl(E_{\hat i\hat j}^{\rm Z}
                   +\I B_{\hat i\hat j}^{\rm Z}\bigr)p^{\hat j}\,,
 \label{kp:eq:explicit-kerr-psi4-EB}
\end{equation}
where all spatial contractions are performed with the Euclidean metric $\delta_{\hat i\hat j}$. Because the background tetrad is stationary, BL coordinate time derivatives act solely on the perturbed Weyl tensor,
\begin{equation}
\partial_t\Psi_4^{\rm K}
=-(\partial_t\delta C_{\hat a\hat b\hat c\hat d}^{\rm Z})
 N^{\hat a}M^{\hat b}N^{\hat c}M^{\hat d}.
\label{kp:eq:kerr-direct-dtpsi4}
\end{equation}

\subsection{Orienting the Incident Wave}
\subsubsection{General incidence direction}
\begin{figure*}[t]
\centering
\includegraphics[width=0.88\textwidth]{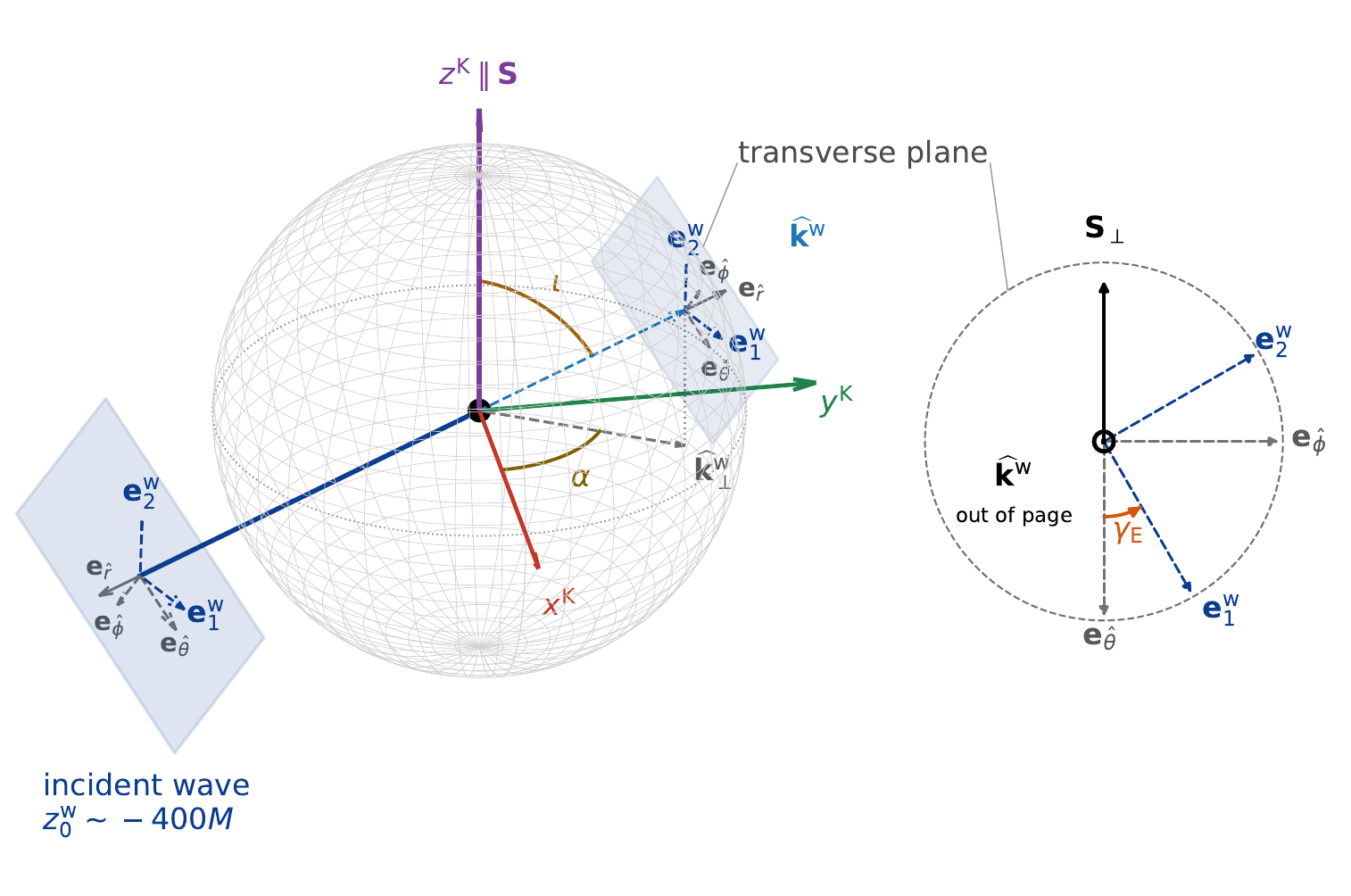}
\caption{Geometric configuration and orientation of the incident gravitational wave packet relative to the Kerr black hole. (Left) Three-dimensional setup showing the Kerr BL Cartesian axes $(x^{\rm K},y^{\rm K},z^{\rm K})$ in red, green, and purple, with $z^{\rm K}$ aligned with the black-hole spin axis $z^{\rm K}\parallel \mathbf S$. The propagation unit vector $\widehat{\mathbf k}^{\rm w}$ is parameterized by the polar inclination $\iota$ from the spin axis and the azimuthal angle $\alpha$ in the equatorial plane from the $x^{\rm K}$-axis to the equatorial projection $\widehat{\mathbf k}^{\rm w}_\perp$ (dashed gray line). The incoming wave packet is launched from the far-field negative half-space ($z_0^{\rm w}\sim -400M$) and propagates along $+\widehat{\mathbf k}^{\rm w}$ toward the origin. In this incoming sector, the propagation direction is anti-parallel to the outward radial unit vector ($\mathbf e_{\hat r}\approx -\widehat{\mathbf k}^{\rm w}$). Shaded light-blue slices indicate transverse wavefront planes carrying the polarization basis $(\mathbf e_1^{\rm w},\mathbf e_2^{\rm w})$ and the local ZAMO triad $(\mathbf e_{\hat r},\mathbf e_{\hat\theta},\mathbf e_{\hat\phi})$. (Right) Transverse wavefront plane viewed from the $+\widehat{\mathbf k}^{\rm w}$ side toward the origin, so the propagation vector points out of the page toward the viewer. The angular basis shown here is evaluated on the outgoing reference axis, where $\mathbf e_{\hat r}=+\widehat{\mathbf k}^{\rm w}$. A positive angle $\gamma_{\rm E}$ rotates this basis counterclockwise into $(\mathbf e_1^{\rm w},\mathbf e_2^{\rm w})$. Conversely, along the opposite (incident) axis, $\mathbf e_{\hat r}=-\widehat{\mathbf k}^{\rm w}$ and $\mathbf e_{\hat\theta}\times\mathbf e_{\hat\phi}=-\widehat{\mathbf k}^{\rm w}$, whereas the fixed wave basis maintains $\mathbf e_1^{\rm w}\times\mathbf e_2^{\rm w}=+\widehat{\mathbf k}^{\rm w}$ on both sides.}
\label{fig:waveframe-orientation}
\end{figure*}

Figure~\ref{fig:waveframe-orientation} illustrates our orientation conventions.
To parameterize an incident wave with arbitrary propagation direction, we introduce the local Cartesian spatial basis $(\mathbf e_{\hat x},\mathbf e_{\hat y},\mathbf e_{\hat z})$ associated with the ZAMO observer. These vectors form an orthonormal spatial frame related to $(\mathbf e_{\hat r},\mathbf e_{\hat\theta},\mathbf e_{\hat\phi})$ in Eq.~\eqref{kp:eq:zamo-frame} by the standard spherical-to-Cartesian rotation, with $\mathbf e_{\hat z}$ aligned along the black-hole spin axis.

From this Cartesian triad, we construct the complex spherical basis (spin-$1$ basis) $\mathbf e_{m}^{\rm Z}$ ($m=-1,0,1$) as
\begin{equation}
\begin{aligned}
 \mathbf e_{+1}^{\rm Z}&=-\frac{\mathbf e_{\hat x}+\I\mathbf e_{\hat y}}{\sqrt2}\,,\\
 \mathbf e_0^{\rm Z}&=\mathbf e_{\hat z}\,,\\
 \mathbf e_{-1}^{\rm Z}&=\frac{\mathbf e_{\hat x}-\I\mathbf e_{\hat y}}{\sqrt2}\,.
\end{aligned}\label{eq:spin_basis}
\end{equation}
For a reference wave propagating along $\mathbf e_{\hat z}$, the transverse complex polarization vector is
\begin{equation}
\boldsymbol\epsilon^{\rm ref}
 =\mathbf e_{\hat x}+\I\mathbf e_{\hat y}
 =-\sqrt2\,\mathbf e_{+1}^{\rm Z}\,.\label{eq:spin_1_z}
\end{equation}
For an incident wave with arbitrary orientation, the polarization vector is obtained via an active $z$-$y$-$z$ Euler rotation,
\begin{equation}
 \mathsf R^{\rm w}(\alpha,\iota,\gamma_{\rm E})
 =R_{\hat z}(\alpha)R_{\hat y}(\iota)R_{\hat z}(\gamma_{\rm E}),
\end{equation}
where Euler angles $(\alpha,\iota)$ specify the propagation direction ($\iota$ being the polar angle from the spin axis, and $\alpha$ the azimuthal angle in the equatorial plane), while $\gamma_{\rm E}$ sets the polarization angle in the transverse plane. The rotated polarization vector is
\begin{equation}
 \boldsymbol\epsilon^{\rm w}
 =\mathsf R^{\rm w}\boldsymbol\epsilon^{\rm ref}
 =\mathbf e_1^{\rm w}+\I\mathbf e_2^{\rm w}\,,\label{eq:rotated_spin}
\end{equation}
where $\mathbf e_1^{\rm w}=\mathsf R^{\rm w}\mathbf e_{\hat x}$ and $\mathbf e_2^{\rm w}=\mathsf R^{\rm w}\mathbf e_{\hat y}$ are the transverse wave-frame triad vectors in Eq.~\eqref{gw:eq:wave-frame}.

Under this rotation, the spin-1 basis transforms via the Wigner $D$-matrix as
\begin{equation}
 \mathsf R^{\rm w}\mathbf e_{m'}^{\rm Z}
=\sum_{m=-1}^{1}\mathbf e_m^{\rm Z}
 D^1_{mm'}(\alpha,\iota,\gamma_{\rm E})\,,
\end{equation}
where we adopt the standard convention
\begin{equation}
 D^j_{mm'}(\alpha,\iota,\gamma_{\rm E})
 =e^{-im\alpha}d^j_{mm'}(\iota)e^{-im'\gamma_{\rm E}}\,.
\label{eq:plane_wave_wigner_D_convention}
\end{equation}
Expanding $\boldsymbol\epsilon^{\rm w}$ in the spin-1 basis gives
\begin{equation}
 \boldsymbol\epsilon^{\rm w}
=\sum_{m=-1}^{1}\epsilon_m^{\rm w}\mathbf e_m^{\rm Z}\,,
 \qquad \epsilon_m^{\rm w}=-\sqrt2\,D^1_{m1}(\alpha,\iota,\gamma_{\rm E})\,.
 \label{kp:eq:wigner-polarization-spherical}
\end{equation}
From Eq.~\eqref{eq:spin_basis}, the ZAMO Cartesian components of $\boldsymbol\epsilon^{\rm w}$ are
\begin{equation}
\begin{aligned}
 \epsilon_{\hat x}^{\rm w}
 &=\frac{\epsilon_{-1}^{\rm w}-\epsilon_{+1}^{\rm w}}{\sqrt2}\,,\\
 \epsilon_{\hat y}^{\rm w}
 &=-\frac{\I(\epsilon_{-1}^{\rm w}+\epsilon_{+1}^{\rm w})}{\sqrt2}\,,\\
 \epsilon_{\hat z}^{\rm w}&=\epsilon_0^{\rm w}\,.
\end{aligned}
 \label{kp:eq:wigner-polarization-cartesian}
\end{equation}
The unit propagation vector $\widehat{\mathbf k}^{\rm w}$ after the active Euler rotation is
\begin{equation}
\begin{aligned}
 \widehat{\mathbf k}^{\rm w}
 &=\mathsf R^{\rm w}\mathbf e_{\hat z}
 =\frac{\I}{2}\boldsymbol\epsilon^{\rm w}
\times\overline{\boldsymbol\epsilon^{\rm w}}\\
&=\sin\iota\cos\alpha\,\mathbf e_{\hat x}
+\sin\iota\sin\alpha\,\mathbf e_{\hat y}
  +\cos\iota\,\mathbf e_{\hat z}\,.
\end{aligned}
 \label{kp:eq:wigner-k}
\end{equation}
Under this rotation, spatial coordinates transform passively as $\mathbf X^{\rm w}=(\mathsf R^{\rm w})^{-1}\mathbf X^{\rm ref}$, and in particular $z^{\rm w}=\widehat{\mathbf k}^{\rm w}\cdot\mathbf X^{\rm ref}$.

The angle $\gamma_{\rm E}$ rotates the transverse polarization basis about $\widehat{\mathbf k}^{\rm w}$ without altering the propagation direction, via
\begin{equation}
\boldsymbol\epsilon^{\rm w}(\alpha,\iota,\gamma_{\rm E})
 =e^{-\I\gamma_{\rm E}}\boldsymbol\epsilon^{\rm w}(\alpha,\iota,0)\,.
\end{equation}
By construction, $\boldsymbol\epsilon^{\rm w}$ satisfies the transversality and normalization conditions
\begin{equation}
 \boldsymbol\epsilon^{\rm w}\cdot\widehat{\mathbf k}^{\rm w}=0\,,
 \qquad
 \boldsymbol\epsilon^{\rm w}\cdot\overline{\boldsymbol\epsilon^{\rm w}}=2\,.
\end{equation}
Projecting $\boldsymbol\epsilon^{\rm w}$ onto the spherical ZAMO basis vectors yields
\begin{equation}
 \epsilon_{\hat i}^{\rm w}(\theta,\phi)
 :=\mathbf e_{\hat i}(\theta,\phi)\cdot\boldsymbol\epsilon^{\rm w}\,,
 \qquad \hat i\in\{\hat r,\hat\theta,\hat\phi\}\,.
 \label{kp:eq:zamo-polarization-components}
\end{equation}
Explicitly, parameterizing the projection in terms of the background Kerr BL frame angles gives
\begin{equation}
\begin{aligned}
 \epsilon_{\hat r}^{\rm w}
 &=\sin\theta\cos\phi\,\epsilon_{\hat x}^{\rm w}
  +\sin\theta\sin\phi\,\epsilon_{\hat y}^{\rm w}
  +\cos\theta\,\epsilon_{\hat z}^{\rm w}\,,\\
 \epsilon_{\hat\theta}^{\rm w}
 &=\cos\theta\cos\phi\,\epsilon_{\hat x}^{\rm w}
  +\cos\theta\sin\phi\,\epsilon_{\hat y}^{\rm w}
  -\sin\theta\,\epsilon_{\hat z}^{\rm w}\,,\\
 \epsilon_{\hat\phi}^{\rm w}
 &=-\sin\phi\,\epsilon_{\hat x}^{\rm w}
  +\cos\phi\,\epsilon_{\hat y}^{\rm w}\,.
\end{aligned}
\end{equation}
Here the superscript `${\rm w}$' specifies the incident wave parameters, and the hatted index denotes the ZAMO spherical basis component, enabling contractions such as $p^{\hat i}\epsilon_{\hat i}^{\rm w}$ to be computed directly.

We now prescribe the local incident Weyl tensor by applying the spatial rotation to Eq.~\eqref{gw:eq:plane-weyl-EM} and identifying the auxiliary wave time with BL coordinate time, $t^{\rm w}=t$. Its self-dual representation in the ZAMO frame is
\begin{equation}
 E_{\hat i\hat j}^{\rm Z}+\I B_{\hat i\hat j}^{\rm Z}
 =-\frac{{\cal A}^{\rm w}W^{\rm w}(\zeta^{\rm w})}{2}
   \epsilon_{\hat i}^{\rm w}\epsilon_{\hat j}^{\rm w}\,.
 \label{kp:eq:EB-helicity-vector}
\end{equation}
Contracting this tensor according to Eq.~\eqref{kp:eq:explicit-kerr-psi4-EB} yields
\begin{equation}
\begin{aligned}
 \Psi_4^{\rm K}
 &={\cal A}^{\rm w}W^{\rm w}(\zeta^{\rm w})
   \bigl(p^{\hat i}\epsilon_{\hat i}^{\rm w}\bigr)^2,\\
 \partial_t\Psi_4^{\rm K}
 &=-c^{\rm w}\,{\cal A}^{\rm w}(W^{\rm w})'(\zeta^{\rm w})
   \bigl(p^{\hat i}\epsilon_{\hat i}^{\rm w}\bigr)^2\,.
\end{aligned}
 \label{kp:eq:psi4-helicity-shortcut}
\end{equation}
Here the superscript `$\rm K$' denotes projection onto the background Kinnersley tetrad in the Kerr BL frame, while `${\rm w}$' denotes dependence on the wave parameters. The incident TT plane wave has a uniform profile across each transverse wavefront. Its Kinnersley projection nevertheless depends on the incidence direction $\widehat{\mathbf k}^{\rm w}$ through the contraction of the polarization vector with the geometry-dependent vector $p^{\hat i}$. For general waveforms, Eq.~\eqref{gw:eq:general-profile-replacement} applies to both relations in Eq.~\eqref{kp:eq:psi4-helicity-shortcut}.

\subsubsection{\texorpdfstring{$\widehat{\mathbf k}^{\rm w}\parallel z^{\rm K}$}{Wave propagation parallel to the Kerr spin axis}}
When the orientation angles vanish ($\alpha=\iota=\gamma_{\rm E}=0$), the wave frame aligns identically with the Kerr BL frame. Introducing the complementary Newman-Penrose scalar $\Psi_0^{\rm K}:=-\delta C(\ell_{\rm K},m_{\rm K},\ell_{\rm K},m_{\rm K})$, the asymptotic contractions yield (with $\mu^{\rm w}=\cos\theta^{\rm w}$)
\begin{equation}
 \Psi_4^{\rm K}\longrightarrow
 \frac18(1+\mu^{\rm w})^2{\cal A}^{\rm w}W^{\rm w}(\zeta^{\rm w})
 e^{2\I\phi^{\rm w}},
 \label{kp:eq:flat-kinnersley-check}
\end{equation}
\begin{equation}
 \Psi_0^{\rm K}\longrightarrow
 \frac12(1-\mu^{\rm w})^2{\cal A}^{\rm w}W^{\rm w}(\zeta^{\rm w})
 e^{2\I\phi^{\rm w}}.
 \label{kp:eq:flat-kinnersley-psi0}
\end{equation}
These limits confirm the amplitude normalization and propagation direction. Along the negative $z^{\rm w}$-axis ($\mu^{\rm w}=-1$), the wave travels inward relative to the outward Kinnersley tetrad ($\widehat{\mathbf k}^{\rm w}=-\mathbf e_{\hat r}$). Hence $\Psi_4^{\rm K}$ vanishes and $\Psi_0^{\rm K}$ carries the signal. Conversely, along the positive $z^{\rm w}$-axis ($\mu^{\rm w}=1$), the roles reverse. $\Psi_0^{\rm K}$ vanishes while $\Psi_4^{\rm K}$ carries the signal. This asymmetry reflects the tetrad-dependent definition of spin-weighted scalars. For instance, a TT plane wave possesses spin weight $-2$ relative to a tetrad oriented along $+\widehat{\mathbf k}^{\rm w}$, but registers as spin weight $+2$ relative to a tetrad oriented along $-\widehat{\mathbf k}^{\rm w}$.

Note that $\Psi_0^{\rm K}$ and $\Psi_4^{\rm K}$ describe the same gravitational perturbation, representing different projections of the Weyl tensor. Consequently, their dynamics are not independent but are linked by the Teukolsky–Starobinsky identities~\cite{TeukolskyPress1974,SasakiTagoshi2003}. 
While either quantity can in principle be evolved, we restrict our subsequent analysis to $\Psi_4^{\rm K}$, as it directly encodes the outgoing gravitational wave signal of primary interest.

\subsection{Initial Hyperboloidal Slice and Wave Support}
At the initial slice $\tau=\tau_0$, the coordinate transformation gives
\begin{equation}
 t_i=M[\tau_0-H(x_i)],\qquad r_i=MR(x_i)\,.
\label{kp:eq:radius-dependent-time}
\end{equation}
Thus each radial grid point $x_i$ on the initial slice corresponds to a distinct BL time $t_i$.
Assuming the initial wave packet front is located at $z_0^{\rm w}<0$ at coordinate time $t=0$, we set
\begin{equation}
 \tau_0=H(x_0),\qquad MR(x_0)=|z_0^{\rm w}|\,,
 \label{kp:eq:auto-tau0}
\end{equation}
which guarantees that the initial slice $\tau=\tau_0$ intersects the wavefront at this event.

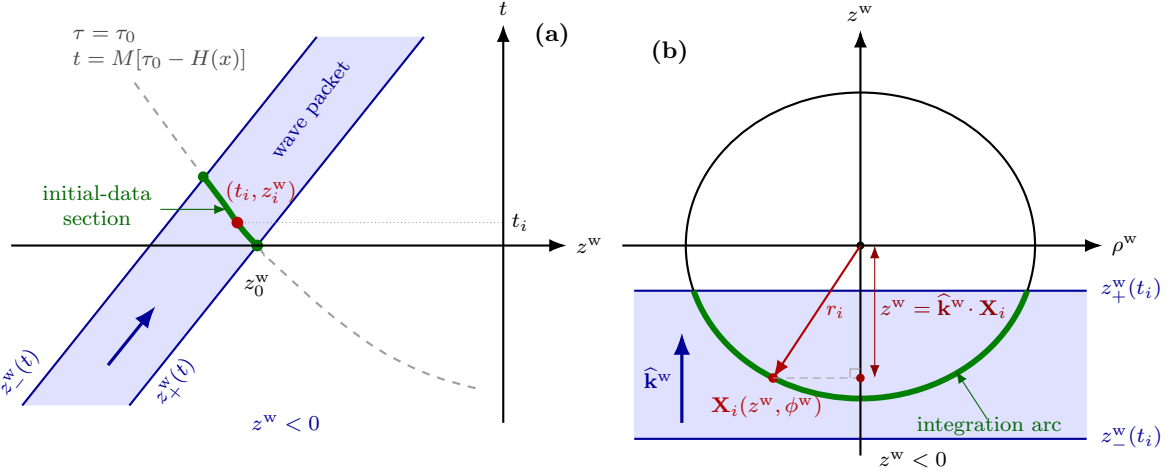
\begin{figure*}[htbp]
\centering
\begin{tikzpicture}[>=Latex,font=\footnotesize,x=1.05cm,y=0.92cm]
  \begin{scope}[shift={(1.10,0)}]
    \node[font=\bfseries,anchor=west] at (0.28,3.02) {(a)};
    \coordinate (Zzero) at (-3.10,0);

    \fill[blue!12]
      (-4.71,-2.30) -- (-6.06,-2.30) -- (-2.35,3.00) -- (-1.00,3.00)
      -- cycle;
    \draw[blue!60!black,thick] (-4.71,-2.30) -- (-1.00,3.00)
      node[pos=0.10,sloped,below,text=blue!60!black]
      {$z_+^{\rm w}(t)$};
    \draw[blue!60!black,thick] (-6.06,-2.30) -- (-2.35,3.00)
      node[pos=0.05,sloped,above,text=blue!60!black]
      {$z_-^{\rm w}(t)$};
    \node[blue!60!black,rotate=53] at (-2.42,1.95)
      {wave packet};
    \draw[blue!60!black,very thick,->] (-4.98,-1.72) -- (-4.38,-0.87);

    \draw[gray!75,thick,dashed]
      plot[smooth,tension=0.8] coordinates {
       (-0.35,-2.05) (-1.15,-1.75) (-2.0,-1.15)
       (-2.9,-0.25) (-3.10,0) (-3.75,0.95) (-4.65,2.35)};
    \node[gray!55!black,align=left,anchor=south east] at (-3.12,2.40)
      {$\tau=\tau_0$\\$t=M[\tau_0-H(x)]$};

    \draw[green!50!black,line width=2.2pt]
      plot[smooth,tension=0.8] coordinates {
        (-3.10,0) (-3.30,0.25) (-3.55,0.65) (-3.78,0.99)};
    \fill[green!45!black] (Zzero) circle (2.2pt);
    \fill[green!45!black] (-3.78,0.99) circle (2pt);
    \node[green!40!black,align=center,anchor=east] at (-4.32,0.58)
      {initial-data \\ section};
    \draw[green!40!black,->] (-4.32,0.52) -- (-3.48,0.52);
    \node[black,anchor=north] at ($(Zzero)+(0,-0.30)$) {$z_0^{\rm w}$};

    \coordinate (Pi) at (-3.35,0.33);
    \draw[densely dotted,gray!75] (Pi) -- (0,0.33)
      node[right,black] {$t_i$};
    \fill[red!70!black] (Pi) circle (2.2pt);
    \node[red!70!black,anchor=south] at ($(Pi)+(0.3,0.15)$)
      {$(t_i,z_i^{\rm w})$};
    \node[blue!60!black,anchor=north west] at (-3.28,-2.32)
      {$z^{\rm w}<0$};

    \draw[->,thick] (-6.20,0) -- (0.80,0) node[right] {$z^{\rm w}$};
    \draw[->,thick] (0,-2.55) -- (0,3.18) node[above] {$t$};
  \end{scope}

  \begin{scope}[shift={(5.60,0)}]
    \node[font=\bfseries,anchor=north west] at (-2.75,3.08) {(b)};
    \fill[blue!12] (-2.85,-2.78) rectangle (2.85,-0.65);
    \draw[blue!60!black,thick] (-2.85,-2.78) -- (2.85,-2.78);
    \draw[blue!60!black,thick] (-2.85,-0.65) -- (2.85,-0.65);
    \draw[black,thick] (0,0) circle (2.2);
    \draw[green!50!black,line width=2.2pt]
      (197.2:2.2) arc[start angle=197.2,end angle=342.8,radius=2.2];
    \draw[->,thick] (-3.0,0) -- (3.05,0)
      node[right] {$\rho^{\rm w}$};
    \draw[->,thick] (0,-3.02) -- (0,3.1) node[above] {$z^{\rm w}$};
    \fill (0,0) circle (1.5pt);
    \draw[red!70!black,thick,->] (0,0) -- (240:2.2)
      node[midway,right] {$r_i$};
    \fill[red!70!black] (240:2.2) circle (1.8pt);
    \node[red!70!black,anchor=north] at ($(240:2.2)+(-0.08,-0.14)$)
      {$\mathbf X_i(z^{\rm w},\phi^{\rm w})$};
    \draw[gray!75,densely dashed] (240:2.2) -- (0,-1.905);
    \fill[red!60!black] (0,-1.905) circle (1.6pt);
    \draw[gray!75] (-0.13,-1.905) -- (-0.13,-1.775) -- (0,-1.775);
    \draw[red!55!black,<->] (0.18,0) -- (0.18,-1.905);
    \node[red!55!black,anchor=north west] at (0.14,-0.57)
      {$z^{\rm w}=\widehat{\mathbf k}^{\rm w}\!\cdot\mathbf X_i$};
    \node[blue!60!black,anchor=west] at (2.92,-2.78)
      {$z_-^{\rm w}(t_i)$};
    \node[blue!60!black,anchor=west] at (2.92,-0.65)
      {$z_+^{\rm w}(t_i)$};
    \node[green!40!black,anchor=west] at (0.6,-2.6)
      {integration arc};
    \draw[green!40!black,->] (1.6,-2.5) -- (1.2,-1.8);
    \draw[blue!60!black,very thick,->] (-2.25,-2.55) -- (-2.25,-1.25);
    \node[blue!60!black,anchor=east] at (-2.25,-1.90)
      {$\widehat{\mathbf k}^{\rm w}$};
    \node[anchor=north west] at (0.12,-2.84) {$z^{\rm w}<0$};
  \end{scope}
\end{tikzpicture}
\caption{Geometric configuration and support of the incoming wave packet on a constant-$\tau$ hypersurface.
(a) Propagation of the wave packet along the positive $z^{\rm w}$-direction. At $t=0$, the wave front is located at 
$z_0^{\rm w}$. Its intersection with the one-parameter family of constant-$\tau$ hypersurfaces determines the initial slice $\tau_0$, where the initial data are specified. 
(b) Support restriction on a spherical shell of radius $r_i$ evaluated at the corresponding BL time $t_i$.
The highlighted green
arc indicates the portion of the shell lying within the wave support, bounded by
$z_{i,-}^{\rm w}$ and $z_{i,+}^{\rm w}$. The full mode decomposition also requires integrating over the azimuthal angle $\phi^{\rm w}$ across this region.}
\label{fig:plane-wave-geometry}
\end{figure*}

On the sphere $r=r_i$, the longitudinal wave coordinate $z^{\rm w}$ and the corresponding retarded phase argument $\zeta_i^{\rm w}$ are given by
\begin{equation}
\begin{aligned}
 z^{\rm w}&=\widehat{\mathbf k}^{\rm w}\cdot\mathbf X_i(\theta,\phi)\,,\\
 \zeta_i^{\rm w}(\theta,\phi)&=z^{\rm w}-c^{\rm w}t_i\,.
\end{aligned}
 \label{kp:eq:q-on-tau}
\end{equation}
If the wave profile $W^{\rm w}(\zeta^{\rm w})$ has compact support on the interval $[\zeta_-^{\rm w},\zeta_+^{\rm w}]$, its spatial extent along the propagation axis at time $t_i$ is bounded by
\begin{equation}
 z_\pm^{\rm w}(t_i)=c^{\rm w}t_i+\zeta_\pm^{\rm w}.
 \label{kp:eq:u-wave-support}
\end{equation}
Restricting this interval to the spherical shell $r=r_i$ yields the effective integration bounds
\begin{equation}
\begin{aligned}
 z_{i,-}^{\rm w}&=\max[-r_i,z_-^{\rm w}(t_i)],\\
 z_{i,+}^{\rm w}&=\min[r_i,z_+^{\rm w}(t_i)].
\end{aligned}
 \label{kp:eq:u-integration-limits}
\end{equation}
If $z_{i,-}^{\rm w}\ge z_{i,+}^{\rm w}$, the intersection is empty, meaning the shell at $r_i$ lies entirely outside the wave packet support at time $t_i$ and contributes zero initial data. As long as the packet remains entirely in the negative half-space $z^{\rm w}<0$, its inner and outer axial boundaries correspond to the radii
\begin{equation}
 r_{\rm in}(t)=-z_+^{\rm w}(t),\qquad
 r_{\rm out}(t)=-z_-^{\rm w}(t).
 \label{kp:eq:negative-z-worldlines}
\end{equation}
The geometry of this hypersurface and the resulting support constraints are illustrated in Fig.~\ref{fig:plane-wave-geometry}.

\subsection{Angular Projection and Spin-Dyad Rotation}
To extract the multipole coefficients of the incident wave on each spherical shell $r = r_i$ at BL time $t_i$, we project the Weyl scalar $\Psi_4^{\rm K}$ onto the spin-weighted spherical harmonics over the 2-sphere $S_i^2$. In the wave frame, the wave packet support forms a planar slab bounded by planes orthogonal to the propagation vector $\widehat{\mathbf k}^{\rm w}$ (see Fig.~\ref{fig:plane-wave-geometry}(b)). We therefore carry out the surface integration in the wave frame and transform the resulting modal coefficients to the Kerr BL frame $(\theta,\phi)$.

Using the orthonormal wave-frame triad $(\mathbf e_1^{\rm w}, \mathbf e_2^{\rm w}, \widehat{\mathbf k}^{\rm w})$ defined in Eq.~\eqref{gw:eq:wave-frame}, any point on the shell $r=r_i$ is parameterized by the longitudinal wave coordinate $z^{\rm w}$ and the azimuthal angle $\phi^{\rm w}$ as
\begin{equation}
\begin{aligned}
 \mathbf X_i(z^{\rm w},\phi^{\rm w})
 ={}&z^{\rm w}\widehat{\mathbf k}^{\rm w}
       +\sqrt{r_i^2-(z^{\rm w})^2}\\
 &\times(\cos\phi^{\rm w}\,\mathbf e_1^{\rm w}
              +\sin\phi^{\rm w}\,\mathbf e_2^{\rm w}).
\end{aligned}
 \label{kp:eq:sphere-u-parametrization}
\end{equation}
Here $z^{\rm w}=r_i\cos\theta^{\rm w}$ measures distance along the propagation direction, and $\phi^{\rm w}\in[0,2\pi)$ is measured from $\mathbf e_1^{\rm w}$ toward $\mathbf e_2^{\rm w}$. The position vector $\mathbf X_i$ is simultaneously a vector in the Kerr BL Cartesian frame. The corresponding Kerr BL spherical angles for $\mathbf X_i$ are obtained via standard Cartesian-to-spherical conversions:
\begin{equation}
 \Omega_i^{\rm K}(z^{\rm w},\phi^{\rm w})
 :=\bigl(\theta_i(z^{\rm w},\phi^{\rm w}),
         \phi_i(z^{\rm w},\phi^{\rm w})\bigr)\,.
 \label{eq:waveframe2kerr}
\end{equation}
Because the wave frame and the Kerr BL Cartesian frame are related by a rigid rotation $\mathsf R^{\rm w}\in\mathrm{SO}(3)$, the solid angle element is rotationally invariant:
\begin{equation}
\dd\Omega = \sin\theta\,\dd\theta\,\dd\phi = \sin\theta^{\rm w}\,\dd\theta^{\rm w}\,\dd\phi^{\rm w}\,.
\end{equation}
Substituting $\dd z^{\rm w}=-r_i\sin\theta^{\rm w}\,\dd\theta^{\rm w}$ and reversing integration limits to integrate in the direction of increasing $z^{\rm w}$, the surface integration over the sphere $S_i^2$ becomes
\begin{equation}
 \int_{S_i^2}f\,\dd\Omega
 =\frac1{r_i}\int_{-r_i}^{r_i}\dd z^{\rm w}
                  \int_0^{2\pi}\dd\phi^{\rm w}\,f(z^{\rm w},\phi^{\rm w})\,,
 \label{kp:eq:u-phi-surface-measure}
\end{equation}
where $f$ is any integrable scalar field.

When projecting onto spin-weighted spherical harmonics, the local angular dyad must be transformed consistently between frames. The standard spin-weighted spherical harmonics ${}_{-2}Y_{jm}(\theta,\phi)$ in the Kerr BL frame are referred to the coordinate tangent basis $(\mathbf e_\theta^{\rm K},\mathbf e_\phi^{\rm K})$. In contrast, spherical harmonics in the wave frame, ${}_{-2}Y_{jm}(\theta^{\rm w},\phi^{\rm w})$, are referred to the wave tangent basis
\begin{equation}
\begin{aligned}
 \mathbf e_\theta^{\rm w}
 &=\cos\theta^{\rm w}
   (\cos\phi^{\rm w}\,\mathbf e_1^{\rm w}+\sin\phi^{\rm w}\,\mathbf e_2^{\rm w})
   -\sin\theta^{\rm w}\,\widehat{\mathbf k}^{\rm w}\,,\\
 \mathbf e_\phi^{\rm w}
 &=-\sin\phi^{\rm w}\,\mathbf e_1^{\rm w}+\cos\phi^{\rm w}\,\mathbf e_2^{\rm w}\,.
\end{aligned}
\end{equation}
From these orthonormal vectors, we form the complex angular dyads in the wave frame and Kerr BL frame, respectively,
\begin{equation}
 \mathbf m_{\rm ang}^{\rm w}
 =\frac{\mathbf e_\theta^{\rm w}+\I\mathbf e_\phi^{\rm w}}{\sqrt2},
 \qquad
 \mathbf m_{\rm ang}^{\rm K}
 =\frac{\mathbf e_\theta^{\rm K}+\I\mathbf e_\phi^{\rm K}}{\sqrt2}.
\end{equation}
At each point on the sphere, the two dyads differ by a local in-plane rotation parameterized by the unimodular phase factor
\begin{equation}
 \chi_i^{\rm w}(z^{\rm w},\phi^{\rm w})
 :=\mathbf m_{\rm ang}^{\rm w}\cdot
       \overline{\mathbf m_{\rm ang}^{\rm K}},
 \qquad |\chi_i^{\rm w}|=1.
 \label{kp:eq:spin-dyad-phase}
\end{equation}
Under a dyad rotation $\mathbf m\to e^{\I\psi}\mathbf m$, a quantity of spin weight $s$ transforms as $\eta\to e^{\I s\psi}\eta$. Because $\Psi_4$ has spin weight $s=-2$, its representation with respect to the wave-frame angular dyad acquires the phase factor $(\chi_i^{\rm w})^{-2}$, giving
\begin{equation}
\begin{aligned}
 \Psi_4^{{\rm K},{\rm w}}&=(\chi_i^{\rm w})^{-2}\Psi_4^{\rm K},\\
 \partial_t\Psi_4^{{\rm K},{\rm w}}
 &=(\chi_i^{\rm w})^{-2}\partial_t\Psi_4^{\rm K}.
\end{aligned}
 \label{kp:eq:spin-dyad-correction}
\end{equation}
Here the superscript `${\rm K}$' indicates that $\Psi_4$ is obtained by contracting the Weyl tensor with the Kinnersley null tetrad in the Kerr BL frame, while `${\rm w}$' denotes that the angular dyad is aligned with the wave frame. Because the background geometry and wave orientation are stationary, $\chi_i^{\rm w}$ is time-independent at any fixed spatial point.

Using the spin-weighted spherical harmonics defined in Eq.~\eqref{eq:swsh-convention}, the field can be expanded in either angular basis at $(t_i,r_i)$ as
\begin{equation}
\begin{aligned}
 \Psi_4^{{\rm K},{\rm w}}
 &=\sum_{jm}R_{jm}^{4,{\rm w}}
                 {}_{-2}Y_{jm}(\theta^{\rm w},\phi^{\rm w}),\\
 \Psi_4^{\rm K}
 &=\sum_{jm}R_{jm}^{4,{\rm K}}{}_{-2}Y_{jm}(\theta,\phi).
\end{aligned}
 \label{kp:eq:psi4-expansion}
\end{equation}
By orthonormality of ${}_{-2}Y_{jm}(\theta^{\rm w},\phi^{\rm w})$, the wave-frame multipole coefficients are
\begin{equation}
 R_{jm}^{4,{\rm w}}
 =\int_{S_i^2}\Psi_4^{{\rm K},{\rm w}}
     {}_{-2}Y_{jm}^*(\theta^{\rm w},\phi^{\rm w})\,\dd\Omega\,,
 \label{kp:eq:dtpsi4-projection}
\end{equation}
and $\partial_tR_{jm}^{4,{\rm w}}$ is obtained by replacing $\Psi_4^{{\rm K},{\rm w}}$ with $\partial_t\Psi_4^{{\rm K},{\rm w}}$.
Substituting $\Psi_4^{\rm K}$ from Eq.~\eqref{kp:eq:psi4-helicity-shortcut} and the dyad phase factor from Eq.~\eqref{kp:eq:spin-dyad-correction}, all angular factors are grouped into the projection kernel
\begin{equation}
\begin{aligned}
 {\cal P}_{jm}^{(i),{\rm w}}(z^{\rm w},\phi^{\rm w})
 :={}&\bigl[p^{\hat j}(r_i,\theta_i)
              \epsilon_{\hat j}^{\rm w}(\theta_i,\phi_i)\bigr]^2\\
 &\times(\chi_i^{\rm w})^{-2}
              {}_{-2}Y_{jm}^*(\theta^{\rm w},\phi^{\rm w})\,.
\end{aligned}
 \label{kp:eq:wave-projection-kernel}
\end{equation}
In this kernel, the Kerr BL angles $(\theta_i,\phi_i)$ entering the Kinnersley projection are evaluated via Eq.~\eqref{eq:waveframe2kerr}, while the harmonic arguments are the wave angles $(\theta^{\rm w},\phi^{\rm w})$ with $\cos\theta^{\rm w}=z^{\rm w}/r_i$. Restricting the $z^{\rm w}$-integration to the compact support $[z_{i,-}^{\rm w}, z_{i,+}^{\rm w}]$ on the shell $r_i$ (Eq.~\eqref{kp:eq:u-integration-limits}), the mode coefficients and their time derivatives reduce to the two-dimensional quadratures
\begin{align}
 R_{jm}^{4,{\rm w}}
 &=\frac{{\cal A}^{\rm w}}{r_i}
   \int_{z_{i,-}^{\rm w}}^{z_{i,+}^{\rm w}}\dd z^{\rm w}
   \int_0^{2\pi}\dd\phi^{\rm w}\,\notag\\[-1mm]
 &\quad\times W^{\rm w}(z^{\rm w}-c^{\rm w}t_i)
                    {\cal P}_{jm}^{(i),{\rm w}},\\
 \partial_tR_{jm}^{4,{\rm w}}
 &=-\frac{c^{\rm w}{\cal A}^{\rm w}}{r_i}
   \int_{z_{i,-}^{\rm w}}^{z_{i,+}^{\rm w}}\dd z^{\rm w}
   \int_0^{2\pi}\dd\phi^{\rm w}\,\notag\\[-1mm]
 &\quad\times (W^{\rm w})'(z^{\rm w}-c^{\rm w}t_i)
                    {\cal P}_{jm}^{(i),{\rm w}}.
 \label{kp:eq:explicit-dtpsi4-projection}
\end{align}
For a general complex profile, the products $\mathcal A^{\rm w}W^{\rm w}$ and $\mathcal A^{\rm w}(W^{\rm w})'$ in Eq.~\eqref{kp:eq:explicit-dtpsi4-projection} are replaced by $\mathcal F^{\rm w}$ and $(\mathcal F^{\rm w})'$ inside the integrals. No constant complex amplitude is factored out in this case.

Finally, once the wave-frame mode coefficients $R_{jm'}^{4,{\rm w}}$ are evaluated, the physical mode coefficients in the Kerr BL frame, $R_{jm}^{4,{\rm K}}$, are obtained via a rotation of the spherical harmonic basis. Under the active Euler rotation $(\alpha,\iota,\gamma_{\rm E})$ relating the two coordinate frames, the transformation is mediated by the standard Wigner $D$-matrices
\begin{equation}
\begin{aligned}
 R_{jm}^{4,{\rm K}}
 &=\sum_{m'=-j}^{j}D^j_{mm'}(\alpha,\iota,\gamma_{\rm E})
                         R_{jm'}^{4,{\rm w}},\\
 \partial_tR_{jm}^{4,{\rm K}}
 &=\sum_{m'=-j}^{j}D^j_{mm'}(\alpha,\iota,\gamma_{\rm E})
                         \partial_tR_{jm'}^{4,{\rm w}}.
\end{aligned}
 \label{kp:eq:wave-to-kerr-modes}
\end{equation}
We highlight the distinct roles played by the two transformations. The local phase factor $(\chi_i^{\rm w})^{-2}$ in Eq.~\eqref{kp:eq:spin-dyad-correction} is a pointwise dyad rotation on the tangent bundle of the 2-sphere that aligns the spin dyad before projection, whereas the Wigner matrix $D^j_{mm'}$ performs the global $\mathrm{SO}(3)$ rotation between the modal representations.

\subsection{Conversion to the Evolution Variables}
\label{kp:sec:conversion-evolution-variables}
The angular projection and frame rotation detailed above yield the multipole coefficients $R_{jm}^{4,{\rm K}}$ and coordinate time derivatives $\partial_t R_{jm}^{4,{\rm K}}$ for the physical Weyl scalar $\Psi_4^{\rm K}$ in the Kerr BL frame. However, the $(1+1)$-dimensional numerical evolution scheme formulated in Sec.~\ref{sec:first-order-system} is solved in terms of the regularized Teukolsky master variables $(\bm\chi, \bm\Pi)$ on the hyperboloidal initial hypersurface $\tau = \tau_0$.

For spin weight $s=-2$, the Teukolsky master field $\psi^{(-2)}$ is related to the physical Weyl scalar by
\begin{equation}
 \psi^{(-2)} = \rho_{\rm NP}^{-4}\Psi_4^{\rm K} = (r - \I a\cos\theta)^4 \Psi_4^{\rm K},
\end{equation}
recalling $\rho_{\rm NP}=-(r-\I a\cos\theta)^{-1}$. To distinguish from the modal representations $R_{jm}^{4,{\rm K}}$ of $\Psi_4^{\rm K}$, we denote the BL spherical harmonic coefficients of $\psi^{(-2)}$ by $R_{jm}^{(-2)}$. In modal space at fixed azimuthal index $m$, multiplication by $\cos\theta$ is represented by the tridiagonal coupling matrix $\mathbf M_m^{(1)}$ defined in Eq.~\eqref{eq:angular-coupling-matrices}. Collecting the coupled harmonic modes into vectors $\bm R_m = (R_{j_{\min}, m}, R_{j_{\min}+1, m}, \dots)^T$, the factor $(r-\I a\cos\theta)^4$ translates into the matrix polynomial
\begin{equation}
\begin{aligned}
 \bm R_m^{(-2)}
 &=[r\Id-\I a\mathbf M_m^{(1)}]^4\bm R_m^{4,{\rm K}}\,,\\
 \partial_t\bm R_m^{(-2)}
 &=[r\Id-\I a\mathbf M_m^{(1)}]^4
                  \partial_t\bm R_m^{4,{\rm K}}\,.
\end{aligned}
 \label{kp:eq:Q-matrix-map}
\end{equation}
Because the background Kerr metric and the angular coupling matrix $\mathbf M_m^{(1)}$ are stationary, the coordinate time derivative commutes with $[r\Id-\I a\mathbf M_m^{(1)}]^4$.

By virtue of the recurrence relation in Eq.~\eqref{eq:cos-recurrence}, each action of $\mathbf M_m^{(1)}$ couples harmonic mode $j$ to $j\pm 1$. The fourth power $[r\Id-\I a\mathbf M_m^{(1)}]^4$ is therefore a banded matrix with half-bandwidth 4, which shifts the multipole index by up to four units ($\Delta j \le 4$). Consequently, if the incoming physical wave packet contains multipolar content up to an initial cutoff $J_{\rm ini}$, the resulting Teukolsky field populates modes up to $J_{\rm ini}+4$. To avoid truncation errors, the numerical evolution cutoff $J_{\rm evolution}$ must satisfy $J_{\rm evolution}\ge J_{\rm ini}+4$. The matrix multiplication in Eq.~\eqref{kp:eq:Q-matrix-map} is evaluated directly in this padded mode space.

Next, we map the BL Teukolsky modes onto the initial hyperboloidal hypersurface $\tau = \tau_0$. At each radial grid point $x_i$, the slice corresponds to the BL radius $r_i = MR(x_i)$ and coordinate time $t_i = M[\tau_0 - H(x_i)]$ (Eq.~\eqref{kp:eq:radius-dependent-time}). Recalling the regularized mode expansion in Eqs.~\eqref{eq:multipole-expansion} and \eqref{eq:R-to-varphi} for $s=-2$, the horizon-regular modal field $\bm\varphi_m^{(-2)}$ is obtained from the BL modes by peeling off the radial weight $\Delta_r^2/r$ and the horizon-regularizing azimuthal phase $\e^{\I mF(r)}$ as
\begin{equation}
\begin{aligned}
 \bm\varphi_m^{(-2)}(\tau_0,x_i)
 &=r_i\Delta_r(r_i)^{-2}\e^{-\I mF(r_i)}
                    \bm R_m^{(-2)}(t_i,r_i),\\
 \left.\partial_\tau\bm\varphi_m^{(-2)}\right|_{\tau_0,x_i}
 &=Mr_i\Delta_r(r_i)^{-2}\e^{-\I mF(r_i)}
                    \partial_t\bm R_m^{(-2)}(t_i,r_i).
\end{aligned}
 \label{kp:eq:native-data}
\end{equation}
The additional factor of $M$ in the second relation arises because $\tau = t/M + H(x)$, so that holding the spatial coordinates $(x,\theta,\phi_\star)$ fixed gives $\left.\partial_\tau\right|_{x,\theta,\phi_\star} = M\left.\partial_t\right|_{r,\theta,\phi}$.

Finally, we apply the static field rescaling defined in Eq.~\eqref{eq:static-antidamping-rescaling}, $\bm\varphi = \e^{q(x)}\bm\chi$ and $\partial_\tau\bm\varphi = \e^{q(x)}\bm\Pi$, to eliminate residual asymptotic antidamping. Inverting this redefinition yields the initial data for the first-order evolution state $\mathbf U = (\bm\chi, \bm\Pi)^T$ on the grid,
\begin{equation}
\begin{aligned}
 \bm\chi(\tau_0,x_i)
 &=\e^{-q(x_i)}\bm\varphi_m^{(-2)}(\tau_0,x_i),\\
 \bm\Pi(\tau_0,x_i)
 &=\e^{-q(x_i)}\left.\partial_\tau\bm\varphi_m^{(-2)}\right|_{\tau_0,x_i}.
\end{aligned}
 \label{kp:eq:initial-chi-pi}
\end{equation}
Together, Eqs.~\eqref{kp:eq:Q-matrix-map}--\eqref{kp:eq:initial-chi-pi} complete the systematic mapping from the physical incoming wave packet to the numerical initial state required by the time integrator.

\section{First-Order Time System and Its Coefficient Blocks}
\label{sec:first-order-system}

For each fixed azimuthal index $m$, we evolve the internally rescaled state vector
\begin{equation}
 \bm\Pi:=\partial_\tau\bm\chi,\qquad
 \mathbf U=
 \begin{pmatrix}\bm\chi\\ \bm\Pi\end{pmatrix}.
\label{eq:first-order-state}
\end{equation}
We define the pointwise angular mass matrix
\begin{equation}
 \mathbf A(x):=\mathbf M-u^2\Id
 =[1-\mu(x)-u(x)^2]\Id+\mu(x)\mathbf M^{(2)}
\label{eq:rstar-angular-mass}
\end{equation}
and the coefficient matrix multiplying $\bm\Pi$ is given by
\begin{equation}
 \widetilde{\mathbf C}
 :=\mathbf B-[u\widetilde p+u']\Id.
\label{eq:rstar-pi-coefficient}
\end{equation}
The two spatial differential operators on the right-hand side are
\begin{equation}
 \mathcal L=\partial_x^2+\widetilde p\,\partial_x
             -\widetilde{\mathbf W},\qquad
 \mathcal N=2u\,\partial_x-\widetilde{\mathbf C}.
\label{eq:continuum-LN}
\end{equation}
The first-order-in-time evolution system then reads
\begin{equation}
 \begin{pmatrix}\Id&\Zero\\ \Zero&\mathbf A\end{pmatrix}
 \partial_\tau
 \begin{pmatrix}\bm\chi\\ \bm\Pi\end{pmatrix}
 =
 \begin{pmatrix}\Zero&\Id\\ \mathcal L&\mathcal N\end{pmatrix}
 \begin{pmatrix}\bm\chi\\ \bm\Pi\end{pmatrix},
\label{eq:first-order-time}
\end{equation}
or, in component form,
\begin{align}
 \partial_\tau\bm\chi&=\bm\Pi,\notag\\
 \mathbf A\,\partial_\tau\bm\Pi
 &=2u\,\partial_x\bm\Pi+\partial_x^2\bm\chi
   +\widetilde p\,\partial_x\bm\chi
   -\widetilde{\mathbf C}\bm\Pi
   -\widetilde{\mathbf W}\bm\chi.
\label{eq:pi-equation}
\end{align}
By Eq.~\eqref{eq:corrected-antidamping}, the Hermitian part of the coefficient multiplying $\bm\Pi$ on the right-hand side is $\widetilde D\Id\le0$, so this term gives a non-positive local contribution to the quadratic balance. Stability of the numerical evolution also depends on the spatial discretization, boundary conditions, CFL restriction, and dissipation described below. At every radial grid point, $\mathbf A(x)$ is a real symmetric, positive-definite pentadiagonal matrix, assembled directly using the finite-$J$ truncation matrix $\mathbf M^{(2)}$.

\section{Discretized Equations and Numerical Algorithms}
\label{sec:numerics}
\subsection{Radial finite differences}
We discretize the spatial domain on a uniform finite-difference grid in the dimensionless tortoise coordinate $x$,
\begin{align}
 x_i&=x_{\min}+i\Delta x,\qquad i=0,\ldots,N_x-1,\notag\\
 \Delta x&=\frac{x_{\max}-x_{\min}}{N_x-1}.
\label{eq:rstar-grid}
\end{align}
The discrete mode vectors $\mathbf U_i = (\bm\chi_i, \bm\Pi_i)^T$ represent the evolved degrees of freedom.

In the computational interior, the first and second spatial derivatives are evaluated with standard fourth-order centered finite-difference stencils,
\begin{align}
 (\partial_x \varphi)_i
 &=\frac{-\varphi_{i+2}+8\varphi_{i+1}-8\varphi_{i-1}+\varphi_{i-2}}{12\Delta x},\notag\\
 (\partial_x^2\varphi)_i
 &=\frac{-\varphi_{i+2}+16\varphi_{i+1}-30\varphi_i+16\varphi_{i-1}-\varphi_{i-2}}
         {12\Delta x^2}.
\label{eq:fourth-order-rstar-stencils}
\end{align}
At the boundary-adjacent points $i=1$ and $i=N_x-2$, we apply second-order centered stencils, while at the boundary points $i=0$ and $i=N_x-1$, second-order one-sided stencils are used.

At each radial grid point $x_i$, we compute the Cholesky factorization of the angular mass matrix once during setup as
\begin{equation}
 \mathbf{A} = \mathbf{L}\mathbf{L}^{\mathsf T},
 \label{eq:banded-cholesky}
\end{equation}
where $\mathbf{L}$ is lower triangular. Because $\mathbf A$ is symmetric positive definite and pentadiagonal, this factorization exists with $L_{qq}>0$ and requires no pivoting. Furthermore, $\mathbf L$ inherits the banded structure with bandwidth 2 (having nonzero entries only on the main diagonal and first two subdiagonals).

Denoting the full spatial right-hand side of Eq.~\eqref{eq:pi-equation} by $\bm{f}$, the linear system $\mathbf{A}\,\partial_\tau\bm{\Pi}=\bm{f}$ is solved in two triangular sweeps,
\begin{equation}
 \mathbf{L}\bm{y}=\bm{f}, \qquad \mathbf{L}^{\mathsf T}\partial_\tau\bm{\Pi}=\bm{y}.
 \label{eq:banded-cholesky-substitution}
\end{equation}
The first is a forward substitution from the lowest to the highest retained multipole mode ($q=0,\ldots,J-j_{\min}$),
\begin{equation}
 y_q = \frac{f_q - L_{q,q-1}y_{q-1} - L_{q,q-2}y_{q-2}}{L_{q,q}},
\end{equation}
followed by a backward substitution in reverse order,
\begin{equation}
 (\partial_\tau\Pi)_q = \frac{y_q - L_{q+1,q}(\partial_\tau\Pi)_{q+1} - L_{q+2,q}(\partial_\tau\Pi)_{q+2}}{L_{q,q}},
 \label{eq:banded-cholesky-recurrence}
\end{equation}
where terms with out-of-range indices are omitted. Each mode couples to at most two adjacent modes, so the computational cost scales strictly linearly with the number of retained multipoles $O(J-j_{\min})$. This approach avoids computing and storing $\mathbf A^{-1}$, which is dense and would incur an $O((J-j_{\min})^2)$ matrix-vector multiplication at every grid point and Runge--Kutta stage.

\subsection{Finite-radius radiation boundaries}
Because the computational domain is truncated at finite tortoise-coordinate values $x_{\min}$ and $x_{\max}$, radiation boundary conditions must be imposed at both endpoints. The radial characteristic speeds of the hyperboloidal foliation, $c_+$ and $c_-$, are given by Eq.~\eqref{eq:hyperboloidal-characteristic-speeds}. At the inner boundary $x_{\min}$, only ingoing radiation (propagating at speed $c_-$) is expected, so we retain only the $c_-$ mode, whereas at the outer boundary $x_{\max}$, only outgoing radiation (propagating at speed $c_+$) is physical and we retain only the $c_+$ mode.

Expressed in terms of the rescaled evolution variables, the resulting one-way Sommerfeld-type condition takes the form \begin{equation} 
\bm{\Pi} = -c_\pm (\partial_x \bm{\chi} + q' \bm{\chi})\,, \label{eq:rstar-radiation-condition} 
\end{equation} 
which reduces to the standard radiation condition $\partial_\tau \bm{\varphi} = -c_\pm \partial_x \bm{\varphi}$ for the regular physical field~$\bm{\varphi}$ once the field redefinition is inverted. Differentiating Eq.~\eqref{eq:rstar-radiation-condition} with respect to the foliation time~$\tau$ yields the boundary evolution equation for~$\bm{\Pi}$
\begin{equation} 
\partial_\tau \bm{\Pi} = -c_\pm (\partial_x \bm{\Pi} + q' \bm{\Pi})\,. \label{eq:rstar-radiation-time-derivative} 
\end{equation}

\subsection{Kreiss--Oliger dissipation}
To suppress spurious high-frequency oscillations that can arise across the $C^2$ slicing transition regions, we apply sixth-order Kreiss--Oliger dissipation to the real and imaginary parts of both $\bm\chi$ and $\bm\Pi$, given by
\begin{align}
 (Q_{\rm KO}\varphi)_i={}&\frac{\epsilon_{\rm KO}}{64\Delta x}
 \bigl(\varphi_{i-3}-6\varphi_{i-2}+15\varphi_{i-1}-20\varphi_i\notag\\
 &\hspace{18mm}+15\varphi_{i+1}-6\varphi_{i+2}+\varphi_{i+3}\bigr).
\label{eq:ko-dissipation}
\end{align}
This operator scales as $O(\Delta x^5)$ for smooth solutions, acting as a high-frequency filter without degrading the fourth-order spatial accuracy in the bulk. In practice, we set $\epsilon_{\rm KO}=0.05$.

It is worth noting that Kreiss–Oliger dissipation acts only on the highest-frequency, grid-scale modes that are poorly resolved and unphysical. Its impact on well-resolved, low-frequency physical modes is suppressed at order $O(\Delta x^5)$ (or higher), thereby leaving the underlying physical solution virtually undisturbed.

\subsection{RK4 evolution and parallel layout}
We integrate the first-order system~\eqref{eq:first-order-time} in time using the classical four-stage, fourth-order Runge--Kutta (RK4) method. At each intermediate stage, we evaluate the spatial operators $\mathcal L\bm\chi$ and $\mathcal N\bm\Pi$, solve the banded linear system $\mathbf A(x_i)\,\partial_\tau\bm\Pi_i=\bm f_i$ pointwise across the radial grid using the pre-factored Cholesky sweeps, and apply the boundary and dissipation operators.

In the method-of-lines framework, RK4 stability requires that $\Delta\tau\lambda$ falls within the RK4 stability domain for all eigenvalues $\lambda$ of the semi-discrete spatial operator. For the advection-dominated wave sector, the spectral radius scales as $\max|c_\pm|/\Delta x$. Since our localized gauge guarantees $\max|c_\pm|\le(1-\beta_{\rm cap})^{-1}=5$ throughout the domain, the uniform CFL condition is
\begin{equation}
 \Delta\tau\le0.4\,\Delta x(1-\beta_{\rm cap}),
 \qquad \beta_{\rm cap}=0.8.
\label{eq:rstar-cfl}
\end{equation}
Unlike non-uniform or stretched meshes where spatial variations in grid spacing distort local dispersion relations and risk artificial reflections at refinement interfaces, the uniform tortoise grid enables homogeneous finite-difference stencils and Kreiss--Oliger dissipation everywhere in the bulk under a single globally uniform CFL limit.

Because different azimuthal sectors $m$ decouple completely, we parallelize the numerical system by distributing independent $m$-sectors across MPI processes, with each process holding the full radial grid for its assigned modes. Within each process, local radial loop operations are parallelized using OpenMP multithreading.
\section{Numerical Results}
\label{sec:numerical-results}
In this section, we present numerical results from our simulations. The incident gravitational-wave packet has a Hanning-windowed waveform profile, $\sin^2(\pi \zeta^{\rm w}/\lambda)$, with length $\lambda/M = 4$. In the wave frame, the packet propagates along the $+z^{\rm w}$ axis with polarization amplitudes $A_+^{\rm w} = A_\times^{\rm w} = 1$. Because the evolution is linear, these unit amplitudes specify a normalization; the small-perturbation regime refers to a physically small overall amplitude multiplying the prescribed initial profiles.
At coordinate time $t=0$, the initial wavefront is located at $z_0^{\rm w}/M = -400$ and intersects the initial slice $\tau = \tau_0$.

The system is evolved for a dimensionless foliation-time interval $\tau_{\rm final}-\tau_0=876$, using a time step $\Delta\tau=0.004$ determined by the CFL condition Eq.~\eqref{eq:rstar-cfl}. This corresponds to $219{,}000$ RK4 steps. The spatial computational domain in the tortoise coordinate spans from $x_{\min}=-300$ to $x_{\max}=800$, with grid spacing $\Delta x=0.05$.

For the background Kerr black hole, we fix the dimensionless spin parameter at $\widehat{a} = 0.7$. The azimuthal angle $\alpha$ of the incident wave direction $\widehat{\mathbf{k}}^{\mathrm{w}}$ can be set to zero by shifting the origin of the Kerr azimuthal coordinate $\phi$, so it does not correspond to a new physical configuration. We therefore vary only the polar angle $\iota$ between $\widehat{\mathbf{k}}^{\mathrm{w}}$ and the black-hole spin axis, considering five values, namely $\iota = 0$, $\pi/3$, $\pi/2$, $2\pi/3$, and $\pi$. The cases $\iota = 0$ and $\iota = \pi$ correspond to aligned and anti-aligned incidence, respectively. The polarization angle is fixed at $\gamma_{\rm E} = \pi/3$.
The physical initial multipole modes span $j_{\min}=\max(2,|m|)$ to $J_{\rm ini}=256$, with $|m|\le256$. Multiplication by $(r-\I a\cos\theta)^4$ in Eq.~\eqref{kp:eq:Q-matrix-map} couples multipoles with $|\Delta j|\le4$, so the evolution cutoff is padded to $J_{\rm evolution}=260$. The azimuthal range remains $|m|\le256$. The Wigner rotations act within each initial $j$, and the subsequent Kerr evolution does not couple different $m$ sectors. For the misaligned configurations, the number of evolved modes is
\begin{equation}
N_{\rm modes}=\sum_{m=-256}^{256}
  [261-\max(2,|m|)]=68{,}097\,.
\end{equation}
\subsection{Hanning profile and Gaussian beams}
\begin{figure*}[!tp]
\begin{center}
\includegraphics[scale=0.4]{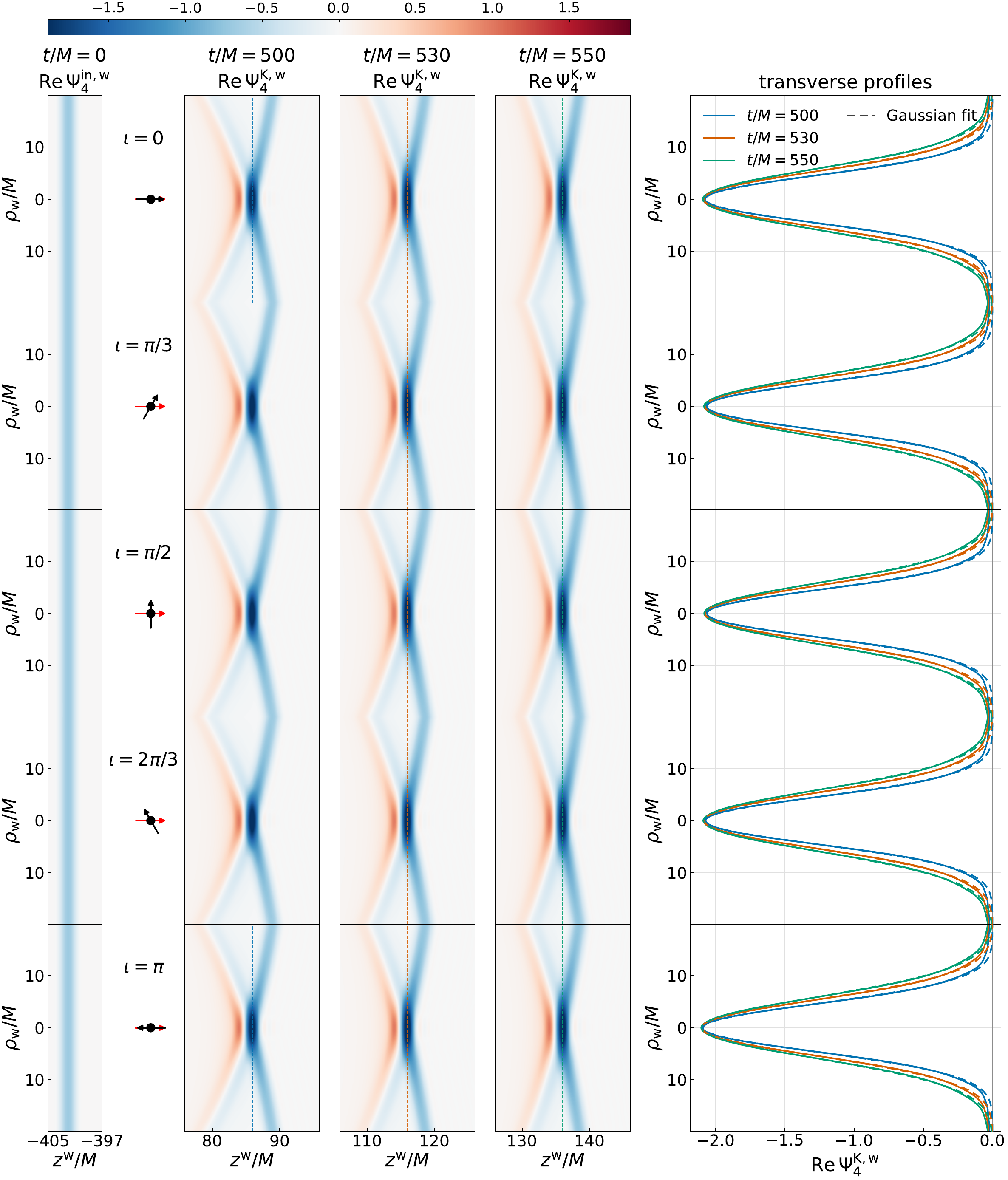}
\end{center}
\caption{Lensing of a Hanning-windowed gravitational wave packet by a Kerr black hole ($\widehat{a} = 0.7$, $\gamma_{\rm E} = \pi/3$) across five polar inclinations $\iota \in \{0, \pi/3, \pi/2, 2\pi/3, \pi\}$ (top to bottom) displayed in the $\widehat{\mathbf{k}}^{\rm w}$--$\mathbf{S}$ plane for noncollinear incidence. The central schematic marks the black hole position, incident wavevector $\widehat{\mathbf{k}}^{\rm w}$ along $+z^{\rm w}$ (red arrow), and spin $\mathbf{S}$ inclined by $\iota$ (black arrow). (Left column) Initial wave packet $\mathrm{Re}(\Psi_4^{\rm in,w})$ at $t/M = 0$. (Middle three columns) Snapshots of the lensed radiation field $\mathrm{Re}\,\Psi_4^{\rm K,w}$ at $t/M = 500$, $530$, and $550$, showing curved wavefronts and diffractive focusing, with vertical dashed lines indicating the longitudinal peak positions. (Right column) Transverse profiles of $\mathrm{Re}\,\Psi_4^{\rm K,w}(\rho_{\rm w})$ along the wavefront peak at $t/M = 500$ (blue), $530$ (orange), and $550$ (green), overlaid with Gaussian fits (dashed black curves).}
\label{fig:wave-scattering-snapshots}
\end{figure*}

Figure~\ref{fig:wave-scattering-snapshots} illustrates the two-dimensional lensing dynamics for incident wave packets across five polar inclinations, namely $\iota = 0$, $\pi/3$, $\pi/2$, $2\pi/3$, and $\pi$ (from top to bottom), with polarization angle $\gamma_{\rm E} = \pi/3$ on a Kerr black hole with $\widehat{a} = 0.7$. For $0<\iota<\pi$, the spatial distributions are displayed in the plane containing the incident wavevector $\widehat{\mathbf{k}}^{\rm w}$ and the black-hole spin vector $\mathbf{S}$. The upper transverse direction follows the projection of $\mathbf S$ orthogonal to $\widehat{\mathbf{k}}^{\rm w}$. At $\iota=0$ and $\pi$, where these vectors are collinear, we retain the continuous limiting section at fixed $\gamma_{\rm E}$. Its wave-frame azimuth is $\phi_{\rm sec}=\pi-\gamma_{\rm E}=2\pi/3$ for the parameters used here. 

To distinguish the two halves of this section, we introduce a signed transverse coordinate $\xi_\perp$ such that $(x^{\rm w},y^{\rm w})=\xi_\perp(\cos\phi_{\rm sec},\sin\phi_{\rm sec})$, related to the cylindrical radius by $\rho_{\rm w}=\sqrt{(x^{\rm w})^2+(y^{\rm w})^2}=|\xi_\perp|\ge0$. The upper ($\xi_\perp>0$) and lower ($\xi_\perp<0$) halves of each panel correspond to the opposite azimuthal directions $\phi_{\rm sec}$ and $\phi_{\rm sec}+\pi$, respectively. Although the vertical axis ticks on both halves display the nonnegative radius $\rho_{\rm w}/M$, comparing the upper and lower wings directly contrasts opposite azimuthal sides at equal radii.

At $t/M = 0$ (left column), the incident planar wave packet possesses a compact Hanning profile centered at $z^{\rm w}/M \approx -402$. Near the incident axis, the propagation direction is approximately $-\mathbf e_{\hat r}$, and the incoming radiation is represented primarily by $\Psi_0^{\rm K}$ in the outward-oriented Kinnersley tetrad. To compare its normalization with the outgoing radiation, we reverse the null directions and include a tetrad boost,
\begin{equation}
\begin{aligned}
 \ell'&=2n_{\rm K},\qquad n'=\ell_{\rm K}/2,\qquad m'=\overline m_{\rm K},\\
 \Psi'_4&=-\delta C(n',\overline m',n',\overline m')
=\frac14\Psi_0^{\rm K}.
\end{aligned}
\label{eq:incident-tetrad-reversal}
\end{equation}
The factor $1/4$ follows from the two factors of $n'=\ell_{\rm K}/2$ in the Weyl contraction. After expressing the transverse basis in the wave frame, we denote this incident radiation field by $\Psi_4^{\rm in,w}$. This convention fixes its normalization relative to the outgoing field. 

As the wave packet traverses the strong-gravity region of the black hole (indicated by the central black dot in the schematic, with red and black arrows denoting $\widehat{\mathbf{k}}^{\rm w}$ and $\mathbf{S}$, respectively), strong gravitational deflection deforms the originally planar wavefront. The middle columns display snapshots of the lensed field $\mathrm{Re}\,\Psi_4^{\rm K,w}$ in the $\widehat{\mathbf{k}}$--$\mathbf{S}$ plane at $t/M = 500$, $530$, and $550$. One clearly observes the development of a curved, fold-like wavefront geometry, accompanied by constructive interference and diffraction fringes along the forward propagation direction behind the black hole. 
To quantify the beam structure, the rightmost column presents the one-dimensional transverse profiles of $\mathrm{Re}\,\Psi_4^{\rm K,w}$ on the two opposite azimuthal sides, labeled by $\rho_{\rm w}/M$, evaluated at the longitudinal peak of the wavefront (marked by the vertical dashed lines in the snapshot panels), alongside their Gaussian fits (dashed curves). In all five cases, the wave amplitude is strongly focused near the forward axis $\rho_{\rm w} \approx 0$, forming a prominent central beam whose transverse profile is accurately characterized by a Gaussian distribution. As the wave packet propagates into the wave zone ($t/M = 500 \to 550$), the beam waist gradually broadens while preserving its Gaussian profile. 

\begin{figure*}[!tp]
\begin{center}
\includegraphics[scale=0.7]{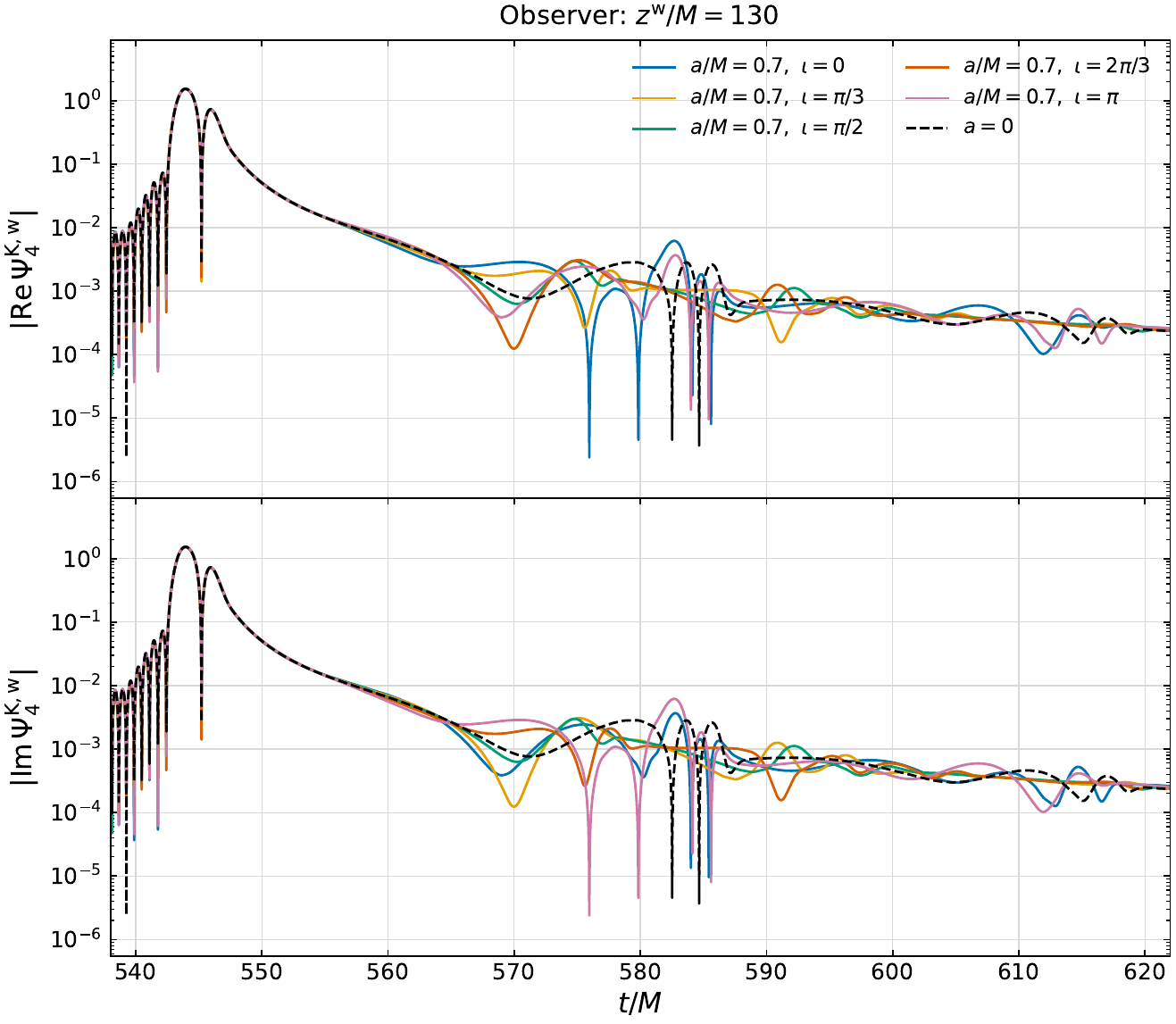}
\end{center}
\caption{Comparison of lensed gravitational waveforms observed at a downstream on-axis location $z^{\rm w}/M = 130$ ($\rho_{\rm w} = 0$) as a function of time $t/M$. The upper and lower panels display the absolute values of the real part $|\mathrm{Re}\,\Psi_4^{\mathrm{K,w}}|$ and imaginary part $|\mathrm{Im}\,\Psi_4^{\mathrm{K,w}}|$ on a logarithmic scale, respectively. Six distinct configurations are shown: five Kerr black hole cases with spin $a/M = 0.7$ at inclinations $\iota = 0$ (blue), $\pi/3$ (yellow), $\pi/2$ (green), $2\pi/3$ (orange), and $\pi$ (pink), alongside the non-rotating Schwarzschild baseline ($a = 0$, dashed black curves). All cases share the incident polarization angle $\gamma_{\rm E} = \pi/3$. While the prompt pulse ($t/M \approx 538$--$555$) exhibits nearly identical wave profiles across all cases, the late-time coda ($t/M \gtrsim 565$) develops rich, inclination-dependent interference notches and diffractive modulations, with the Schwarzschild benchmark clearly separating from the spinning Kerr counterparts.}
\label{fig:wave-scattering-observer}
\end{figure*}

To examine the temporal structure and late-time coda of the lensed signal in detail, Fig.~\ref{fig:wave-scattering-observer} presents the time evolution of the wave-frame Weyl scalar $\Psi_4^{\mathrm{K,w}}$ recorded by a fixed downstream observer situated on the forward propagation axis at $(x^{\rm w},y^{\rm w},z^{\rm w})=(0,0,130M)$. The absolute values of the real part, $|\mathrm{Re}\,\Psi_4^{\mathrm{K,w}}|$ (upper panel), and of the imaginary part, $|\mathrm{Im}\,\Psi_4^{\mathrm{K,w}}|$ (lower panel), are plotted on a logarithmic scale over the time interval $t/M \in [538, 622]$. In addition to the five Kerr cases with spin $a/M = 0.7$ spanning inclinations from aligned ($\iota = 0$) to anti-aligned ($\iota = \pi$), we include the non-rotating Schwarzschild black hole ($a = 0$) as a reference benchmark.

During the prompt stage ($t/M \approx 538$--$555$), the main wave packet arrives with peak amplitude exceeding unity near $t/M \approx 544$. Remarkably, all six waveforms coincide almost indistinguishably throughout this initial pulse, indicating that the coherent forward beam is dominated by direct transmission through the curved exterior and remains largely insensitive to the background spin. In contrast, differences emerge in the subsequent lensing coda ($t/M \gtrsim 565$). The non-spinning Schwarzschild baseline ($a = 0$) exhibits a distinctly different interference pattern compared to the Kerr configurations, clearly revealing the impact of frame dragging and ergosphere geometry on the lensed wave tails. Furthermore, varying the inclination angle $\iota$ for the Kerr background systematically shifts the positions of the interference minima and alters the relative amplitudes between the real and imaginary components. The aligned ($\iota = 0$) and anti-aligned ($\iota = \pi$) configurations exhibit distinct coda dynamics due to differential spin-orbit coupling between prograde and retrograde trajectories, while intermediate inclinations smoothly interpolate between these extremes. 

\subsection{Realistic binary black hole coalescence waveforms}
To connect our time-domain framework directly to astrophysical observations, we simulate the strong-field gravitational lensing of a realistic chirping waveform emitted by a coalescing binary black hole (BBH). In this setup, the incident wave is generated by a non-spinning, equal-mass binary with component masses $m_1 = m_2 = 22.5\,M_{\odot}$ (yielding a total binary mass of $45\,M_{\odot}$) in a face-on orientation. This incident wave packet impinges on an isolated black hole lens of mass $M = 157.5\,M_{\odot}$ with polarization angle $\gamma_{\rm E} = 0$. The two strain polarizations are generated using the surrogate model \texttt{NRHybSur3dq8}~\cite{Varma:2018mmi}. Their second time derivatives define the general complex curvature profile $\mathcal F^{\rm w}$ in Eq.~\eqref{gw:eq:general-complex-curvature}. The corresponding Kinnersley scalar and its time derivative are obtained by the tetrad projection described above. At coordinate time $t=0$, the initial wavefront is located at $z_0^{\rm w}/M = -400$, intersecting the initial slice $\tau = \tau_0$, the same as before. The resulting radiation field is recorded by a downstream observer located on the optical axis at $z^{\rm w}/M = 130$ in the wave frame. The plus ($h_+$) and cross ($h_\times$) polarizations are recovered by double integration.

\begin{figure*}[!tp]
\begin{center}
\includegraphics[width=0.92\textwidth]{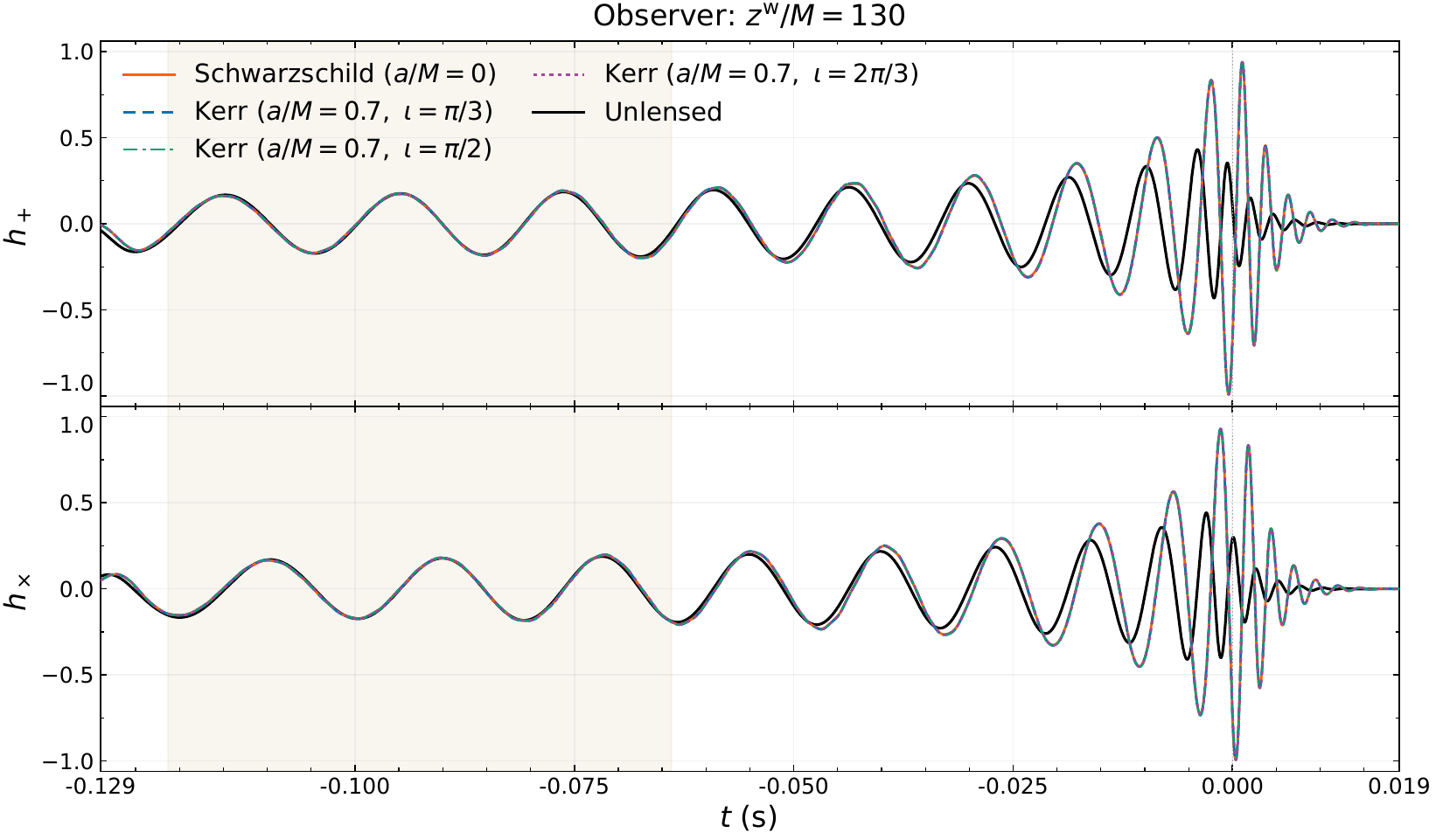}
\end{center}
\caption{Gravitational-wave strains from a binary black hole coalescence lensed by a Schwarzschild black hole ($a/M = 0$, orange solid curves) and a Kerr black hole ($a/M = 0.7$) at various incident angles relative to the Kerr spin axis $\iota = \pi/3\,,\pi/2\,,2\pi/3$ (blue dashed, green dash-dotted and purple dotted curves), compared with the unlensed incident signal (black solid curves). The source is a face-on, non-spinning, equal-mass binary with component masses of $22.5\,M_\odot$ (total mass $45\,M_\odot$). Both lenses have a mass of $M = 157.5\,M_\odot$, and the incident polarization angle is $\gamma_{\rm E} = 0$. The upper and lower panels display the plus ($h_+$) and cross ($h_\times$) polarizations, respectively, as recorded by an on-axis downstream observer at $z^{\mathrm{w}}/M = 130$. Time $t$ is measured in seconds relative to the merger peak at $t = 0$. The lensed strains are normalized to unit peak amplitude, whereas the unlensed signal is rescaled to match the lensed waveforms within the early-inspiral interval $t \in [-0.118, -0.063]\,\mathrm{s}$ (shaded region). While lensing leaves the early inspiral nearly unaltered, it induces substantial amplitude and phase modulations during the merger and ringdown phases. For the inclinations shown, the normalized Kerr and Schwarzschild strains nearly overlap in both polarizations during inspiral and merger on this plotting scale.}
\label{fig:realistic-bbh-lensing}
\end{figure*}

Figure~\ref{fig:realistic-bbh-lensing} compares the lensed GW strain (orange solid curves) with the unlensed incident waveform (black solid curves) for both $h_+$ and $h_\times$ polarizations as functions of time $t$ in seconds. The waveform amplitudes are normalized such that the maximum peak of the lensed signal reaches unity. The unlensed waveform is normalized by matching its first few cycles to the lensed signal during the early inspiral stage, as delineated by the shaded region over $t \in [-0.118, -0.063]\,\text{s}$. 

Throughout this early inspiral window, the lensed and unlensed waveforms track each other with near-perfect fidelity. This is because the wavelength of the incident waves is much larger than the gravitational radius of the lensing black hole. The signal propagates without noticeable distortion. However, as the binary accelerates toward merger, the orbital frequency increases and the emitted gravitational wavelength decreases, eventually becoming comparable to the scale of the lens. In this strong-field diffraction regime, the lensed ringdown exhibits substantial modifications in amplitude, instantaneous frequency, and phase, accompanied by a pronounced, modulated late-time coda.

Figure~\ref{fig:realistic-bbh-lensing} further shows that, for rotating Kerr black holes, the lensed waveforms during the inspiral and merger phases are nearly indistinguishable from those of the non-spinning Schwarzschild baseline, regardless of the inclination angle of the incident gravitational wave relative to the BH's spin axis. To quantify these differences, we use mismatch, defined as $1-{\rm FF}$, where $\rm FF$ is the fitting factor (FF)~\cite{PhysRevD.52.605}
\begin{equation}
\mathrm{FF}\equiv \max_{t_0}\,|z(t_0)|  \,.
\end{equation}
$z(t_0)$ is the complex overlap defined as
\begin{equation}
\begin{aligned}
z(t_0)&\equiv\frac{4}{\sigma(h_{\rm Kerr}) \sigma(h_{\rm Schw})}
\!\int_0^{\infty} \!\frac{\tilde{h}_{\rm Kerr} \overline{\tilde{h}}_{\rm Schw}}{S_n} \\
&\quad \times e^{2\pi i f t_0}\,\mathrm{d}f \,,
\end{aligned}
\label{eq:z_complex}
\end{equation}
$S_n(f)$ is the one-sided power spectral density (PSD) of the detector noise. The normalization $\sigma(\cdot)$ is defined by
\begin{equation}
\sigma^2(h)=4\int_0^{\infty} \frac{|\tilde{h}(f)|^2}{S_n(f)}\, \mathrm{d}f\,.
\end{equation}
When evaluating Eq.~(\ref{eq:z_complex}), we integrate over the frequency range from $20\,\text{Hz}$ to $1024\,\text{Hz}$ using the Advanced LIGO (aLIGO) sensitivity curve projected for the O4 observing run\footnote{\url{https://dcc.ligo.org/T2200043-v3/public}}.
\begin{table}[tbp]
	\centering
	\begin{tabular}{c|c|c}
\hline
Inclination & $h_{+}$ & $h_{\times}$ \\
\hline
\(\pi/3\) & \(1.814\times10^{-5}\) & \(1.815\times10^{-5}\)   \\
\hline
\(\pi/2\) & \(2.139\times10^{-5}\) & \(2.139\times10^{-5}\)   \\
\hline
\(2\pi/3\) & \(1.322\times10^{-5}\) & \(1.323\times10^{-5}\) \\
\hline
\end{tabular}
\caption{Mismatches between Kerr- and Schwarzschild-lensed waveforms}
\label{tab:my-table}
\end{table}
Table~\ref{tab:my-table} presents the mismatches between Kerr- and Schwarzschild-lensed waveforms across various inclinations and polarizations. These values are typically below $2.2\times10^{-5}$, indicating that the two waveforms are practically indistinguishable at current LIGO sensitivity. This finding is consistent with our earlier analysis of a Hanning-windowed gravitational-wave packet across different lens spins. The early-time and peak segments of the forward-lensed signal are largely insensitive to the spin of the lens, as these regimes are dominated by rays with large impact parameters traversing the exterior spacetime.

\section{Discussions}
\label{sec:discussion}
In this work, we generalized our previous $(1+1)$D time-domain mode-sum framework for gravitational-wave lensing from a Schwarzschild background~\cite{gg33-r64q} to the full Kerr spacetime. In Boyer--Lindquist coordinates, the Teukolsky equation contains local antidamping terms that drive error growth at large radii, leading to numerical instabilities. We overcome this issue by introducing a localized hyperboloidal foliation that bounds the height-function slope to $|u|\le 0.8$, paired with a static field redefinition $\bm{\varphi}=e^{q(x)}\bm{\chi}$. This combination also preserves bounded characteristic speeds on an extended, uniform tortoise-coordinate grid, thereby enabling the use of an efficient, explicit RK4 evolution scheme.

Furthermore, we construct the initial incident GWs from a plane-wave Weyl tensor, projecting the incident Weyl scalars through a sequence of ZAMO-to-Kinnersley transformations. By combining local spin-dyad rotations with global Wigner $D$-matrix transformations, we establish an initial-data prescription for arbitrary incidence and polarization angles $(\iota,\gamma_{\rm E})$. To handle the arbitrary non-axisymmetric configurations, we developed an efficient parallel scheme capable of concurrently evolving over $68,000$ coupled $(j,m)$ modes. 

Based on our code, we performed simulations of both Hanning-windowed wave packets and realistic binary-coalescence waveforms lensed by Schwarzschild and Kerr backgrounds. For Hanning-windowed wave packets, our simulations reveal a coherent picture of strong-field wave-optics lensing. Gravitational deflection focuses the transmitted field into a prominent forward beam with a Gaussian transverse profile, demonstrating that the black hole acts as a diffraction-limited thick gravitational lens rather than an ideal point-mass deflector. Frame dragging breaks the equatorial symmetry for misaligned incidence, imprinting inclination-dependent transverse shear in the beam. The observer waveforms show a clear two-stage temporal structure. The prompt pulse is nearly spin-independent across all simulated inclinations, since it is dominated by wide-impact-parameter rays in the curved exterior. The subsequent lensing coda, by contrast, is exquisitely sensitive to the black-hole spin, with inclination-dependent interference notches arising from wave components that grazed the photon sphere and accumulated spin-dependent phase shifts distinguishing prograde from retrograde trajectories. These features are absent from both geometric optics, which predicts an unphysical singular caustic, and thin-lens models, which omit the strong-field effects entirely.

In a realistic binary coalescence scenario, strong-field lensing leaves the early inspiral largely unaffected while substantially modifying the amplitude, instantaneous frequency, and phase of the post-merger ringdown. For rotating Kerr black holes, the lensed inspiral and merger waveforms remain virtually indistinguishable from their non-spinning Schwarzschild counterparts. The mismatches between Kerr- and Schwarzschild-lensed waveforms across various
inclinations and polarizations are typically  below $2.2\times10^{-5}$, which means that the two waveforms are practically indistinguishable at current LIGO sensitivity.
This spin insensitivity arises because the lensed waveforms are dominated by null geodesics with large impact parameters that probe only the exterior spacetime. Consequently, our findings demonstrate that a Schwarzschild background provides an accurate and sufficient approximation for modeling strong-field lensing with current ground-based detectors.

Unlike in the Kerr spacetime, the $j$-modes in the Schwarzschild background are mutually decoupled. This enables highly efficient parallelization, which substantially reduces the computational cost of generating training data across a broad region of parameter space. Furthermore, when combined with surrogate modeling~\cite{PhysRevX.4.031006}, this framework enables rapid waveform generation and evaluation, supporting dedicated searches in both archival and future GW data. A detailed analysis will be presented in our follow-up papers. 

\vspace{5mm}
\noindent \textbf{Acknowledgments}
This work is supported by the National Key R\&D Program of China (Grant No.~2021YFC2203002) and the National Natural Science Foundation of China (Grant No.~12475058).

\section*{Appendix: Spin-Weighted Spherical Harmonic Conventions}
Throughout this paper, we adopt the standard convention for the Wigner small-$d$ matrix,
\begin{equation}
 d^j_{m'm}(\theta)
 \equiv\langle jm'|\exp(-\I\theta J_y)|jm\rangle ,
\label{eq:wigner-small-d-definition}
\end{equation}
where $J_y$ is the $y$-component of the angular momentum operator. Under this convention,
the full spin-weighted spherical harmonics are defined as
\begin{align}
 {}_sY_{jm}(\theta,\phi)
 &=(-1)^m\sqrt{\frac{2j+1}{4\pi}}\,
   d^j_{-s,m}(\theta)e^{\I m\phi}                                    \notag\\
 &\equiv e^{\I m\phi}\,{}_sy_{jm}(\theta)\,.
\label{eq:swsh-convention}
\end{align}
The reduced angular functions ${}_sy_{jm}(\theta)$ contain the polar dependence through
\begin{equation}
 {}_sy_{jm}(\theta)
 =(-1)^m\sqrt{\frac{2j+1}{4\pi}}\,d^j_{-s,m}(\theta)\,.
\label{eq:reduced-swsh-definition}
\end{equation}
They satisfy the eigenvalue equation
\begin{equation}
 \Bigl[-\frac{1}{\sin\theta}\partial_\theta
   (\sin\theta\,\partial_\theta)
 +\frac{(m+s\cos\theta)^2}{\sin^2\theta}-s\Bigr]{}_sy_{jm}
 =\lambda_j{}_sy_{jm},
\label{eq:angular-eigenvalue}
\end{equation}
with eigenvalues $\lambda_j=(j-s)(j+s+1)$ as given in Eq.~\eqref{eq:lambda-def}.
For a fixed spin weight $s$, ${}_sY_{jm}(\theta,\phi)$ satisfies orthonormality over the two-sphere,
\begin{equation}
 \int_{S^2}{}_sY_{\ell m}^{*}{}_sY_{jm'}\,\dd\Omega
 =\delta_{\ell j}\delta_{mm'}\,.
\label{eq:swsh-orthogonality}
\end{equation}
Note, however, that harmonics with different spin weights $s\neq s'$ are generally not orthogonal.
As an explicit example, consider
\begin{equation}
{}_0Y_{11}(\theta,\phi)
=Y_{11}(\theta,\phi)=-\sqrt{\frac{3}{8\pi}}\sin\theta\,e^{\I\phi},
\end{equation}
and
\begin{equation}
{}_1Y_{11}(\theta,\phi)=-\sqrt{\frac{3}{16\pi}}(1-\cos\theta)e^{\I\phi}.
\end{equation}
Evaluating their inner product yields a non-vanishing overlap,
\begin{align}
\int_{S^2} {}_0Y_{11}^{*}\,{}_1Y_{11}\,\dd\Omega
&=
\frac{3}{8\sqrt{2}\pi}
\int_0^{2\pi}\dd\phi
\int_0^\pi
\sin^2\theta\,(1-\cos\theta)\,\dd\theta \nonumber \\
&=\frac{3\pi}{8\sqrt{2}}
\neq 0.
\end{align}

\bibliography{bibliography}

@article{NonlinearRingdown1,
  title = {Nonlinear Effects in Black Hole Ringdown},
  author = {Cheung, Mark Ho-Yeuk and Baibhav, Vishal and Berti, Emanuele and Cardoso, Vitor and Carullo, Gregorio and Cotesta, Roberto and Del Pozzo, Walter and Duque, Francisco and Helfer, Thomas and Shukla, Estuti and Wong, Kaze W. K.},
  journal = {Phys. Rev. Lett.},
  volume = {130},
  issue = {8},
  pages = {081401},
  numpages = {6},
  year = {2023},
  month = {Feb},
  publisher = {American Physical Society},
  doi = {10.1103/PhysRevLett.130.081401},
  url = {https://link.aps.org/doi/10.1103/PhysRevLett.130.081401},
  eprint = {2208.07374},
  archivePrefix = {arXiv},
  primaryClass = {gr-qc}
}

@article{NonlinearRingdown2,
  author  = {Mitman, Keefe and Lagos, Macarena and Stein, Leo C. and Ma, Sizheng
             and Hui, Lam and Chen, Yanbei and Deppe, Nils
             and H\'ebert, Fran{\c c}ois and Kidder, Lawrence E. and Moxon, Jordan
             and Scheel, Mark A. and Teukolsky, Saul A. and Throwe, William
             and Vu, Nils L.},
  title   = {Nonlinearities in Black Hole Ringdowns},
  journal = {Phys. Rev. Lett.},
  volume  = {130},
  pages   = {081402},
  year    = {2023},
  doi     = {10.1103/PhysRevLett.130.081402},
  eprint  = {2208.07380},
  archivePrefix = {arXiv},
  primaryClass = {gr-qc}
}

@ARTICLE{He2022PRD,
       author = {{He}, Jian-hua and {Wu}, Zhenyu},
        title = "{Simulating gravitational waves passing through the spacetime of a black hole}",
      journal = {Phys. Rev. D},
         year = 2022,
        month = dec,
       volume = {106},
       number = {12},
          eid = {124037},
        pages = {124037},
          doi = {10.1103/PhysRevD.106.124037},
       eprint = {2208.01621},
archivePrefix = {arXiv},
 primaryClass = {gr-qc},
       adsurl = {https://ui.adsabs.harvard.edu/abs/2022PhRvD.106l4037H}
}

@article{Yin2024PRL,
  title = {Detectability of Single Spinless Stellar-Mass Black Holes through Gravitational Lensing of Gravitational Waves with Advanced LIGO},
  author = {Yin, Chengjiang and He, Jian-hua},
  journal = {Phys. Rev. Lett.},
  volume = {132},
  issue = {1},
  pages = {011401},
  numpages = {6},
  year = {2024},
  month = {Jan},
  publisher = {American Physical Society},
  doi = {10.1103/PhysRevLett.132.011401},
  url = {https://link.aps.org/doi/10.1103/PhysRevLett.132.011401},
  eprint = {2312.12451},
  archivePrefix = {arXiv},
  primaryClass = {astro-ph.HE}
}

@article{Dafermos:2016uzj,
    author = "Dafermos, Mihalis and Holzegel, Gustav and Rodnianski, Igor",
    title = "{The linear stability of the Schwarzschild solution to gravitational perturbations}",
    eprint = "1601.06467",
    archivePrefix = "arXiv",
    primaryClass = "gr-qc",
    doi = "10.4310/acta.2019.v222.n1.a1",
    journal = "Acta Math.",
    volume = "222",
    number = "1",
    pages = "1--214",
    year = "2019"
}

@article{PhysRevD.74.084028,
  title = {Numerical study of the quasinormal mode excitation of Kerr black holes},
  author = {Dorband, Ernst Nils and Berti, Emanuele and Diener, Peter and Schnetter, Erik and Tiglio, Manuel},
  journal = {Phys. Rev. D},
  volume = {74},
  issue = {8},
  pages = {084028},
  numpages = {18},
  year = {2006},
  month = {Oct},
  publisher = {American Physical Society},
  doi = {10.1103/PhysRevD.74.084028},
  url = {https://link.aps.org/doi/10.1103/PhysRevD.74.084028},
  eprint = {gr-qc/0608091},
  archivePrefix = {arXiv},
  primaryClass = {gr-qc}
}

@article{Andersson,
  title = {Excitation of Schwarzschild black-hole quasinormal modes},
  author = {Andersson, Nils},
  journal = {Phys. Rev. D},
  volume = {51},
  issue = {2},
  pages = {353--363},
  numpages = {0},
  year = {1995},
  month = {Jan},
  publisher = {American Physical Society},
  doi = {10.1103/PhysRevD.51.353},
  url = {https://link.aps.org/doi/10.1103/PhysRevD.51.353}
}

@misc{Dafermos:2021cbw,
    author = "Dafermos, Mihalis and Holzegel, Gustav and Rodnianski, Igor and Taylor, Martin",
    title = "{The non-linear stability of the Schwarzschild family of black holes}",
    eprint = "2104.08222",
    archivePrefix = "arXiv",
    primaryClass = "gr-qc",
    month = "4",
    year = "2021"
}

@article{Suyama:2005mx,
      author         = "Suyama, Teruaki and Takahashi, Ryuichi and Michikoshi,
                        Shugo",
      title          = "{Wave propagation in a weak gravitational field and the
                        validity of the thin lens approximation}",
      journal        = "Phys. Rev. D",
      volume         = "72",
      year           = "2005",
      pages          = "043001",
      doi            = "10.1103/PhysRevD.72.043001",
      eprint         = "astro-ph/0505023",
      archivePrefix  = "arXiv",
      primaryClass   = "astro-ph",
      reportNumber   = "KUNS-1969",
      SLACcitation   = "%%CITATION = ASTRO-PH/0505023;%%"
}

@article{PhysRevLett.80.1138,
  title = {Gravitational Lensing of Gravitational Waves from Inspiraling Binaries by a Point Mass Lens},
  author = {Nakamura, Takahiro T.},
  journal = {Phys. Rev. Lett.},
  volume = {80},
  issue = {6},
  pages = {1138--1141},
  numpages = {0},
  year = {1998},
  month = {Feb},
  publisher = {American Physical Society},
  doi = {10.1103/PhysRevLett.80.1138}
}

@article{PhysRevD.98.083005,
  title = {Discovering intermediate-mass black hole lenses through gravitational wave lensing},
  author = {Lai, Kwun-Hang and Hannuksela, Otto A. and Herrera-Mart\'{\i}n, Antonio and Diego, Jose M. and Broadhurst, Tom and Li, Tjonnie G. F.},
  journal = {Phys. Rev. D},
  volume = {98},
  issue = {8},
  pages = {083005},
  numpages = {7},
  year = {2018},
  month = {Oct},
  publisher = {American Physical Society},
  doi = {10.1103/PhysRevD.98.083005},
  eprint = {1801.07840},
  archivePrefix = {arXiv},
  primaryClass = {gr-qc}
}

@article{Ruffa_1999,
  author  = {Ruffa, Anthony A.},
  title   = {Gravitational Lensing of Gravitational Waves},
  journal = {Astrophys. J. Lett.},
  volume  = {517},
  pages   = {L31--L33},
  year    = {1999},
  doi     = {10.1086/312015}
}

@article{DePaolis:2002tw,
      author         = "De Paolis, Francesco and Ingrosso, G. and Nucita, A. A.
                        and Qadir, A.",
      title          = "{A Note on gravitational wave lensing}",
      booktitle      = "{Conference on Astronomical Telescopes and Instrumenation
                        Waikoloa, Hawaii, August 22-28, 2002}",
      journal        = "Astron. Astrophys.",
      volume         = "394",
      year           = "2002",
      pages          = "749-752",
      doi            = "10.1051/0004-6361:20021258",
      eprint         = "astro-ph/0209149",
      archivePrefix  = "arXiv",
      primaryClass   = "astro-ph",
      reportNumber   = "UNILE-1090902",
      SLACcitation   = "%%CITATION = ASTRO-PH/0209149;%%"
}

@article{Liao:2019aqq,
    author = "Liao, Kai and Biesiada, Marek and Fan, Xi-Long",
    title = "{The wave nature of continuous gravitational waves from microlensing}",
    reportNumber = "LIGO-G1802118,LIGO-P1800330, LIGO-G1802118, LIGO-P1800330",
    doi = "10.3847/1538-4357/ab1087",
    journal = "Astrophys. J.",
    volume = "875",
    number = "2",
    pages = "139",
    year = "2019",
  eprint = {1903.06612},
  archivePrefix = {arXiv},
  primaryClass = {gr-qc}
}

@article{Dai:2018enj,
    author = "Dai, Liang and Li, Shun-Sheng and Zackay, Barak and Mao, Shude and Lu, Youjun",
    title = "{Detecting Lensing-Induced Diffraction in Astrophysical Gravitational Waves}",
    eprint = "1810.00003",
    archivePrefix = "arXiv",
    primaryClass = "gr-qc",
    doi = "10.1103/PhysRevD.98.104029",
    journal = "Phys. Rev. D",
    volume = "98",
    number = "10",
    pages = "104029",
    year = "2018"
}

@article{PhysRevD.90.062003,
  title = {Gravitational lensing effects on parameter estimation in gravitational wave detection with advanced detectors},
  author = {Cao, Zhoujian and Li, Li-Fang and Wang, Yan},
  journal = {Phys. Rev. D},
  volume = {90},
  issue = {6},
  pages = {062003},
  numpages = {12},
  year = {2014},
  month = {Sep},
  publisher = {American Physical Society},
  doi = {10.1103/PhysRevD.90.062003}
}

@article{Christian,
  title = {Detecting stellar lensing of gravitational waves with ground-based observatories},
  author = {Christian, Pierre and Vitale, Salvatore and Loeb, Abraham},
  journal = {Phys. Rev. D},
  volume = {98},
  issue = {10},
  pages = {103022},
  numpages = {7},
  year = {2018},
  month = {Nov},
  publisher = {American Physical Society},
  doi = {10.1103/PhysRevD.98.103022},
  eprint = {1802.02586},
  archivePrefix = {arXiv},
  primaryClass = {astro-ph.HE}
}

@article{Meena:2019ate,
       author = {{Meena}, Ashish Kumar and {Bagla}, Jasjeet Singh},
        title = "{Gravitational lensing of gravitational waves: wave nature and prospects for detection}",
      journal = {Mon. Not. Roy. Astron. Soc.},
         year = 2020,
        month = feb,
       volume = {492},
       number = {1},
        pages = {1127-1134},
          doi = {10.1093/mnras/stz3509},
       adsurl = {https://ui.adsabs.harvard.edu/abs/2020MNRAS.492.1127M},
  eprint = {1903.11809},
  archivePrefix = {arXiv},
  primaryClass = {astro-ph.CO}
}

@ARTICLE{2021PhRvD.103j4055W,
       author = {{Wang}, Yijun and {Lo}, Rico K.~L. and {Li}, Alvin K.~Y. and {Chen}, Yanbei},
        title = "{Identifying type II strongly lensed gravitational-wave images in third-generation gravitational-wave detectors}",
      journal = {Phys. Rev. D},
         year = 2021,
        month = may,
       volume = {103},
       number = {10},
          eid = {104055},
        pages = {104055},
          doi = {10.1103/PhysRevD.103.104055},
       adsurl = {https://ui.adsabs.harvard.edu/abs/2021PhRvD.103j4055W},
  eprint = {2101.08264},
  archivePrefix = {arXiv},
  primaryClass = {gr-qc}
}

@article{Hongsheng:2018ibg,
  author        = {Zhang, Hongsheng and Fan, Xilong},
  title         = {Poisson--Arago spot for gravitational waves},
  journal       = {Sci. China Phys. Mech. Astron.},
  volume        = {64},
  pages         = {120462},
  year          = {2021},
  doi           = {10.1007/s11433-021-1764-y},
  eprint        = {1809.06511},
  archivePrefix = {arXiv},
  primaryClass  = {gr-qc}
}

@article{PhysRevD.5.2419,
  title = {Nonspherical Perturbations of Relativistic Gravitational Collapse. I. Scalar and Gravitational Perturbations},
  author = {Price, Richard H.},
  journal = {Phys. Rev. D},
  volume = {5},
  issue = {10},
  pages = {2419--2438},
  numpages = {0},
  year = {1972},
  month = {May},
  publisher = {American Physical Society},
  doi = {10.1103/PhysRevD.5.2419},
  url = {https://link.aps.org/doi/10.1103/PhysRevD.5.2419}
}

@article{PhysRevD.110.124011,
  title = {Spin wave optics for gravitational waves lensed by a Kerr black hole},
  author = {Kubota, Kei-ichiro and Arai, Shun and Motohashi, Hayato and Mukohyama, Shinji},
  journal = {Phys. Rev. D},
  volume = {110},
  issue = {12},
  pages = {124011},
  numpages = {17},
  year = {2024},
  month = {Dec},
  publisher = {American Physical Society},
  doi = {10.1103/PhysRevD.110.124011},
  url = {https://link.aps.org/doi/10.1103/PhysRevD.110.124011},
  eprint = {2408.03289},
  archivePrefix = {arXiv},
  primaryClass = {gr-qc}
}

@article{PhysRevD.110.044054,
  title = {Wave optics lensing of gravitational waves: Theory and phenomenology of triple systems in the LISA band},
  author = {Pijnenburg, Martin and Cusin, Giulia and Pitrou, Cyril and Uzan, Jean-Philippe},
  journal = {Phys. Rev. D},
  volume = {110},
  issue = {4},
  pages = {044054},
  numpages = {25},
  year = {2024},
  month = {Aug},
  publisher = {American Physical Society},
  doi = {10.1103/PhysRevD.110.044054},
  url = {https://link.aps.org/doi/10.1103/PhysRevD.110.044054},
  eprint = {2404.07186},
  archivePrefix = {arXiv},
  primaryClass = {gr-qc}
}

@article{6h6r-46cd,
  title = {Lensing and wave optics in the strong field of a black hole},
  author = {Chan, Juno C. L. and Dyson, Conor and Garcia, Matilde and Redondo-Yuste, Jaime and Vujeva, Luka},
  journal = {Phys. Rev. D},
  volume = {112},
  issue = {6},
  pages = {064009},
  numpages = {16},
  year = {2025},
  month = {Sep},
  publisher = {American Physical Society},
  doi = {10.1103/6h6r-46cd},
  url = {https://link.aps.org/doi/10.1103/6h6r-46cd},
  eprint = {2502.14073},
  archivePrefix = {arXiv},
  primaryClass = {gr-qc}
}

@article{PhysRevD.111.103539,
  title = {Gravitational lensing of waves: Novel methods for wave-optics phenomena},
  author = {Villarrubia-Rojo, Hector and Savastano, Stefano and Zumalac\'arregui, Miguel and Choi, Lyla and Goyal, Srashti and Dai, Liang and Tambalo, Giovanni},
  journal = {Phys. Rev. D},
  volume = {111},
  issue = {10},
  pages = {103539},
  numpages = {22},
  year = {2025},
  month = {May},
  publisher = {American Physical Society},
  doi = {10.1103/PhysRevD.111.103539},
  url = {https://link.aps.org/doi/10.1103/PhysRevD.111.103539},
  eprint = {2409.04606},
  archivePrefix = {arXiv},
  primaryClass = {gr-qc}
}

@article{Kanai:2013rga,
    author = "Kanai, Ken-ichiro and Nambu, Yasusada",
    title = "{Viewing Black Holes by Waves}",
    eprint = "1303.5520",
    archivePrefix = "arXiv",
    primaryClass = "gr-qc",
    doi = "10.1088/0264-9381/30/17/175002",
    journal = "Class. Quant. Grav.",
    volume = "30",
    pages = "175002",
    year = "2013"
}

@ARTICLE{marquez_divergence_1972,
       author = {{Marquez}, L.},
        title = "{On the Divergence of the Rutherford Scattering Amplitude in Terms of Coulomb Phase Shifts}",
      journal = {American Journal of Physics},
         year = 1972,
        month = oct,
       volume = {40},
       number = {10},
        pages = {1420-1427},
          doi = {10.1119/1.1986863},
          url = {https://doi.org/10.1119/1.1986863}
}

@incollection{andersson_scattering_2001,
  author    = {Andersson, N. and Jensen, B. P.},
  title     = {Scattering by Black Holes},
  booktitle = {Scattering: Scattering and Inverse Scattering in Pure and Applied Science},
  editor    = {Pike, E. R. and Sabatier, P. C.},
  publisher = {Academic Press},
  address   = {London},
  year      = {2002},
  eprint    = {gr-qc/0011025},
  archivePrefix = {arXiv},
  pages = {1607--1626},
  doi = {10.1016/B978-012613760-6/50085-1}
}

@article{Nambu:2015aea,
  author        = {Nambu, Yasusada and Noda, Sousuke},
  title         = {Wave optics in black hole spacetimes: the {Schwarzschild} case},
  journal       = {Class. Quant. Grav.},
  volume        = {33},
  pages         = {075011},
  year          = {2016},
  doi           = {10.1088/0264-9381/33/7/075011},
  eprint        = {1502.05468},
  archivePrefix = {arXiv},
  primaryClass  = {gr-qc}
}

@article{Pijnenburg:2022pug,
  author        = {Pijnenburg, Martin and Cusin, Giulia and Pitrou, Cyril
                   and Uzan, Jean-Philippe},
  title         = {Rutherford scattering of quantum and classical fields},
  journal       = {Am. J. Phys.},
  volume        = {92},
  pages         = {597--605},
  year          = {2024},
  doi           = {10.1119/5.0175025},
  eprint        = {2211.14113},
  archivePrefix = {arXiv},
  primaryClass  = {math-ph}
}

@article{xsxj-9vdw,
  title = {Rigorous calculation of scalar scattering in the Schwarzschild background: The convergence of the partial-wave series and the Poisson spot},
  author = {Li, Zhao and Zhao, Wen},
  journal = {Phys. Rev. D},
  volume = {112},
  issue = {8},
  pages = {083030},
  numpages = {17},
  year = {2025},
  month = {Oct},
  publisher = {American Physical Society},
  doi = {10.1103/xsxj-9vdw},
  url = {https://link.aps.org/doi/10.1103/xsxj-9vdw},
  eprint = {2508.17253},
  archivePrefix = {arXiv},
  primaryClass = {gr-qc}
}

@article{PhysRevD.84.104002,
  title = {Wave propagation and quasinormal mode excitation on Schwarzschild spacetime},
  author = {Dolan, Sam R. and Ottewill, Adrian C.},
  journal = {Phys. Rev. D},
  volume = {84},
  issue = {10},
  pages = {104002},
  numpages = {14},
  year = {2011},
  month = {Nov},
  publisher = {American Physical Society},
  doi = {10.1103/PhysRevD.84.104002},
  url = {https://link.aps.org/doi/10.1103/PhysRevD.84.104002},
  eprint = {1106.4318},
  archivePrefix = {arXiv},
  primaryClass = {gr-qc}
}

@ARTICLE{1973ApJ...185..635T,
       author = {{Teukolsky}, Saul A.},
        title = "{Perturbations of a Rotating Black Hole. I. Fundamental Equations for Gravitational, Electromagnetic, and Neutrino-Field Perturbations}",
      journal = {Astrophys. J.},
         year = 1973,
        month = oct,
       volume = {185},
        pages = {635-648},
          doi = {10.1086/152444},
       adsurl = {https://ui.adsabs.harvard.edu/abs/1973ApJ...185..635T}
}

@ARTICLE{2017arXiv170200786A,
       author = {{Amaro-Seoane}, Pau and {Audley}, Heather and {Babak}, Stanislav and {Baker}, John and {Barausse}, Enrico and {Bender}, Peter and {Berti}, Emanuele and {Binetruy}, Pierre and {Born}, Michael and {Bortoluzzi}, Daniele and {Camp}, Jordan and {Caprini}, Chiara and {Cardoso}, Vitor and {Colpi}, Monica and {Conklin}, John and {Cornish}, Neil and {Cutler}, Curt and {Danzmann}, Karsten and {Dolesi}, Rita and {Ferraioli}, Luigi and {Ferroni}, Valerio and {Fitzsimons}, Ewan and {Gair}, Jonathan and {Gesa Bote}, Lluis and {Giardini}, Domenico and {Gibert}, Ferran and {Grimani}, Catia and {Halloin}, Hubert and {Heinzel}, Gerhard and {Hertog}, Thomas and {Hewitson}, Martin and {Holley-Bockelmann}, Kelly and {Hollington}, Daniel and {Hueller}, Mauro and {Inchauspe}, Henri and {Jetzer}, Philippe and {Karnesis}, Nikos and {Killow}, Christian and {Klein}, Antoine and {Klipstein}, Bill and {Korsakova}, Natalia and {Larson}, Shane L and {Livas}, Jeffrey and {Lloro}, Ivan and {Man}, Nary and {Mance}, Davor and {Martino}, Joseph and {Mateos}, Ignacio and {McKenzie}, Kirk and {McWilliams}, Sean T and {Miller}, Cole and {Mueller}, Guido and {Nardini}, Germano and {Nelemans}, Gijs and {Nofrarias}, Miquel and {Petiteau}, Antoine and {Pivato}, Paolo and {Plagnol}, Eric and {Porter}, Ed and {Reiche}, Jens and {Robertson}, David and {Robertson}, Norna and {Rossi}, Elena and {Russano}, Giuliana and {Schutz}, Bernard and {Sesana}, Alberto and {Shoemaker}, David and {Slutsky}, Jacob and {Sopuerta}, Carlos F. and {Sumner}, Tim and {Tamanini}, Nicola and {Thorpe}, Ira and {Troebs}, Michael and {Vallisneri}, Michele and {Vecchio}, Alberto and {Vetrugno}, Daniele and {Vitale}, Stefano and {Volonteri}, Marta and {Wanner}, Gudrun and {Ward}, Harry and {Wass}, Peter and {Weber}, William and {Ziemer}, John and {Zweifel}, Peter},
        title = "{Laser Interferometer Space Antenna}",
      journal = {arXiv e-prints},
         year = 2017,
        month = feb,
          eid = {arXiv:1702.00786},
        pages = {arXiv:1702.00786},
          doi = {10.48550/arXiv.1702.00786},
archivePrefix = {arXiv},
       eprint = {1702.00786},
 primaryClass = {astro-ph.IM},
       adsurl = {https://ui.adsabs.harvard.edu/abs/2017arXiv170200786A}
}

@ARTICLE{2010CQGra..27s4002P,
       author = {{Punturo}, M. and {Abernathy}, M. and {Acernese}, F. and {Allen}, B. and {Andersson}, N. and {Arun}, K. and {Barone}, F. and {Barr}, B. and {Barsuglia}, M. and {Beker}, M. and {Beveridge}, N. and {Birindelli}, S. and {Bose}, S. and {Bosi}, L. and {Braccini}, S. and {Bradaschia}, C. and {Bulik}, T. and {Calloni}, E. and {Cella}, G. and {Chassande Mottin}, E. and {Chelkowski}, S. and {Chincarini}, A. and {Clark}, J. and {Coccia}, E. and {Colacino}, C. and {Colas}, J. and {Cumming}, A. and {Cunningham}, L. and {Cuoco}, E. and {Danilishin}, S. and {Danzmann}, K. and {De Luca}, G. and {De Salvo}, R. and {Dent}, T. and {De Rosa}, R. and {Di Fiore}, L. and {Di Virgilio}, A. and {Doets}, M. and {Fafone}, V. and {Falferi}, P. and {Flaminio}, R. and {Franc}, J. and {Frasconi}, F. and {Freise}, A. and {Fulda}, P. and {Gair}, J. and {Gemme}, G. and {Gennai}, A. and {Giazotto}, A. and {Glampedakis}, K. and {Granata}, M. and {Grote}, H. and {Guidi}, G. and {Hammond}, G. and {Hannam}, M. and {Harms}, J. and {Heinert}, D. and {Hendry}, M. and {Heng}, I. and {Hennes}, E. and {Hild}, S. and {Hough}, J. and {Husa}, S. and {Huttner}, S. and {Jones}, G. and {Khalili}, F. and {Kokeyama}, K. and {Kokkotas}, K. and {Krishnan}, B. and {Lorenzini}, M. and {L{\"u}ck}, H. and {Majorana}, E. and {Mandel}, I. and {Mandic}, V. and {Martin}, I. and {Michel}, C. and {Minenkov}, Y. and {Morgado}, N. and {Mosca}, S. and {Mours}, B. and {M{\"u}ller{\textendash}Ebhardt}, H. and {Murray}, P. and {Nawrodt}, R. and {Nelson}, J. and {Oshaughnessy}, R. and {Ott}, C.~D. and {Palomba}, C. and {Paoli}, A. and {Parguez}, G. and {Pasqualetti}, A. and {Passaquieti}, R. and {Passuello}, D. and {Pinard}, L. and {Poggiani}, R. and {Popolizio}, P. and {Prato}, M. and {Puppo}, P. and {Rabeling}, D. and {Rapagnani}, P. and {Read}, J. and {Regimbau}, T. and {Rehbein}, H. and {Reid}, S. and {Rezzolla}, L. and {Ricci}, F. and {Richard}, F. and {Rocchi}, A. and {Rowan}, S. and {R{\"u}diger}, A. and {Sassolas}, B. and {Sathyaprakash}, B. and {Schnabel}, R. and {Schwarz}, C. and {Seidel}, P. and {Sintes}, A. and {Somiya}, K. and {Speirits}, F. and {Strain}, K. and {Strigin}, S. and {Sutton}, P. and {Tarabrin}, S. and {Th{\"u}ring}, A. and {van den Brand}, J. and {van Leewen}, C. and {van Veggel}, M. and {van den Broeck}, C. and {Vecchio}, A. and {Veitch}, J. and {Vetrano}, F. and {Vicere}, A. and {Vyatchanin}, S. and {Willke}, B. and {Woan}, G. and {Wolfango}, P. and {Yamamoto}, K.},
        title = "{The Einstein Telescope: a third-generation gravitational wave observatory}",
      journal = {Classical and Quantum Gravity},
         year = 2010,
        month = oct,
       volume = {27},
       number = {19},
          eid = {194002},
        pages = {194002},
          doi = {10.1088/0264-9381/27/19/194002},
       adsurl = {https://ui.adsabs.harvard.edu/abs/2010CQGra..27s4002P}
}

@article{PhysRevD.52.605,
  title = {Search templates for gravitational waves from precessing, inspiraling binaries},
  author = {Apostolatos, Theocharis A.},
  journal = {Phys. Rev. D},
  volume = {52},
  issue = {2},
  pages = {605--620},
  numpages = {0},
  year = {1995},
  month = {Jul},
  publisher = {American Physical Society},
  doi = {10.1103/PhysRevD.52.605}
}

@article{gg33-r64q,
  title = {Scalar quasinormal modes in black hole gravitational lensing},
  author = {Yin, Chengjiang and Lin, Zihao and He, Jian-hua},
  journal = {Phys. Rev. D},
  volume = {113},
  issue = {2},
  pages = {024045},
  numpages = {13},
  year = {2026},
  month = {Jan},
  publisher = {American Physical Society},
  doi = {10.1103/gg33-r64q},
  url = {https://link.aps.org/doi/10.1103/gg33-r64q},
  eprint = {2601.14303},
  archivePrefix = {arXiv},
  primaryClass = {gr-qc}
}

@article{PhysRevX.4.031006,
  title = {Fast Prediction and Evaluation of Gravitational Waveforms Using Surrogate Models},
  author = {Field, Scott E. and Galley, Chad R. and Hesthaven, Jan S. and Kaye, Jason and Tiglio, Manuel},
  journal = {Phys. Rev. X},
  volume = {4},
  issue = {3},
  pages = {031006},
  numpages = {21},
  year = {2014},
  month = {Jul},
  publisher = {American Physical Society},
  doi = {10.1103/PhysRevX.4.031006},
  url = {https://link.aps.org/doi/10.1103/PhysRevX.4.031006},
  eprint = {1308.3565},
  archivePrefix = {arXiv},
  primaryClass = {gr-qc}
}

@article{Varma:2018mmi,
    author = "Varma, Vijay and Field, Scott E. and Scheel, Mark A. and Blackman, Jonathan and Kidder, Lawrence E. and Pfeiffer, Harald P.",
    title = "{Surrogate model of hybridized numerical relativity binary black hole waveforms}",
    doi = "10.1103/PhysRevD.99.064045",
    journal = "Phys. Rev. D",
    volume = "99",
    number = "6",
    pages = "064045",
    year = "2019",
  eprint = {1812.07865},
  archivePrefix = {arXiv},
  primaryClass = {gr-qc}
}

@article{LIGOScientific:2025cwb,
    author = "Abac, A. G. and others",
    collaboration = "LIGO Scientific, VIRGO, KAGRA",
    title = "{GWTC-4.0: Searches for Gravitational-Wave Lensing Signatures}",
    journal = {arXiv e-prints},
    eprint = "2512.16347",
    archivePrefix = "arXiv",
    primaryClass = "gr-qc",
    reportNumber = "LIGO-P2500419",
    month = "12",
    year = "2025",
  note = {Accepted for publication in The Astrophysical Journal}
}

@article{Markakis:2023pfh,
    author = "Markakis, Charalampos and Bray, Sean and Zengino{\u{g}}lu, An{\i}l",
    title = "{Symmetric integration of the 1+1 Teukolsky equation on hyperboloidal foliations of Kerr spacetimes}",
    journal = {arXiv e-prints},
    eprint = "2303.08153",
    archivePrefix = "arXiv",
    primaryClass = "gr-qc",
    month = "3",
    year = "2023"
}

@article{PhysRevD.61.024026,
  title = {Late time decay of scalar, electromagnetic, and gravitational perturbations outside rotating black holes},
  author = {Barack, Leor},
  journal = {Phys. Rev. D},
  volume = {61},
  issue = {2},
  pages = {024026},
  numpages = {27},
  year = {2000},
  month = {Jan},
  publisher = {American Physical Society},
  doi = {10.1103/PhysRevD.61.024026},
  url = {https://link.aps.org/doi/10.1103/PhysRevD.61.024026},
  eprint = {gr-qc/9908005},
  archivePrefix = {arXiv},
  primaryClass = {gr-qc}
}

@article{PhysRevD.95.104033,
  title = {Time-domain metric reconstruction for self-force applications},
  author = {Barack, Leor and Giudice, Paco},
  journal = {Phys. Rev. D},
  volume = {95},
  issue = {10},
  pages = {104033},
  numpages = {22},
  year = {2017},
  month = {May},
  publisher = {American Physical Society},
  doi = {10.1103/PhysRevD.95.104033},
  url = {https://link.aps.org/doi/10.1103/PhysRevD.95.104033},
  eprint = {1702.04204},
  archivePrefix = {arXiv},
  primaryClass = {gr-qc}
}

@article{ansorgSpectralDecompositionBlackhole2016,
  title = {Spectral Decomposition of Black-Hole Perturbations on Hyperboloidal Slices},
  author = {Ansorg, Marcus and Macedo, Rodrigo Panosso},
  year = {2016},
  month = jun,
  journal = {Physical Review D},
  volume = {93},
  number = {12},
  eprint = {1604.02261},
  pages = {124016},
  issn = {2470-0010, 2470-0029},
  doi = {10.1103/PhysRevD.93.124016},
  archiveprefix = {arxiv},
  langid = {english}
}

@article{fieldGPUAcceleratedMixedPrecisionWENO2021,
  author  = {Field, Scott E. and Gottlieb, Sigal and Grant, Zachary J.
             and Isherwood, Leah F. and Khanna, Gaurav},
  title   = {A {GPU}-Accelerated Mixed-Precision {WENO} Method for Extremal
             Black Hole and Gravitational Wave Physics Computations},
  journal = {Commun. Appl. Math. Comput.},
  volume  = {5},
  pages   = {97--115},
  year    = {2023},
  doi     = {10.1007/s42967-021-00129-2},
  eprint = {2010.04760},
  archivePrefix = {arXiv},
  primaryClass = {math.NA}
}

@article{harmsNumericalSolutionTeukolsky2013,
  title = {Numerical Solution of the 2+1 {{Teukolsky}} Equation on a Hyperboloidal and Horizon Penetrating Foliation of {{Kerr}} and Application to Late-Time Decays},
  author = {Harms, Enno and Bernuzzi, Sebastiano and Bruegmann, Bernd},
  year = {2013},
  month = jun,
  journal = {Classical and Quantum Gravity},
  volume = {30},
  number = {11},
  eprint = {1301.1591},
  primaryclass = {gr-qc},
  pages = {115013},
  issn = {0264-9381, 1361-6382},
  doi = {10.1088/0264-9381/30/11/115013},
  archiveprefix = {arxiv}
}

@article{raczNumericalInvestigationLatetime2011,
  title = {Numerical Investigation of the Late-Time {{Kerr}} Tails},
  author = {Racz, Istvan and Toth, Gabor Zsolt},
  year = {2011},
  month = oct,
  journal = {Classical and Quantum Gravity},
  volume = {28},
  number = {19},
  eprint = {1104.4199},
  pages = {195003},
  issn = {0264-9381, 1361-6382},
  doi = {10.1088/0264-9381/28/19/195003},
  archiveprefix = {arxiv},
  langid = {english}
}

@article{shlapentokh-rothmanBoundednessDecayTeukolsky2020,
  author  = {Shlapentokh-Rothman, Yakov and Teixeira da Costa, Rita},
  title   = {Boundedness and Decay for the {Teukolsky} Equation on {Kerr}
             in the Full Subextremal Range $|a|<M$: Frequency Space Analysis},
  journal = {Arch. Ration. Mech. Anal.},
  volume  = {250},
  pages   = {78},
  year    = {2026},
  doi     = {10.1007/s00205-026-02173-9},
  eprint  = {2007.07211},
  archivePrefix = {arXiv},
  primaryClass = {gr-qc}
}

@article{vishveshwara_scattering_1970,
  title = {Scattering of {{Gravitational Radiation}} by a {{Schwarzschild Black-hole}}},
  author = {Vishveshwara, C. V.},
  year = {1970},
  month = aug,
  journal = {Nature},
  volume = {227},
  number = {5261},
  pages = {936--938},
  issn = {1476-4687},
  doi = {10.1038/227936a0}
}

@article{zenginoglu_hyperboloidal_2008,
  title = {Hyperboloidal Foliations and Scri-Fixing},
  author = {Zengino{\u g}lu, An{\i}l},
  year = {2008},
  month = jul,
  journal = {Classical and Quantum Gravity},
  volume = {25},
  number = {14},
  pages = {145002},
  issn = {0264-9381, 1361-6382},
  doi = {10.1088/0264-9381/25/14/145002},
  langid = {english},
  eprint = {0712.4333},
  archivePrefix = {arXiv},
  primaryClass = {gr-qc}
}

@article{zenginoglu_hyperboloidal_2011,
  title = {Hyperboloidal Layers for Hyperbolic Equations on Unbounded Domains},
  author = {Zengino{\u g}lu, An{\i}l},
  year = {2011},
  month = mar,
  journal = {Journal of Computational Physics},
  volume = {230},
  number = {6},
  pages = {2286--2302},
  issn = {00219991},
  doi = {10.1016/j.jcp.2010.12.016},
  langid = {english},
  eprint = {1008.3809},
  archivePrefix = {arXiv},
  primaryClass = {math.NA}
}

@article{Zenginoglu:2008wc,
  title = {A {{Hyperboloidal}} Study of Tail Decay Rates for Scalar and {{Yang-Mills}} Fields},
  author = {Zenginoglu, Anil},
  journal = {Class. Quant. Grav.},
  year = {2008},
  volume = {25},
  number = {17},
  eprint = {0803.2018},
  primaryclass = {gr-qc},
  pages = {175013},
  doi = {10.1088/0264-9381/25/17/175013},
  archiveprefix = {arxiv}
}

@article{zenginogluIntermediateBehaviorKerr2014,
  title = {Intermediate Behavior of {{Kerr}} Tails},
  author = {Zengino{\u g}lu, An{\i}l and Khanna, Gaurav and Burko, Lior M.},
  year = {2014},
  month = mar,
  journal = {General Relativity and Gravitation},
  volume = {46},
  number = {3},
  eprint = {1208.5839},
  primaryclass = {gr-qc},
  pages = {1672},
  issn = {0001-7701, 1572-9532},
  doi = {10.1007/s10714-014-1672-8},
  archiveprefix = {arxiv}
}

@article{zenginogluSpacelikeMatchingNull2009,
  title = {Spacelike Matching to Null Infinity},
  author = {Zenginoglu, Anil and Tiglio, Manuel},
  year = {2009},
  month = jul,
  journal = {Physical Review D},
  volume = {80},
  number = {2},
  eprint = {0906.3342},
  primaryclass = {gr-qc},
  pages = {024044},
  issn = {1550-7998, 1550-2368},
  doi = {10.1103/PhysRevD.80.024044},
  archiveprefix = {arxiv}
}

@article{Long:2021ufh,
    author = "Long, Oliver and Barack, Leor",
    title = "{Time-domain metric reconstruction for hyperbolic scattering}",
    eprint = "2105.05630",
    archivePrefix = "arXiv",
    primaryClass = "gr-qc",
    doi = "10.1103/PhysRevD.104.024014",
    journal = "Phys. Rev. D",
    volume = "104",
    number = "2",
    pages = "024014",
    year = "2021"
}

@ARTICLE{2019arXiv191013452P,
       author = {{Panosso Macedo}, Rodrigo},
        title = "{Hyperboloidal framework for the Kerr spacetime}",
      journal = {Class. Quantum Grav.},
         year = {2020},
        month = {mar},
          eid = {065019},
        pages = {065019},
          doi = {10.1088/1361-6382/ab6e3e},
archivePrefix = {arXiv},
       eprint = {1910.13452},
 primaryClass = {gr-qc},
volume = {37},
  number = {6},
  url = {https://doi.org/10.1088/1361-6382/ab6e3e}
}

@ARTICLE{2014JCoPh.276..357P,
       author = {{Panosso Macedo}, Rodrigo and {Ansorg}, Marcus},
        title = "{Axisymmetric fully spectral code for hyperbolic equations}",
      journal = {Journal of Computational Physics},
         year = 2014,
        month = nov,
       volume = {276},
        pages = {357-379},
          doi = {10.1016/j.jcp.2014.07.040},
archivePrefix = {arXiv},
       eprint = {1402.7343},
 primaryClass = {physics.comp-ph},
       adsurl = {https://ui.adsabs.harvard.edu/abs/2014JCoPh.276..357P}
}

@article{TheVirgo:2014hva,
    author = "Acernese, F. and others",
    collaboration = "VIRGO",
    title = "{Advanced Virgo: a second-generation interferometric gravitational wave detector}",
    eprint = "1408.3978",
    archivePrefix = "arXiv",
    primaryClass = "gr-qc",
    doi = "10.1088/0264-9381/32/2/024001",
    journal = "Class. Quant. Grav.",
    volume = "32",
    number = "2",
    pages = "024001",
    year = "2015"
}

@article{Evans:2021gyd,
    author = "Evans, Matthew and others",
    title = "{A Horizon Study for Cosmic Explorer: Science, Observatories, and Community}",
    journal = {arXiv e-prints},
    eprint = "2109.09882",
    archivePrefix = "arXiv",
    primaryClass = "astro-ph.IM",
    month = "9",
    year = "2021"
}

@article{Ezquiaga:2020gdt,
  author        = {Ezquiaga, Jose Mar{\'\i}a and Holz, Daniel E. and Hu, Wayne and Lagos, Macarena and Wald, Robert M.},
  title         = {Phase effects from strong gravitational lensing of gravitational waves},
  journal       = {Phys. Rev. D},
  volume        = {103},
  pages         = {064047},
  year          = {2021},
  doi           = {10.1103/PhysRevD.103.064047},
  eprint        = {2008.12814},
  archivePrefix = {arXiv},
  primaryClass  = {astro-ph.CO}
}

@article{Oguri:2019fix,
    author = "Oguri, Masamune",
    title = "{Strong gravitational lensing of explosive transients}",
    eprint = "1907.06830",
    archivePrefix = "arXiv",
    primaryClass = "astro-ph.CO",
    doi = "10.1088/1361-6633/ab4fc5",
    journal = "Rept. Prog. Phys.",
    volume = "82",
    number = "12",
    pages = "126901",
    year = "2019"
}

@article{Tambalo:2022plm,
  author        = {Tambalo, Giovanni and Zumalac\'arregui, Miguel and Dai, Liang
                   and Cheung, Mark Ho-Yeuk},
  title         = {Gravitational wave lensing as a probe of halo properties and dark matter},
  journal       = {Phys. Rev. D},
  volume        = {108},
  pages         = {103529},
  year          = {2023},
  doi           = {10.1103/PhysRevD.108.103529},
  eprint        = {2212.11960},
  archivePrefix = {arXiv},
  primaryClass  = {astro-ph.CO}
}

@book{Futterman:1988ni,
    author = "Futterman, J. A. H. and Handler, F. A. and Matzner, R. A.",
    title = "{Scattering from Black Holes}",
    isbn = "978-0-521-32986-1",
    publisher = "Cambridge University Press",
    address = "Cambridge, England",
    series = "Cambridge Monographs on Mathematical Physics",
    year = "1988",
  doi = {10.1017/CBO9780511735615}
}

@article{Crispino:2009xt,
  author        = {Crispino, Luis C. B. and Dolan, Sam R. and Oliveira, Ednilton S.},
  title         = {Electromagnetic wave scattering by {Schwarzschild} black holes},
  journal       = {Phys. Rev. Lett.},
  volume        = {102},
  pages         = {231103},
  year          = {2009},
  doi           = {10.1103/PhysRevLett.102.231103},
  eprint        = {0905.3339},
  archivePrefix = {arXiv},
  primaryClass  = {gr-qc}
}

@article{LIGOScientific:2014pky,
    author = "Aasi, J. and others",
    collaboration = "LIGO Scientific",
    title = "{Advanced LIGO}",
    eprint = "1411.4547",
    archivePrefix = "arXiv",
    primaryClass = "gr-qc",
    doi = "10.1088/0264-9381/32/7/074001",
    journal = "Class. Quant. Grav.",
    volume = "32",
    pages = "074001",
    year = "2015"
}

@article{Agol:2001hb,
  author        = {Agol, Eric and Kamionkowski, Marc},
  title         = {X-rays from isolated black holes in the {Milky} {Way}},
  journal       = {Mon. Not. Roy. Astron. Soc.},
  volume        = {334},
  pages         = {553--562},
  year          = {2002},
  doi           = {10.1046/j.1365-8711.2002.05523.x},
  eprint        = {astro-ph/0109539},
  archivePrefix = {arXiv},
  primaryClass  = {astro-ph}
}

@article{Agol:2002microlensing,
  author        = {Agol, Eric and Kamionkowski, Marc and Koopmans, L\'eon V. E.
                   and Blandford, Roger D.},
  title         = {Finding Black Holes with Microlensing},
  journal       = {Astrophys. J. Lett.},
  volume        = {576},
  pages         = {L131--L135},
  year          = {2002},
  doi           = {10.1086/343758},
  eprint        = {astro-ph/0203257},
  archivePrefix = {arXiv},
  primaryClass  = {astro-ph}
}

@article{Dolan:2008scattering,
  author  = {Dolan, Sam R.},
  title   = {Scattering and absorption of gravitational plane waves by rotating black holes},
  journal = {Class. Quant. Grav.},
  volume  = {25},
  pages   = {235002},
  year    = {2008},
  doi     = {10.1088/0264-9381/25/23/235002},
  eprint  = {0801.3805},
  archivePrefix = {arXiv},
  primaryClass  = {gr-qc}
}

@article{siciliaBlackHoleMass2022,
  author        = {Sicilia, Alex and Lapi, Andrea and Boco, Lumen and Spera, Mario
                   and Di Carlo, Ugo N. and Mapelli, Michela and Shankar, Francesco
                   and Alexander, David M. and Bressan, Alessandro and Danese, Luigi},
  title         = {The Black Hole Mass Function Across Cosmic Times. {I}.
                   Stellar Black Holes and Light Seed Distribution},
  journal       = {Astrophys. J.},
  volume        = {924},
  pages         = {56},
  year          = {2022},
  doi           = {10.3847/1538-4357/ac34fb},
  eprint        = {2110.15607},
  archivePrefix = {arXiv},
  primaryClass  = {astro-ph.GA}
}

@book{1975weoc.book.....F,
    author = "Friedlander, F. G.",
    title = "{The Wave Equation on a Curved Space-Time}",
publisher = "Cambridge University Press",
    address = "Cambridge, England",
    year = "1975"
}

@article{LIGOScientific:2021izm,
  author        = {Abbott, R. and others},
  collaboration = {LIGO Scientific, Virgo},
  title         = {Search for Lensing Signatures in the Gravitational-Wave
                   Observations from the First Half of {LIGO}--{Virgo}'s Third
                   Observing Run},
  journal       = {Astrophys. J.},
  volume        = {923},
  pages         = {14},
  year          = {2021},
  doi           = {10.3847/1538-4357/ac23db},
  eprint        = {2105.06384},
  archivePrefix = {arXiv},
  primaryClass  = {gr-qc}
}

@article{LIGOScientific:2023bwz,
  author        = {Abbott, R. and others},
  collaboration = {LIGO Scientific, Virgo, KAGRA},
  title         = {Search for Gravitational-lensing Signatures in the Full Third
                   Observing Run of the {LIGO}--{Virgo} Network},
  journal       = {Astrophys. J.},
  volume        = {970},
  pages         = {191},
  year          = {2024},
  doi           = {10.3847/1538-4357/ad3e83},
  eprint        = {2304.08393},
  archivePrefix = {arXiv},
  primaryClass  = {gr-qc}
}

@article{Whiting1989ModeStability,
  author = {Whiting, Bernard F.},
  title = {Mode stability of the {Kerr} black hole},
  journal = {J. Math. Phys.},
  volume = {30},
  number = {6},
  pages = {1301--1305},
  year = {1989},
  doi = {10.1063/1.528308}
}

@book{KlainermanSzeftel2020Schwarzschild,
  author = {Klainerman, Sergiu and Szeftel, J{\'e}r{\'e}mie},
  title = {Global Nonlinear Stability of {Schwarzschild} Spacetime under Polarized Perturbations},
  series = {Annals of Mathematics Studies},
  volume = {210},
  publisher = {Princeton University Press},
  address = {Princeton, NJ},
  year = {2020},
  doi = {10.1515/9780691218526},
  eprint = {1711.07597},
  archivePrefix = {arXiv},
  primaryClass = {gr-qc}
}

@article{KlainermanSzeftel2023Kerr,
  author = {Klainerman, Sergiu and Szeftel, J{\'e}r{\'e}mie},
  title = {{Kerr} stability for small angular momentum},
  journal = {Pure Appl. Math. Q.},
  volume = {19},
  number = {3},
  pages = {791--1678},
  year = {2023},
  doi = {10.4310/pamq.2023.v19.n3.a1},
  eprint = {2104.11857},
  archivePrefix = {arXiv},
  primaryClass = {math.AP}
}

@article{GiorgiKlainermanSzeftel2024Kerr,
  author = {Giorgi, Elena and Klainerman, Sergiu and Szeftel, J{\'e}r{\'e}mie},
  title = {Wave equations estimates and the nonlinear stability of slowly rotating {Kerr} black holes},
  journal = {Pure Appl. Math. Q.},
  volume = {20},
  number = {7},
  pages = {2865--3849},
  year = {2024},
  doi = {10.4310/pamq.241128023033},
  eprint = {2205.14808},
  archivePrefix = {arXiv},
  primaryClass = {math.AP}
}

@article{Shen2023GCM,
  author = {Shen, Dawei},
  title = {Construction of {GCM} hypersurfaces in perturbations of {Kerr}},
  journal = {Ann. PDE},
  volume = {9},
  number = {1},
  pages = {11},
  year = {2023},
  doi = {10.1007/s40818-023-00152-x},
  eprint = {2205.12336},
  archivePrefix = {arXiv},
  primaryClass = {math.AP}
}

@misc{ShlapentokhRothmanTeixeira2023Physical,
  author = {Shlapentokh-Rothman, Yakov and {Teixeira da Costa}, Rita},
  title = {Boundedness and decay for the {Teukolsky} equation on {Kerr} in the full subextremal range $|a|<{M}$: physical space analysis},
  year = {2023},
  eprint = {2302.08916},
  archivePrefix = {arXiv},
  primaryClass = {gr-qc},
  url = {https://arxiv.org/abs/2302.08916},
  note = {To appear in Analysis \& PDE}
}

@misc{HaefnerHintzVasy2025FullKerr,
  author = {H{\"a}fner, Dietrich and Hintz, Peter and Vasy, Andr{\'a}s},
  title = {Linear stability of {Kerr} black holes in the full subextremal range},
  year = {2025},
  eprint = {2506.21183},
  archivePrefix = {arXiv},
  primaryClass = {gr-qc},
  url = {https://arxiv.org/abs/2506.21183}
}

@misc{Hintz2026NonlinearKerr,
  author = {Hintz, Peter},
  title = {Nonlinear stability of subextremal {Kerr} black holes},
  year = {2026},
  eprint = {2606.28253},
  archivePrefix = {arXiv},
  primaryClass = {gr-qc},
  url = {https://arxiv.org/abs/2606.28253v2}
}

@book{Ringstrom2009Cauchy,
  author = {Ringstr{\"o}m, Hans},
  title = {The {Cauchy} Problem in General Relativity},
  series = {ESI Lectures in Mathematics and Physics},
  publisher = {European Mathematical Society},
  address = {Z{\"u}rich},
  year = {2009},
  doi = {10.4171/053}
}

@article{AnderssonBackdahlBlueMa2022RadiationGauge,
  author = {Andersson, Lars and B{\"a}ckdahl, Thomas and Blue, Pieter and Ma, Siyuan},
  title = {Nonlinear Radiation Gauge for Near {Kerr} Spacetimes},
  journal = {Commun. Math. Phys.},
  volume = {396},
  number = {1},
  pages = {45--90},
  year = {2022},
  doi = {10.1007/s00220-022-04461-3},
  eprint = {2108.03148},
  archivePrefix = {arXiv},
  primaryClass = {gr-qc}
}

@article{9bxg-sqn5,
  title = {Strong-field gravitational-wave lensing in the Kerr background},
  author = {Saketh, M. V. S. and Ghosh, Rajes and Mishra, Anuj},
  journal = {Phys. Rev. D},
  volume = {113},
  issue = {8},
  pages = {084056},
  numpages = {24},
  year = {2026},
  month = {Apr},
  publisher = {American Physical Society},
  doi = {10.1103/9bxg-sqn5},
  url = {https://link.aps.org/doi/10.1103/9bxg-sqn5},
  eprint = {2511.23110},
  archivePrefix = {arXiv},
  primaryClass = {gr-qc}
}

@article{KAGRA:2020tym,
    author = "Akutsu, T. and others",
    collaboration = "KAGRA",
    title = "{Overview of KAGRA: Detector design and construction history}",
    eprint = "2005.05574",
    archivePrefix = "arXiv",
    primaryClass = "physics.ins-det",
    doi = "10.1093/ptep/ptaa125",
    journal = "Prog. Theor. Exp. Phys.",
    volume = "2021",
    number = "5",
    pages = "05A101",
    year = "2021"
}

@article{SasakiTagoshi2003,
  author  = {Sasaki, Misao and Tagoshi, Hideyuki},
  title   = {Analytic Black Hole Perturbation Approach to Gravitational Radiation},
  journal = {Living Rev. Rel.},
  volume  = {6},
  pages   = {6},
  year    = {2003},
  doi     = {10.12942/lrr-2003-6},
  eprint  = {gr-qc/0306120},
  archivePrefix = {arXiv}
}

@article{TeukolskyPress1974,
  author        = {Teukolsky, Saul A. and Press, William H.},
  title         = {Perturbations of a Rotating Black Hole. {III}. Interaction of the Hole with Gravitational and Electromagnetic Radiation},
  journal       = {The Astrophysical Journal},
  volume        = {193},
  pages         = {443--461},
  year          = {1974},
  doi           = {10.1086/153180}
}

@article{PhysRevD.56.3395,
  title = {Dynamics of perturbations of rotating black holes},
  author = {Krivan, William and Laguna, Pablo and Papadopoulos, Philippos and Andersson, Nils},
  journal = {Phys. Rev. D},
  volume = {56},
  issue = {6},
  pages = {3395--3404},
  numpages = {0},
  year = {1997},
  month = {Sep},
  publisher = {American Physical Society},
  doi = {10.1103/PhysRevD.56.3395},
  url = {https://link.aps.org/doi/10.1103/PhysRevD.56.3395},
  eprint = {gr-qc/9702048},
  archivePrefix = {arXiv},
  primaryClass = {gr-qc}
}

@article{csukasNumericalInvestigationDynamics2019,
  author  = {Csuk\'as, K\'aroly and R\'acz, Istv\'an and T\'oth, G\'abor Zsolt},
  title   = {Numerical investigation of the dynamics of linear spin $s$ fields on a {Kerr} background: Late-time tails of spin $s=\pm1,\pm2$ fields},
  journal = {Phys. Rev. D},
  volume  = {100},
  pages   = {104025},
  year    = {2019},
  doi     = {10.1103/PhysRevD.100.104025},
  eprint  = {1905.09082},
  archivePrefix = {arXiv},
  primaryClass = {gr-qc}
}

\end{document}